\documentclass[12pt]{article}
\usepackage{epsfig,amsmath}                                           

\begin{document}
 
\vspace*{-0.50in}
\begin{flushright}
Fermilab Pub-01/033-T \\
hep-ph/0104208 \\
April~2001 \\
\end{flushright}
\vspace*{0.90in}
\begin{center}
{\Large{\bf Excited Heavy-Light Systems and Hadronic Transitions}} \\
\vspace*{.50in}
{\large{M. Di Pierro and E. Eichten \\
\vspace*{.15in}
Fermilab, Batavia, IL 60510, USA }}\\
\end{center}
\vspace*{.45in}
\begin{abstract}
A detailed study of orbital and radial excited 
states in $D, D_s, B$ and $B_s$ systems is performed.
The chiral quark model provides the framework for the
calculation of pseudoscalar meson 
($\pi, K, ...$) hadronic transitions
among heavy-light excited and ground states.
To calculate the excited states masses and wavefunctions,
we must resort to a relativistic quark model. 
Our model includes the leading order corrections 
in $1/m_{(c,b)}$ (e.g. mixing).
Numerical results for masses and 
light hadronic transition rates are 
compared to existing experimental data.
The effective coupling of the 
chiral quark model can be determined by comparing 
with independent results from lattice simulations 
($g^8_A = 0.53 \pm 0.11$) or fitting to known
widths ($g^8_A = 0.82 \pm 0.09$).

\end{abstract}    

\newpage
\def\sla{/\!\!\!}

\section{Introduction}

Although heavy quark spectroscopy is now a rather 
mature subject, a number of interesting issues remain.
In  particular, the detailed properties of the
excitation spectrum of heavy-light mesons 
($D, D_s, B, B_s$) and their light hadronic transitions
are yet to be fully understood.  
Experimentally, much of this excitation spectrum remains 
to be observed. Only the ground state S-waves  
and a few of the $j_l = 3/2$ P-waves are presently well established.
However, many of these states will be accessible in the present 
and future B factories: CLEO, BaBar, Belle, CDF, D0, BTeV and LHC~B.
Besides furthering our understanding of QCD dynamics, 
the detailed study of these excited states may have
practical benefits.  For example, 
triggering on excited states may provide an efficient method of 
same side B tagging in hadron colliders.   
Tagging is essential to the study of CP violation in the 
B system.

The theoretical tools available to determine the 
properties of 
excited states in heavy-light systems
include Heavy Quark Effective Theory (HQET) and
low energy chiral effective theory~\cite{ChengBurdman}.
Unfortunately, these tools are not sufficient to determine 
the detailed properties of these states.
Lattice gauge theory is the only existing technique 
which allows the systematic study of all the aspects of QCD 
in heavy-light systems.  
Detailed studies of the P-wave excited heavy-light states 
within the quenched approximation already exist~\cite{latHQ}.
Future lattice studies will provide more insight 
into the nature of QCD dynamics as well as the masses
and static properties of heavy-light hadrons.

It is clear that a model to estimate the hadronic transitions
from excited state to ground states would also be very useful.
Such a formalism has been developed and applied
extensively to transitions in heavy-heavy
$(\overline Q Q')$ systems~\cite{hadtrans}.
However, heavy-light mesons are more difficult because the
light quark is subject to the full nonperturbative QCD dynamics.
One possible approach to providing a framework 
for these hadronic transitions is to use the chiral quark model~\cite{GM84}. 
This has been suggested by Goity and Roberts~\cite{GR}.

In this paper we closely follow the work done in 
Refs.~\cite{GR, roberts, EHQ}.
To compute the masses and wavefunctions of the excited
states we use a Dirac equation for the light quark  
in the potential generated by 
the heavy quark (including first order corrections in the heavy quark 
expansion and mixing effects).  We then use these masses and 
wavefunctions to compute the hadronic decay amplitudes of 
excited heavy mesons in the context of a chiral quark 
model~\cite{GM84}.  

The main differences between the present and preceding works are in the
choice for the parameters of the chromoelectric potential 
and in the inclusion of mixing effects both in the spectrum and 
in the decay amplitudes. 
Moreover we use our results for the radial wavefunctions of 
the excited mesons to
make a comparison with recent lattice results. 
From the comparison we extract an estimation for $g_A^8$,
the effective coupling of the quark to the pseudoscalar mesons.
We find $g_A^8 = 0.53 \pm 0.11$

We present numerical results for the low-lying spectrum (excited 
states up to the 3S states). We also compute the pseudoscalar 
meson hadronic transitions for these states as a function of the
chiral quark model effective coupling constant. 
Comparing our results with recent experimental width measurements we
estimate this effective coupling $g^8_A = 0.82 \pm 0.09$. 

In Section 2, we discuss our determination of the spectrum of excited states.
Our notation, the choice of the potential, inclusion of mixing and 
other order $1/m_h$ corrections are explained.  
Details of the masses and wavefunctions 
are presented for the low-lying excitation spectrum.
Comparison is made with present experimental data.
Our treatment of hadronic decays is described in Section 3.
The analytic results are summarized in 
Eqs.~(\ref{amplitude}-\ref{use_amplitude}). 
Explicit expressions for the coupling coefficients appearing in these 
equations are given in Appendix~A.  
Also in Section 3, details of the partial rates for the 1S and 1P states in
the 
$D, D_s, B$ and $B_s$ systems are presented.
Again comparison is made with present experimental data.
A complete list of remaining results for masses and
partial decay widths is reported in Appendix~B.

\section{Spectrum}

\subsection{Basic Model and Notation}

The general Hamiltonian of the heavy-light system can be 
expanded in powers of ($1/m_h$) 
\begin{equation}
{\cal H}={\cal H}^{(0)}+\frac 1{m_h}{\cal H}^{(1)}+\frac 1{m_h^2}{\cal
H}^{(2)}
  +...  \label{expansion}
\end{equation}
However, even within the heavy quark limit, 
the general form of the zeroth order Hamiltonian, ${\cal H}^0$, still involves
the full nonperturbative QCD dynamics for the remaining degrees of freedom 
(including light quark pair creation and gluonic degrees of freedom). At 
present it can not be solved analytically. We 
are forced to resort to use a relativistic potential model for 
${\cal H}^0$.

We model the most general heavy-light meson 
(in the ${D, D_s, B, B_s}$ family), H, as a bound
state of a light quark ($q$) and a heavy quark ($h$). The heavy quark is
treated as a static source of chromoelectric field and the only
quantum number associated with it is its spin. The light quark is treated
relativistically and its state is described by the wavefunction 
$\psi_{n,\ell ,j,m}(r,\theta ,\varphi )$.  In analogy with the hydrogen atom,
we introduce the following quantum numbers:

\begin{itemize}
\item  $n$, the number associated with the radial excitations;

\item  $\ell $, the orbital angular momentum;

\item  $j$, the total angular momentum of the light quark;

\item  $m$, the component of $j$ along the $\widehat{z}$
axis;

\item  $J$, the total angular momentum of the system;

\item  $M$, the component of $J$ along the $\widehat{z}$
axis;

\item  $S$, the spin of the heavy quark along the 
$\widehat{z}$ axis;
\end{itemize}

The parameters of our model are the masses of the light 
quarks ($m_q$ for $q = u, d {\rm ~or~} s$), the masses of the heavy quarks ($m_h$ for $h = c
{\rm ~or~}b$) and the chromoelectric potential of the heavy quark ($V(r)$).

The total wavefunction of the system can be decomposed as follows

\begin{equation}
\Psi _{n,\ell ,j,J,M}(r,\theta ,\varphi )=\sum_{S\in \left\{ -\frac{1}{2},+%
\frac{1}{2}\right\} }C_{j,m;\frac{1}{2},S}^{J,M}\psi _{n,\ell ,j,m}(r,\theta
,\varphi ) \otimes \xi _{S}  \label{notation0}
\end{equation}
where $C_{j,m;\frac{1}{2},S}^{J,M}$ are the usual Clebsh-Gordan coefficients
and $\xi _{S}$ is a two component spinor representing the heavy quark. 
Eq.~(\ref {notation0}) is a solution of the following eigenvalue problem

\begin{equation}
{\cal H}\Psi _{n,\ell ,j,J,M}=E_{n,\ell ,j,J}\Psi _{n,\ell ,j,J,M}
\label{dirac}
\end{equation}
where ${\cal H}$ is the Hamiltonian of the system. The energy levels in
Eq.~(\ref{dirac}) do not depend on $M$ because of rotational invariance.

We rewrite Eq.~(\ref{notation0}) introducing the most general parameterization
for the four spin components of the light quark wavefunction

\begin{equation}
\Psi _{n,\ell ,j,J,M}(r,\theta ,\varphi )=\hspace{-0.1cm}\sum_{S\in \left\{
-\frac 12,+\frac 12\right\} }C_{j,m;\frac 12,S}^{J,M}\left( 
\begin{array}{l}
if_{n,\ell ,j}^0(r)k_{\ell ,j,m}^{+}\,\,Y_{m-\frac 12}^\ell (\theta ,\varphi )
\\ 
if_{n,\ell ,j}^0(r)k_{\ell ,j,m}^{-}\,\,Y_{m+\frac 12}^\ell (\theta ,\varphi )
\\ 
\,f_{n,\ell ,j}^1(r)k_{2j-\ell ,j,m}^{+}Y_{m-\frac 12}^{2j-\ell }(\theta
,\varphi ) \\ 
\,f_{n,\ell ,j}^1(r)k_{2j-\ell ,j,m}^{-}Y_{m+\frac 12}^{2j-\ell }(\theta
,\varphi )
\end{array}
\right) \otimes \xi _S  \label{notation}
\end{equation}
Here $Y_m^\ell (\theta ,\varphi )$ are spherical harmonics that encode the
angular dependence while $f_{n,\ell ,j}^0(r)$, $f_{n,\ell ,j}^1(r)$ are real
functions that encode that radial dependence. $k_{\ell ,j,m}^{+}$ and $%
k_{\ell ,j,m}^{-}$ are fixed, up to an overall phase, by imposing a
normalization condition. Our choice of the phase is such that 
\begin{equation}
k_{\ell ,j,m}^{\pm }=\left\{ 
\begin{array}{ll}
+\sqrt{\frac{\ell \pm m+\frac 12}{2\ell +1}} & \text{for }j=\ell +\frac12 \\ 
\pm \sqrt{\frac{\ell \mp m+\frac 12}{2\ell +1}} & \text{for }j=\ell -\frac12
\end{array}
\right. \label{kappas}
\end{equation}

\subsection{Choice of the potential}

Within our basic framework,  
${\cal H}^{(0)}$ is given by the relativistic Dirac Hamiltonian 
\begin{equation}
{\cal H}^{(0)}=\gamma ^0(-i \sla \partial +m_q)+V(r)  \label{dirac2}
\end{equation}
and the rotational-invariant potential is the sum of a constant factor ($M_h$%
), a scalar part ($V_s$) and (the zeroth component of) a vector part ($V_v$) 
\begin{equation}
V(r)=M_h+\gamma ^0V_s(r)+V_v(r)  \label{potential}
\end{equation}
The constant $M_h$ is a an overall energy shift that depends on the heavy
quark flavor and, in general, it is not equal to $m_h$,
as often assumed in the literature.  For this reason we consider $m_h$ and $%
M_h$ two independent parameters of the model.

Asymptotic freedom suggests that at short distances the potential is
dominated by a vector part that asymptotically approaches a
Coulomb potential, $V\sim V_v\sim 1/r$. On the other hand lattice
simulations indicate that at large distances the potential is confining,
scalar and asymptotically linear, $V\sim V_s\sim r$.

The naive assumption about a short distance Coulomb-like divergent behavior
of the potential is doomed because it gives rise to ultraviolet divergences
(as discussed in Ref.~\cite{isgur} and Ref.~\cite{lepage}). It this context
the divergence arises in the $1/m_h$ correction to the energy and it is due
to the inconsistency of a static point-like source (the heavy quark) within a
relativistic framework. One solution is assuming that the heavy quark is
static but not point-like, therefore the potential that it generates is a
convolution of the Coulomb-like potential and the square of the heavy quark
wavefunction (peaked around the center of mass of the system and smeared
within some small length scale $\lambda ^{-1}$).

More generally, one is allowed to cure this divergence by regulating the
potential close to the origin (on a length scale of the order $\lambda ^{-1}$%
). Different choices for the regulator are allowed and they do not affect
the physics we want to describe, providing that $\lambda ^{-1}$ is small
enough. The values of the parameters that appear in the Hamiltonian, on the
contrary, depend on this choice since they run with $\lambda $. In fact, to
obtain the same spectrum, different choices for the regulator imply
different fitting parameters.

We chose to regulate the vector potential by assuming a Gaussian
shape for the wavefunction of the heavy quark, $\Phi (x)=\exp (-x^2\lambda
^2/2)$, and with this choice 
\begin{equation}
V_v(r)=-\frac43 \int \left| \Phi (x)\right| ^2\frac{\alpha _s}
{\left| \mathbf{r}-%
\mathbf{x}\right| }\text{d}^3x=-\frac43\frac{\alpha _s}r\text{erf}(\lambda r)
\label{potential1}
\end{equation}
For the scalar potential we assume a simple linear form 
\begin{equation}
V_s(r)=br+c  \label{potential2}
\end{equation}
We observe that $c$ is not a physical parameter since it can be absorbed
into the definition of $m_q$. For this reason $c$ will be omitted from now
on.

Summarizing, the nine parameters of our model are 
\begin{equation}
\alpha _s,\lambda ,b,m_u,m_s,m_c,M_c,m_b,M_b  \label{params}
\end{equation}
where $m_u \equiv m_d$ and $m_s$ are mass parameters for the light $u,d$ 
and $s$ quarks respectively, equivalent to constituent quark masses 
shifted by the constant amount $c$ of eq.~(\ref{potential2}), which is
undetermined in our model. $m_c$ is the mass of the $c$ quark with $M_c$ the
corresponding energy shift. Analogously for the $b$ quark.

\subsection{$1/m_{h}$ correction}

For any given set of input parameters we solve the eigenvalue 
problem, Eq.~(\ref{dirac}), using the
Hamiltonian of Eq.~(\ref{dirac2}) and the potential specified by Eqs.~(\ref
{potential}), (\ref{potential1}) and (\ref{potential2}). In this way we
determine the radial wavefunctions $f_{n,\ell ,j}^0$ and 
$f_{n,\ell ,j}^1$
associated to the energy levels $E_{n,\ell ,j}^{(0)}$. We then compute 
$1/m_h$ corrections to the energy levels 
in first order perturbation theory 
\begin{equation}
E_{n,\ell ,j,J}=E_{n,\ell ,j}^{(0)}+\frac 1{m_h}\delta E_{n,\ell ,j,J}^{(1)}
\label{e1}
\end{equation}
where, ignoring for the moment the mixing of the states, 
\begin{equation}
\delta E_{n,\ell ,j,J}^{(1)}=\sum_M\int \Psi _{n,\ell
,j,J,M}^{\dagger}(x){\cal H}^{(1)}\Psi _{n,\ell ,j,J,M}(x)\text{d}^3x  \label{e2}
\end{equation}
The analytical expression for ${\cal H}^{(1)}$ has been derived in Ref.~\cite
{roberts} using the Bethe-Salpeter formalism. In terms of the radial 
wavefunctions (after the analytical integration of the angular part), 
we rewrite $\delta E^{(1)}$ as a sum of three contributions 
\begin{equation}
\delta E_{n,\ell ,j,J}^{(1)}=A_{n,\ell ,j,J}+B_{n,\ell ,j,J}+C_{n,\ell ,j,J}
\label{e3}
\end{equation}
These terms are respectively:

\begin{itemize}
\item  The kinetic energy
\begin{eqnarray}
A_{n,\ell ,j,J}=-\frac12 \int_0^\infty \Big[ && \!\!\!\!\!\!\!\!
f_{n,\ell ,j}^0\left( \partial
_r^2+\frac 2r\partial _r-\frac{\ell ^2+\ell }{r^2}\right) f_{n,\ell ,j}^0 \\
&& \!\!\!\!\!\!\!\!
f_{n,\ell ,j}^1\left( \partial
_r^2+\frac 2r\partial _r-\frac{\bar \ell ^2+\bar \ell }{r^2}\right) f_{n,\ell
,j}^1 
\Big] r^2 \text{d}r  \label{i1}
\end{eqnarray}
with $\bar \ell = 2j-\ell$.

\item  A shift due to spin-orbit interaction
\begin{equation}
B_{n,\ell ,j,J}=\!\! \int_0^\infty \!\!\! V_v \!\! \left[ f_{n,\ell ,j}^1\left( \partial_r \! - \! \frac \ell r\right) f_{n,\ell ,j}^0 \!-\! f_{n,\ell ,j}^0\left( \partial _r \! + \! \frac{\ell +2}r\right) f_{n,\ell ,j}^1\right] \! r^2\text{d}r  \label{i2}
\end{equation}
for $j=\ell +\frac 12$, or 
\begin{equation}
B_{n,\ell ,j,J}=\!\! \int_0^\infty \!\!\! V_v \!\! \left[ f_{n,\ell ,j}^1\left( \partial _r \! + \! \frac{\ell +1}r\right) f_{n,\ell ,j}^0 \!-\! f_{n,\ell ,j}^0\left( \partial _r \!-\!\frac{\ell -1}r\right) f_{n,\ell ,j}^1\right] \! r^2\text{d}r  \label{i3}
\end{equation}

for $j=\ell -\frac 12$.

\item  The hyperfine splitting  
\begin{equation}
C_{n,\ell ,j,J}=\left( -1\right) ^{J-\ell}\frac{2j+1}{2J+1}\int_0^\infty
(\partial _rV_v)f_{n,\ell ,j}^0f_{n,\ell ,j}^1~r^2\text{d}r  \label{i4}
\end{equation}
\end{itemize}

\subsection{Mixing}

In Eqs.(\ref{e1}) and (\ref{e2}) we assumed that the Hamiltonian was
diagonal. This is not the case because the $\frac 1{m_h}$ interaction term
in the Hamiltonian mixes states. In general correction terms can mix 
any states with the same total angular momentum, $J$, and parity, $P$,
However, there are only two types of sizable mixings. 
Large mixing can occur for pairs of states $\Psi _{n,\ell ,j,J,M}$ 
and $\Psi _{n^{\prime},\ell^{\prime},j^{\prime },J,M}$ with: (1)
$n=n'$, $\ell=\ell^{\prime}$ and $j+\frac12=j'-\frac12=\ell=J$ (i.e.
mixing within a given $n$ and $l$ multiplet); or 
(2) $n+1=n'$, $\ell +2=\ell^{\prime}$ and $j+\frac12=j'-\frac12=\ell=J$ 
(e.g. S-D mixing).
The off-diagonal term in the Hamiltonian that mixes such pairs of states
has the form 
\begin{eqnarray}
\epsilon \!\! &=& \!\!\! \frac 1{m_h}\sum_M\int \Psi^\dagger _{n,\ell
,j,J,M}(x){\cal H}^{(1)}\Psi
_{n^{\prime},\ell^{\prime} ,j^{\prime },J,M}(x)\text{d}^3x  \label{delta} \\
&=& \!\!\! \left( -1\right) ^{J-\ell}\frac 1{m_h}\frac{\sqrt{J (J +1)}}{2J +1}
 \!\! \int_0^\infty \!\!\! (\partial _rV_v)
    \left[ f_{n,\ell ,j}^0f_{n^{\prime},\ell^{\prime},j^{\prime }}^1+f_{n,\ell
,j}^1
           f_{n^{\prime},\ell^{\prime},j^{\prime }}^0\right] \! r^2\text{d}r
\nonumber
\end{eqnarray}
and it induces a mixing in the wavefunctions and in the energy levels
\begin{eqnarray}
\left( 
\begin{array}{l}
\psi_{n,\ell,j,m} \\ 
\psi_{n',\ell^{\prime},j',m}
\end{array}
\right)^{\text{phys.}} &=&
\left( 
\begin{array}{cc}
1 & +\frac{\epsilon}{2\Delta} \\
-\frac{\epsilon}{2\Delta} & 1 
\end{array}
\right)
\left( 
\begin{array}{l}
\psi_{n,\ell,j,m} \\ 
\psi_{n',\ell^{\prime},j',m}
\end{array}
\right) + {\mathcal O}(\epsilon^2) \\
\left( 
\begin{array}{l}
E_{n,\ell ,j,J} \\ 
E_{n',\ell^{\prime} ,j^{\prime },J}
\end{array}
\right)^{\text{phys.}} &=& \left( 
\begin{array}{l}
E_{n,\ell ,j,J}+\frac{\epsilon^2}{2\Delta} \\
E_{n',\ell^{\prime} ,j^{\prime },J}-\frac{\epsilon^2}{2\Delta} 
\end{array}
\right) +{\mathcal O}(\epsilon^2)   \label{mixing}
\end{eqnarray}
with $\Delta =(E_{n,\ell ,j,J}-E_{n',\ell^{\prime},j',J})/2$. 

We find that the effect of mixing is generally 
negligible except for the $P$ waves, where the mixing among 
wavefunctions can be of the order of 10\%. 
In Tables~\ref{table_spectrum1}-\ref{table_spectrum4} 
we report the value of $\phi = \frac{100}2 \epsilon/\Delta$ for each
excited state. It measures, in percent, the contribution of the mixing to
the wavefunction.


\begin{table}
\begin{center}
\begin{tabular}{|lrrrr|} \hline
$H~(n^j L_J)$ & $m_{\text{exp.}}$ & $E^0$ & $E^{\text{phys.}}$ & $\phi(\%)$
\\ 
\hline
$D~(1^{\frac12}S_0)$ & $1.865$ & $1.895$ & $1.868$ &        \\ 
$D~(1^{\frac12}S_1)$ & $2.007$ & $1.895$ & $2.005$ &        \\ 
$D~(1^{\frac12}P_0)$ &         & $2.282$ & $2.377$ &        \\ 
$D~(1^{\frac32}P_1)$ & $2.422$ & $2.253$ & $2.417$ & -10.92 \\ 
$D~(1^{\frac32}P_2)$ & $2.459$ & $2.253$ & $2.460$ &        \\ 
$D~(1^{\frac12}P_1)$ &         & $2.282$ & $2.490$ &  10.92 \\ 
$D~(2^{\frac12}S_0)$ &         & $2.447$ & $2.589$ &        \\ 
$D~(2^{\frac12}S_1)$ &         & $2.447$ & $2.692$ &  2.17  \\ 
$D~(1^{\frac52}D_2)$ &         & $2.504$ & $2.775$ & -5.41  \\ 
$D~(1^{\frac32}D_1)$ &         & $2.553$ & $2.795$ & -2.17  \\ 
$D~(1^{\frac52}D_3)$ &         & $2.504$ & $2.799$ &        \\ 
$D~(1^{\frac32}D_2)$ &         & $2.553$ & $2.833$ &  5.41  \\ 
$D~(2^{\frac12}P_0)$ &         & $2.683$ & $2.949$ &        \\ 
$D~(2^{\frac32}P_1)$ &         & $2.679$ & $2.995$ & -10.70 \\ 
$D~(2^{\frac32}P_2)$ &         & $2.679$ & $3.035$ &  1.79  \\ 
$D~(2^{\frac12}P_1$ &         & $2.683$ & $3.045$ &  10.70 \\ 
$D~(1^{\frac72}F_3)$ &         & $2.709$ & $3.074$ & -3.17  \\ 
$D~(1^{\frac72}F_4)$ &         & $2.709$ & $3.091$ &        \\ 
$D~(1^{\frac52}F_2)$ &         & $2.760$ & $3.101$ & -1.79  \\ 
$D~(1^{\frac52}F_3)$ &         & $2.760$ & $3.123$ &  3.17  \\ 
$D~(3^{\frac12}S_0)$ &         & $2.823$ & $3.141$ &        \\ 
$D~(3^{\frac12}S_1)$ &         & $2.823$ & $3.226$ &        \\ 
\hline \end{tabular}
\end{center}
\caption{Tabulated spectrum for $D$ mesons.
$E^0$ denotes the lowest order energies. $E^{\rm phys.}$ includes all
the order $1/m_h$ corrections.  (All units in GeV).
The mixings between lowest order states
are denoted by $\phi$.
\label{table_spectrum1}}
\end{table}

\begin{table} \begin{center}
\begin{tabular}{|lrrrr|} \hline
$H~(n^j L_J)$ & $m_{\text{exp.}}$ & $E^0$ & $E^{\text{phys.}}$ & $\phi(\%)$
\\ 
\hline
$D_s~(1^{\frac12}S_0)$ & $1.969$ & $1.988$ & $1.965$ &        \\
$D_s~(1^{\frac12}S_1)$ & $2.112$ & $1.988$ & $2.113$ &        \\ 
$D_s~(1^{\frac12}P_0)$ &         & $2.374$ & $2.487$ &        \\ 
$D_s~(1^{\frac32}P_1)$ & $2.535$ & $2.353$ & $2.535$ & -11.62 \\ 
$D_s~(1^{\frac32}P_2)$ & $2.573$ & $2.353$ & $2.581$ &        \\ 
$D_s~(1^{\frac12}P_1)$ &         & $2.374$ & $2.605$ &  11.62 \\ 
$D_s~(2^{\frac12}S_0)$ &         & $2.540$ & $2.700$ &        \\ 
$D_s~(2^{\frac12}S_1)$ &         & $2.540$ & $2.806$ &  1.97  \\ 
$D_s~(1^{\frac52}D_2)$ &         & $2.606$ & $2.900$ & -6.11  \\ 
$D_s~(1^{\frac32}D_1)$ &         & $2.648$ & $2.913$ & -1.97  \\ 
$D_s~(1^{\frac52}D_3)$ &         & $2.606$ & $2.925$ &        \\ 
$D_s~(1^{\frac32}D_2)$ &         & $2.648$ & $2.953$ &  6.11  \\ 
$D_s~(2^{\frac12}P_0)$ &         & $2.777$ & $3.067$ &        \\ 
$D_s~(2^{\frac32}P_1)$ &         & $2.775$ & $3.114$ & -10.58 \\ 
$D_s~(2^{\frac32}P_2)$ &         & $2.775$ & $3.157$ &  1.81  \\ 
$D_s~(2^{\frac12}P_1)$ &         & $2.777$ & $3.165$ &  10.58 \\ 
$D_s~(1^{\frac72}F_3)$ &         & $2.812$ & $3.203$ & -3.60  \\ 
$D_s~(1^{\frac72}F_4)$ &         & $2.812$ & $3.220$ &        \\ 
$D_s~(1^{\frac52}F_2)$ &         & $2.857$ & $3.224$ & -1.81  \\ 
$D_s~(1^{\frac52}F_3)$ &         & $2.857$ & $3.247$ &  3.60  \\ 
$D_s~(3^{\frac12}S_0)$ &         & $2.917$ & $3.259$ &        \\ 
$D_s~(3^{\frac12}S_1)$ &         & $2.917$ & $3.345$ &        \\ 
\hline \end{tabular}
\end{center}
\caption{Tabulated spectrum for $D_s$ mesons. (All units in GeV).   
Notation as in Table~\ref{table_spectrum1}.
\label{table_spectrum2}}
\end{table}

\begin{table} \begin{center}
\begin{tabular}{|lrrrr|} \hline
$H~(n^j L_J)$ & $m_{\text{exp.}}$ & $E^0$ & $E^{\text{phys.}}$ & $\phi(\%)$
\\ 
\hline
$B~(1^{\frac12}S_0)$ & $5.279$ & $5.288$ & $5.279$ &        \\
$B~(1^{\frac12}S_1)$ & $5.325$ & $5.288$ & $5.324$ &        \\ 
$B~(1^{\frac32}P_1)$ &         & $5.646$ & $5.700$ & -6.00  \\ 
$B~(1^{\frac12}P_0)$ &         & $5.675$ & $5.706$ &        \\ 
$B~(1^{\frac32}P_2)$ &         & $5.646$ & $5.714$ &        \\ 
$B~(1^{\frac12}P_1)$ &         & $5.675$ & $5.742$ &  6.00  \\ 
$B~(2^{\frac12}S_0)$ &         & $5.840$ & $5.886$ &        \\ 
$B~(2^{\frac12}S_1)$ &         & $5.840$ & $5.920$ &  0.69  \\ 
$B~(1^{\frac52}D_2)$ &         & $5.897$ & $5.985$ & -1.96  \\ 
$B~(1^{\frac52}D_3)$ &         & $5.897$ & $5.993$ &        \\ 
$B~(1^{\frac32}D_1)$ &         & $5.946$ & $6.025$ & -0.69  \\ 
$B~(1^{\frac32}D_2)$ &         & $5.946$ & $6.037$ &  1.96  \\ 
$B~(2^{\frac12}P_0)$ &         & $6.076$ & $6.163$ &        \\ 
$B~(2^{\frac32}P_1)$ &         & $6.072$ & $6.175$ & -9.11  \\ 
$B~(2^{\frac32}P_2)$ &         & $6.072$ & $6.188$ &  0.50  \\ 
$B~(2^{\frac12}P_1)$ &         & $6.076$ & $6.194$ &  9.11  \\ 
$B~(1^{\frac72}F_3)$ &         & $6.102$ & $6.220$ & -0.99  \\ 
$B~(1^{\frac72}F_4)$ &         & $6.102$ & $6.226$ &        \\ 
$B~(1^{\frac52}F_2)$ &         & $6.153$ & $6.264$ & -0.50 \\ 
$B~(1^{\frac52}F_3)$ &         & $6.153$ & $6.271$ &  0.99  \\ 
$B~(3^{\frac12}S_0)$ &         & $6.216$ & $6.320$ &        \\ 
$B~(3^{\frac12}S_1)$ &         & $6.216$ & $6.347$ &        \\ 
\hline \end{tabular}
\end{center}
\caption{Tabulated spectrum for $B$ mesons. (All units in GeV).
Notation as in Table~\ref{table_spectrum1}.
\label{table_spectrum3}}
\end{table}

\begin{table} \begin{center}
\begin{tabular}{|lrrrr|} \hline
$H~(n^j L_J)$ & $m_{\text{exp.}}$ & $E^0$ & $E^{\text{phys.}}$ & $\phi(\%)$
\\ 
\hline
$B_s~(1^{\frac12}S_0)$ & $5.369$ & $5.381$ & $5.373$ &        \\ 
$B_s~(1^{\frac12}S_1)$ & $5.417$ & $5.381$ & $5.421$ &        \\ 
$B_s~(1^{\frac12}P_0)$ &         & $5.767$ & $5.804$ &        \\ 
$B_s~(1^{\frac32}P_1)$ &         & $5.746$ & $5.805$ & -7.19  \\ 
$B_s~(1^{\frac32}P_2)$ &         & $5.746$ & $5.820$ &        \\ 
$B_s~(1^{\frac12}P_1)$ &         & $5.767$ & $5.842$ &  7.19  \\ 
$B_s~(2^{\frac12}S_0)$ &         & $5.933$ & $5.985$ &        \\ 
$B_s~(2^{\frac12}S_1)$ &         & $5.933$ & $6.019$ &  0.64  \\ 
$B_s~(1^{\frac52}D_2)$ &         & $5.999$ & $6.095$ & -2.31  \\ 
$B_s~(1^{\frac52}D_3)$ &         & $5.999$ & $6.103$ &        \\ 
$B_s~(1^{\frac32}D_1)$ &         & $6.041$ & $6.127$ & -0.64  \\ 
$B_s~(1^{\frac32}D_2)$ &         & $6.041$ & $6.140$ &  2.31  \\ 
$B_s~(2^{\frac12}P_0)$ &         & $6.170$ & $6.264$ &        \\ 
$B_s~(2^{\frac32}P_1)$ &         & $6.168$ & $6.278$ & -9.81  \\ 
$B_s~(2^{\frac32}P_2)$ &         & $6.168$ & $6.292$ &  0.51  \\ 
$B_s~(2^{\frac12}P_1)$ &         & $6.170$ & $6.296$ &  9.81  \\ 
$B_s~(1^{\frac72}F_3)$ &         & $6.205$ & $6.332$ & -1.15  \\ 
$B_s~(1^{\frac72}F_4)$ &         & $6.205$ & $6.337$ &        \\ 
$B_s~(1^{\frac52}F_2)$ &         & $6.250$ & $6.369$ & -0.51  \\ 
$B_s~(1^{\frac52}F_3)$ &         & $6.250$ & $6.376$ &  1.15  \\ 
$B_s~(3^{\frac12}S_0)$ &         & $6.310$ & $6.421$ &        \\ 
$B_s~(3^{\frac12}S_1)$ &         & $6.310$ & $6.449$ &        \\ 
\hline \end{tabular}
\end{center}
\caption{Tabulated spectrum for $B_s$ mesons. (All units in GeV).
Notation as in Table~\ref{table_spectrum1}.
\label{table_spectrum4}}
\end{table}


\begin{figure}
\begin{center}
\begin{tabular}{rr}
\epsfxsize=6.5cm
\epsfysize=7cm
\epsfbox{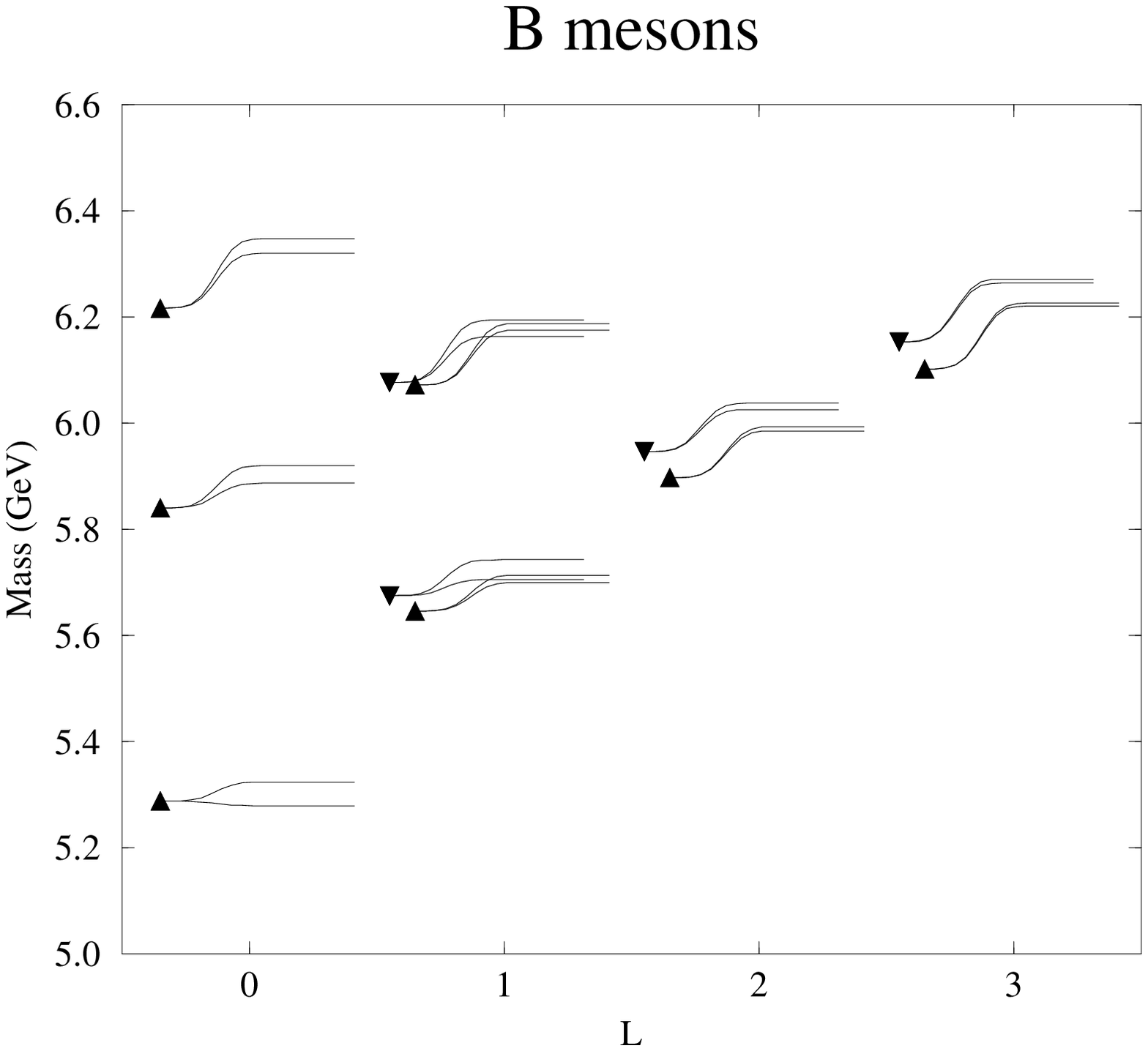} &
\epsfxsize=6.5cm
\epsfysize=7cm
\epsfbox{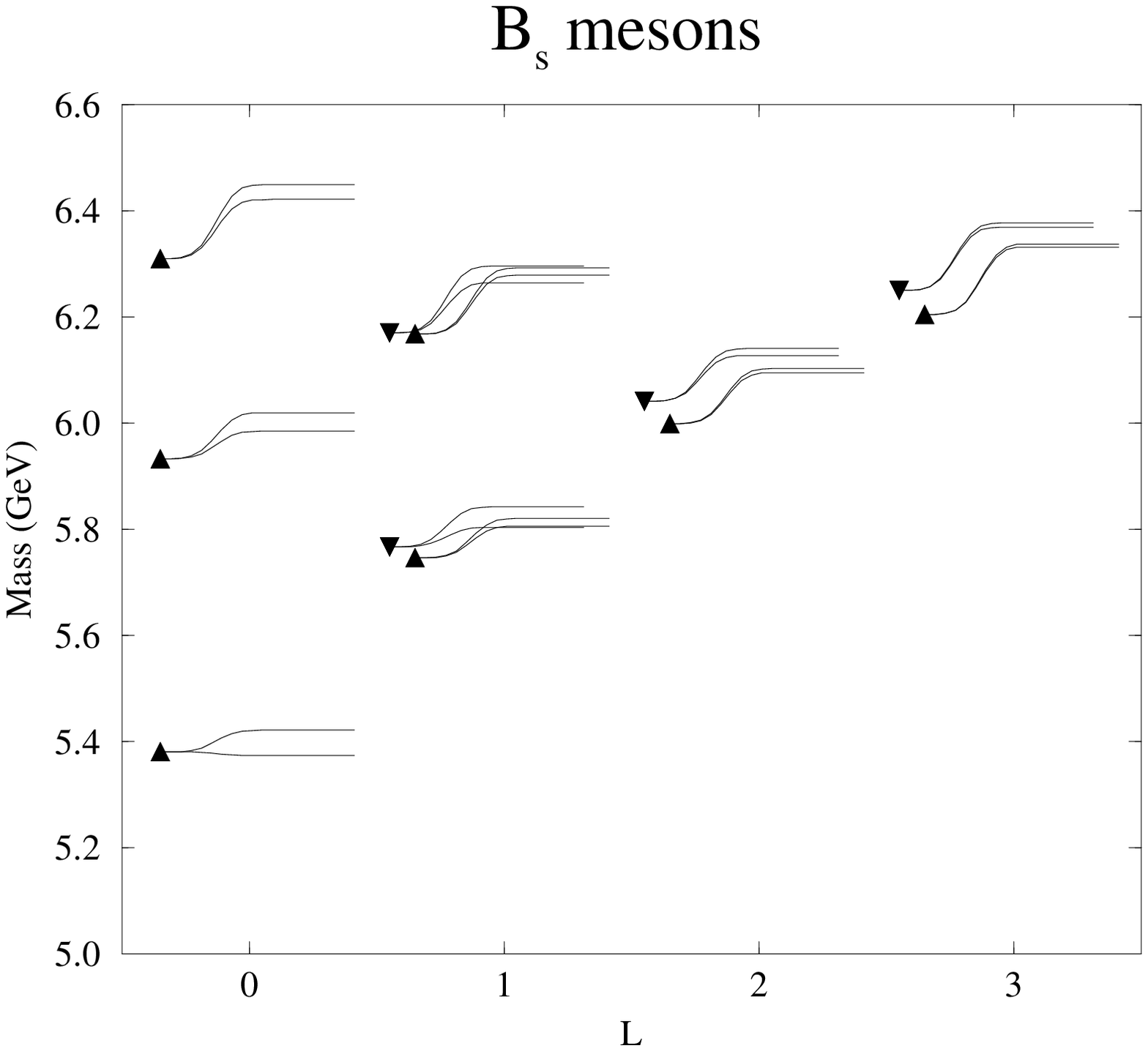} \\
\epsfxsize=6.5cm
\epsfysize=7cm
\epsfbox{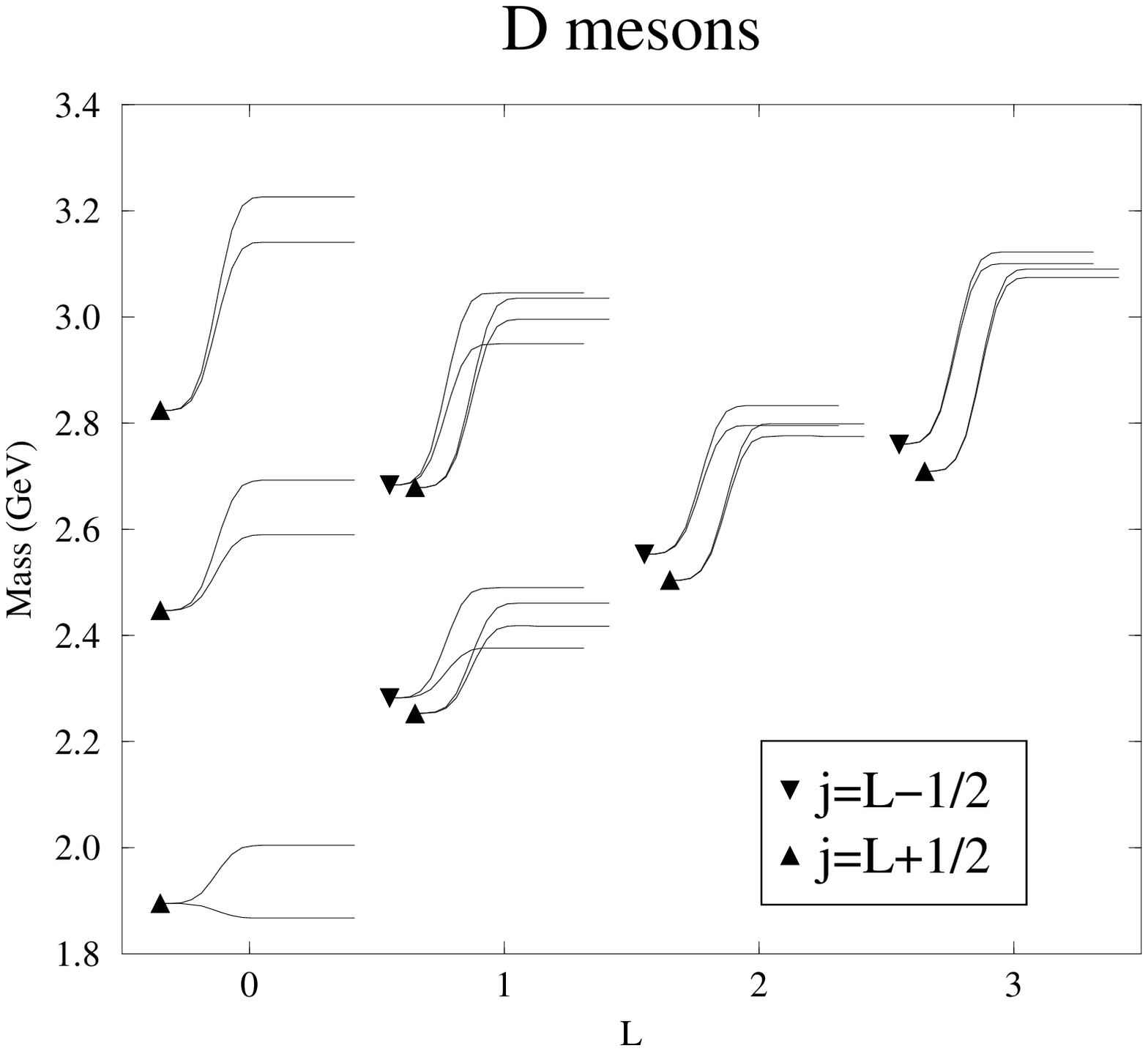} &
\epsfxsize=6.5cm
\epsfysize=7cm
\epsfbox{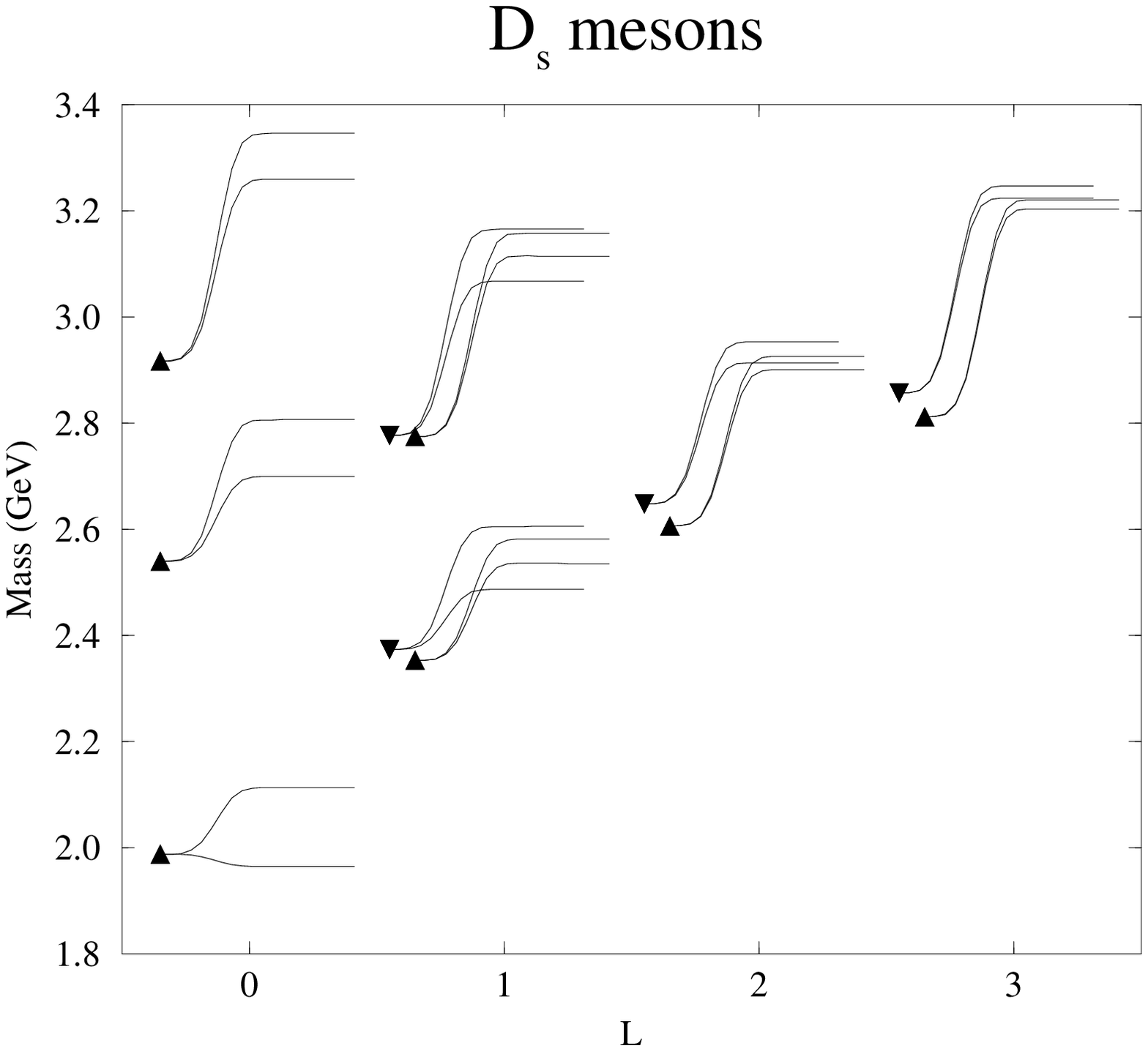} 
\end{tabular}
\end{center}
\caption{Computed spectrum of excited states in the $\{D,D_s,B,B_s\}$ family.
The plot shows the spectrum before and after $1/m_h$ corrections (including 
mixing). These corrections are responsible for the hyperfine splitting. 
The horizontal axis is the orbital angular momentum of the meson ($\ell$).
For each value of $\ell$ and $j$
there is a doublet of states ($J=j-\frac12$ and $J=j+\frac12$, 
with lower and higher energy respectively).\label{fig_spectrum}}
\end{figure}  

\begin{figure}
\begin{center}
\begin{tabular}{rrr}
\epsfxsize=4.5cm
\epsfysize=3.5cm
\hskip -5mm \epsfbox{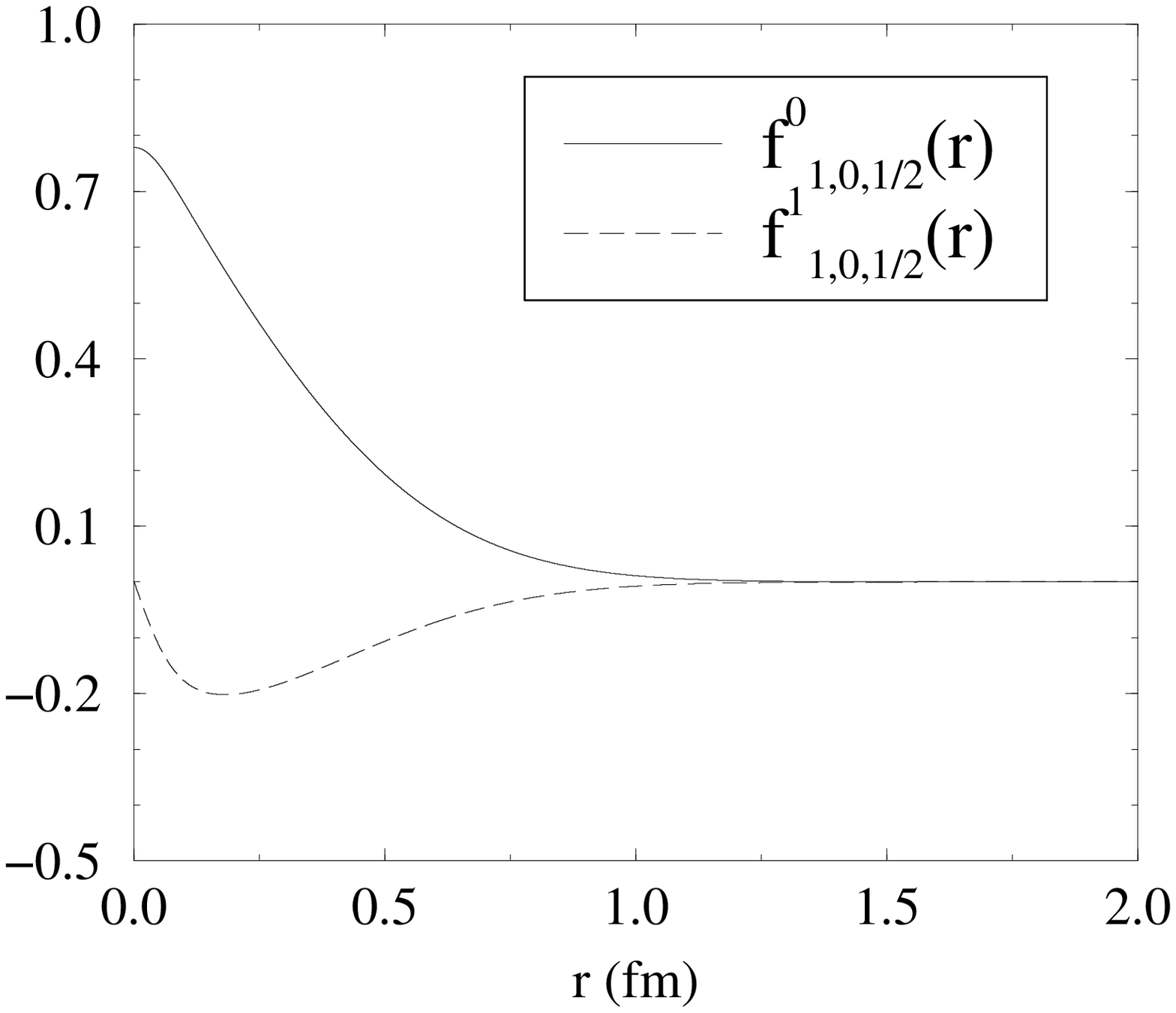} &  
\epsfxsize=4.5cm
\epsfysize=3.5cm
\hskip -5mm \epsfbox{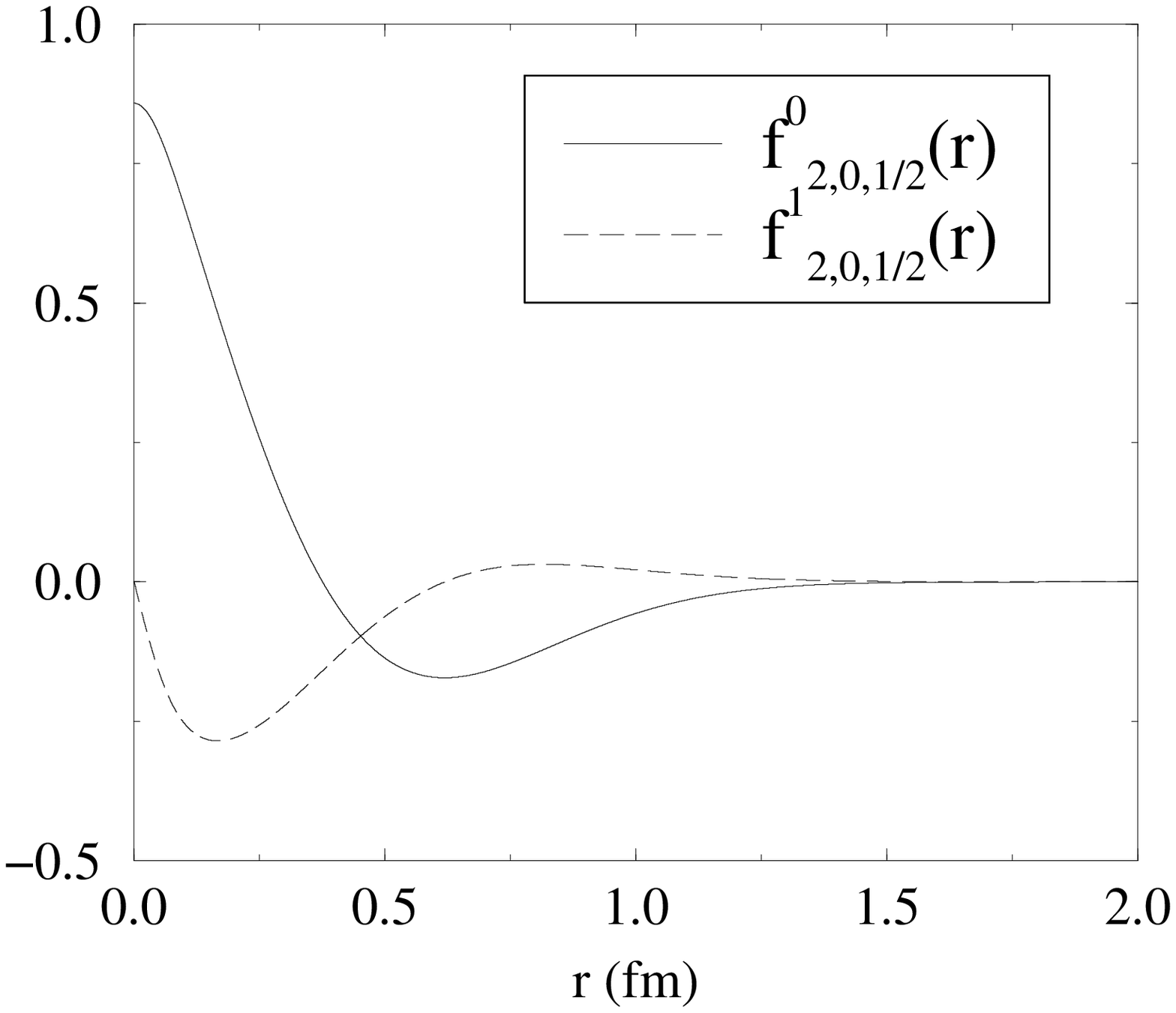} &
\epsfxsize=4.5cm
\epsfysize=3.5cm
\hskip -5mm \epsfbox{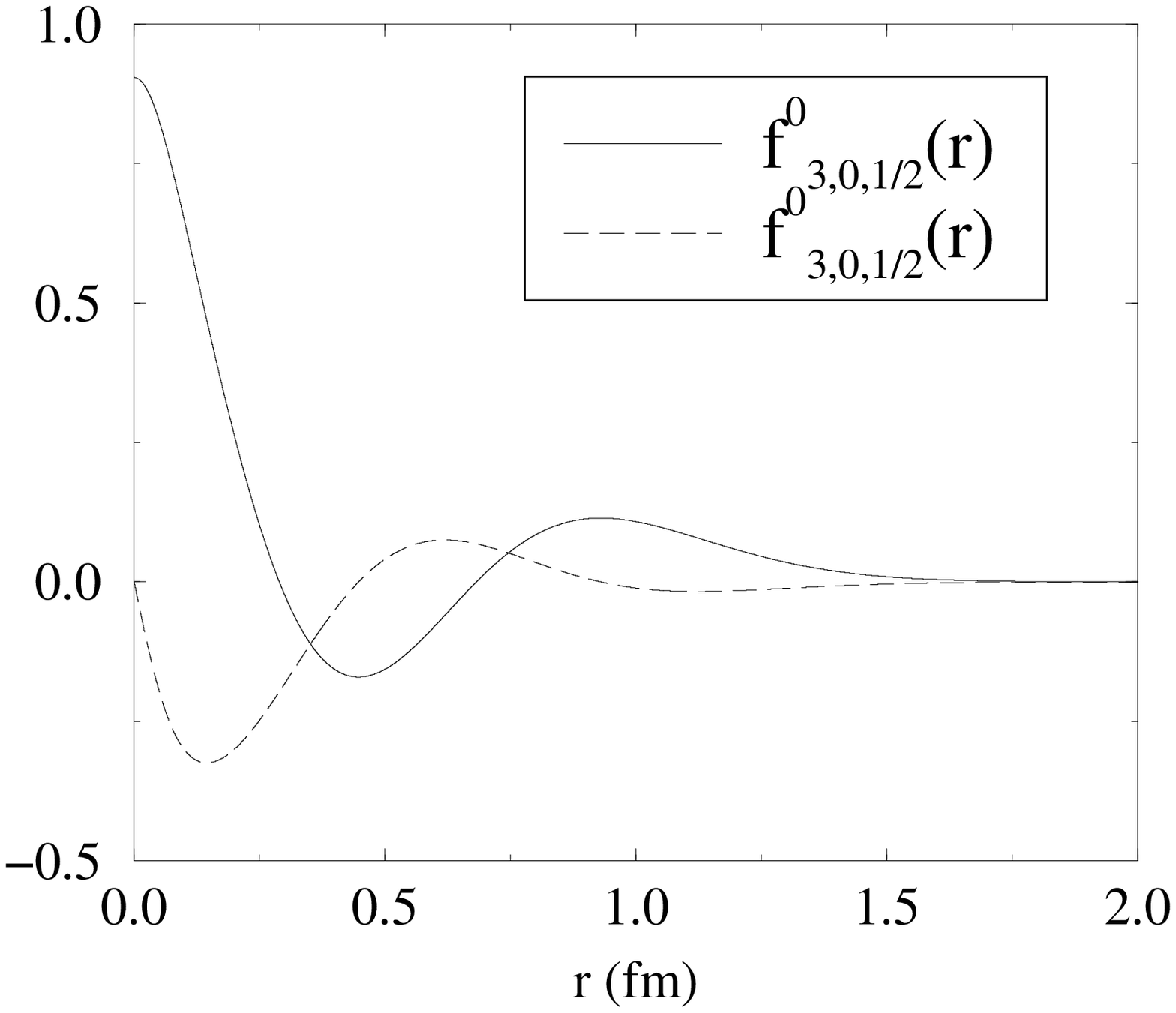} \\
\epsfxsize=4.5cm
\epsfysize=3.5cm
\hskip -5mm \epsfbox{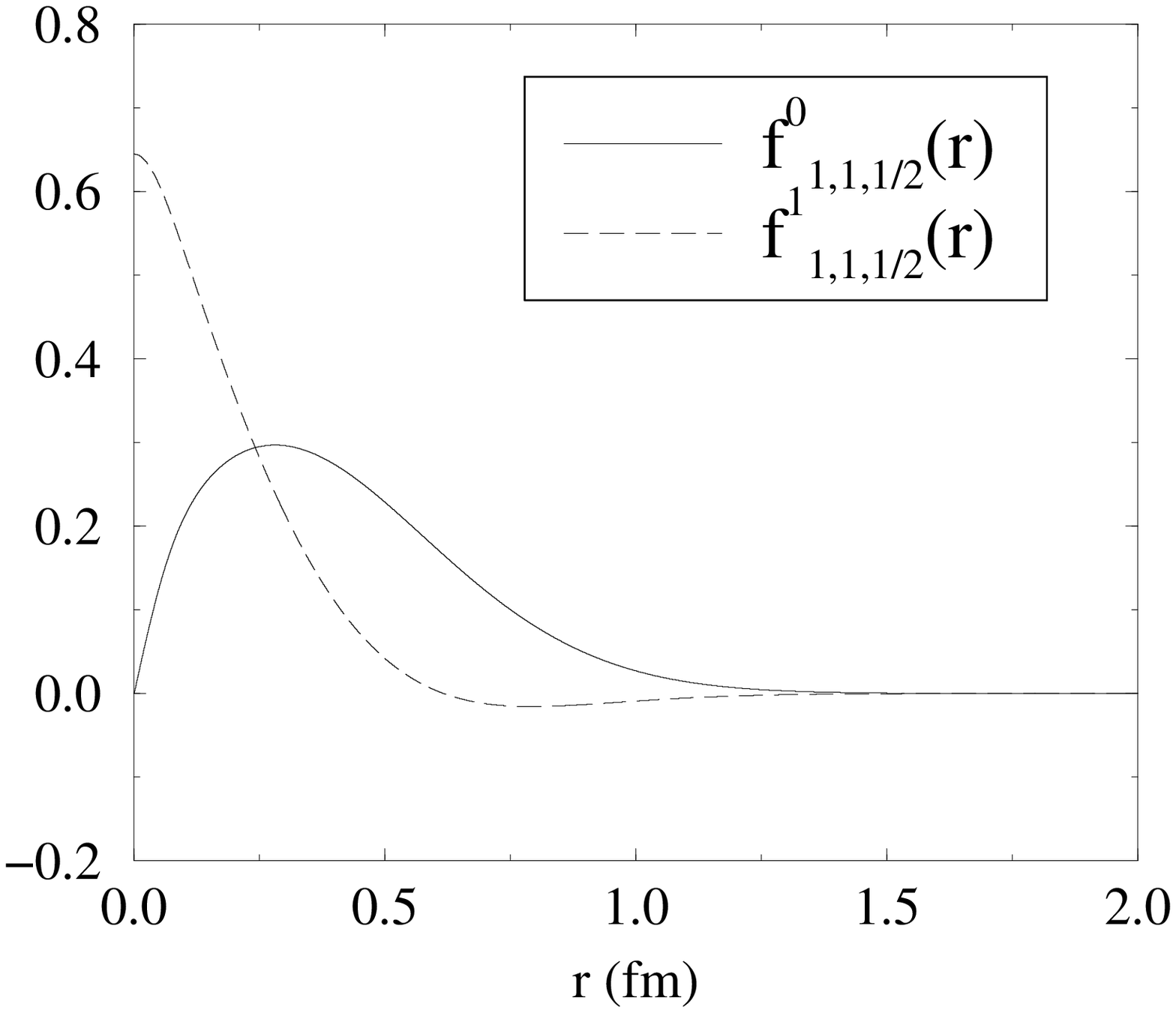} &
\epsfxsize=4.5cm
\epsfysize=3.5cm
\hskip -5mm \epsfbox{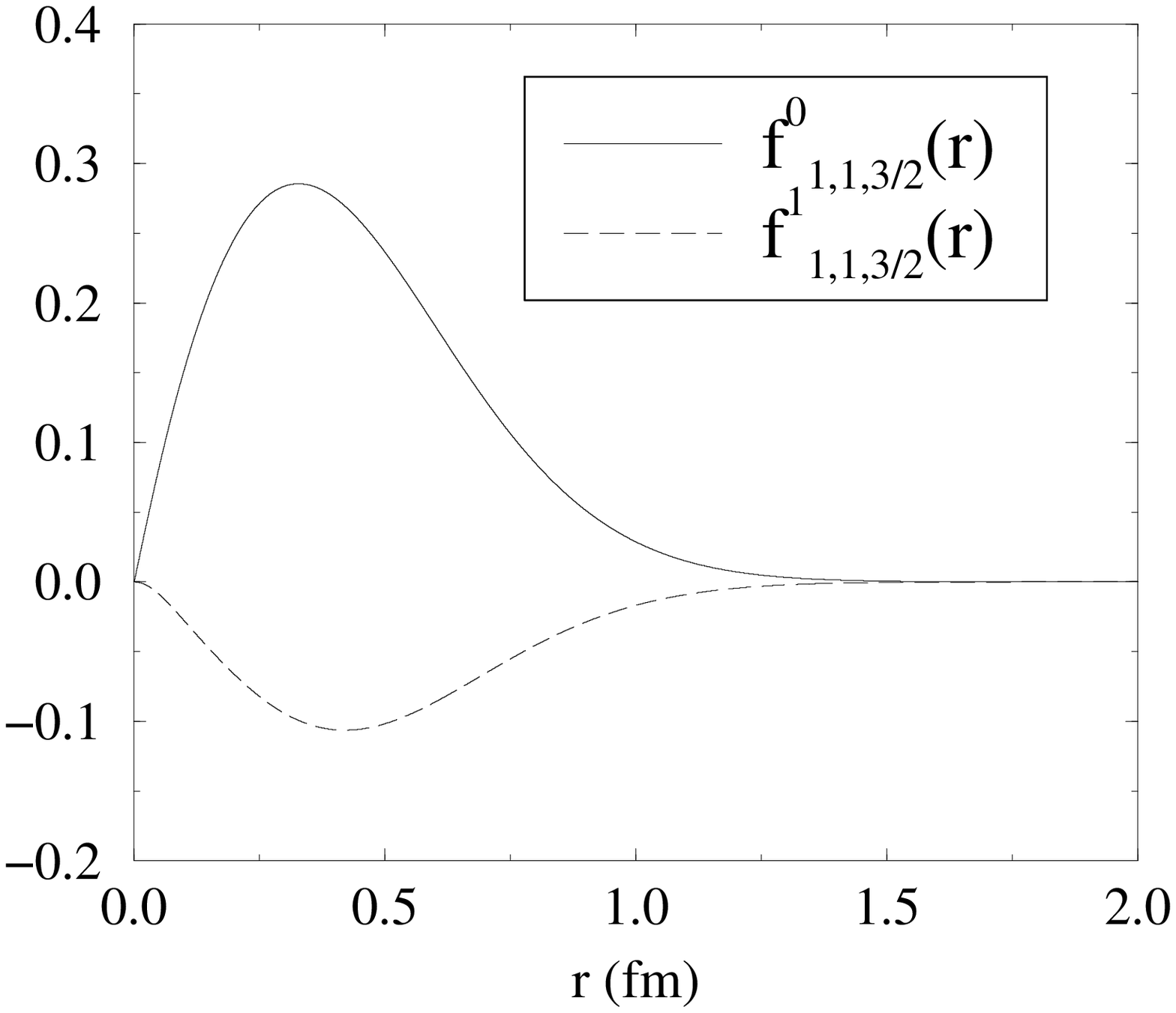} &
\epsfxsize=4.5cm
\epsfysize=3.5cm
\hskip -5mm \epsfbox{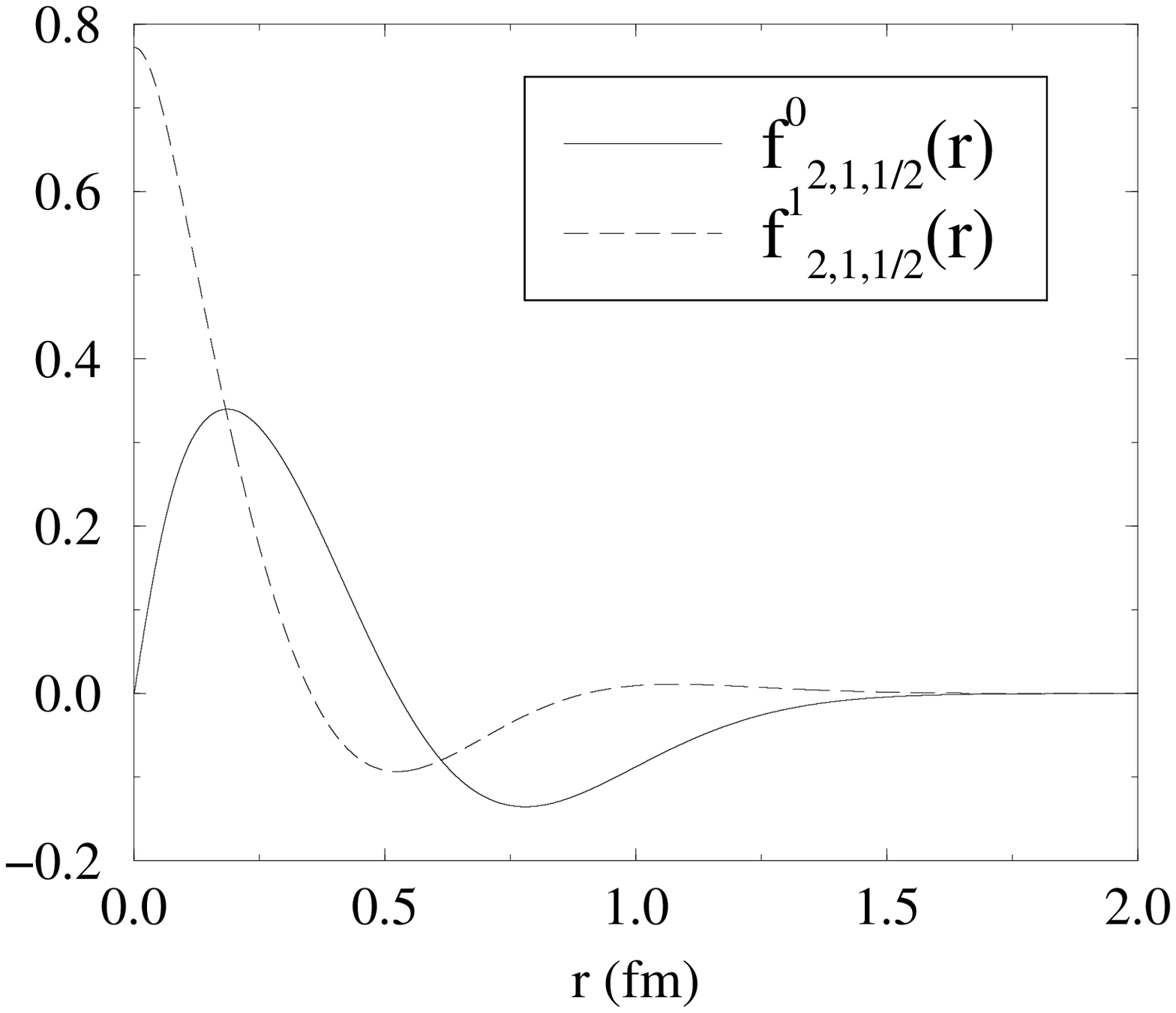} \\
\epsfxsize=4.5cm
\epsfysize=3.5cm
\hskip -5mm \epsfbox{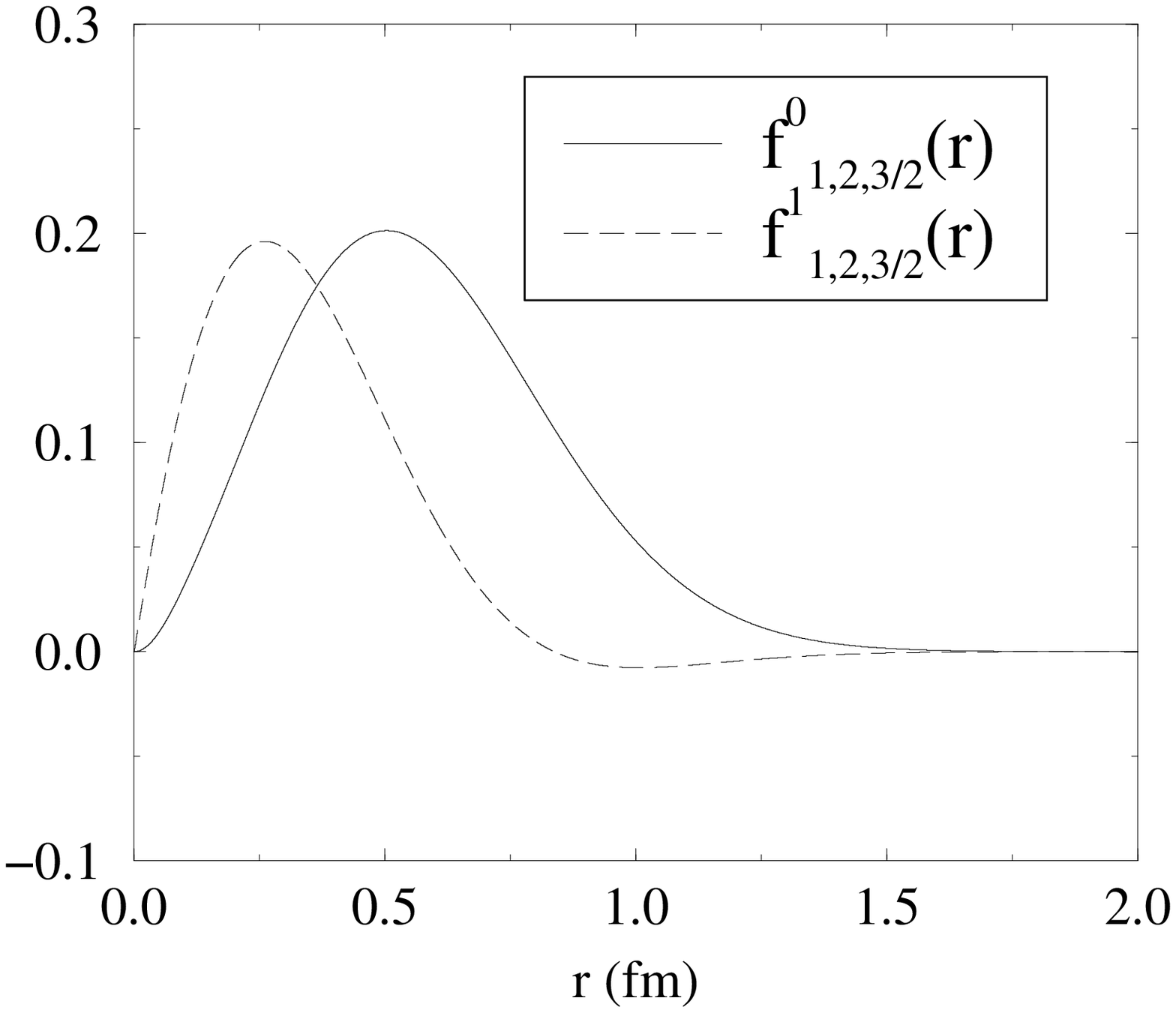} &
\epsfxsize=4.5cm
\epsfysize=3.5cm
\hskip -5mm \epsfbox{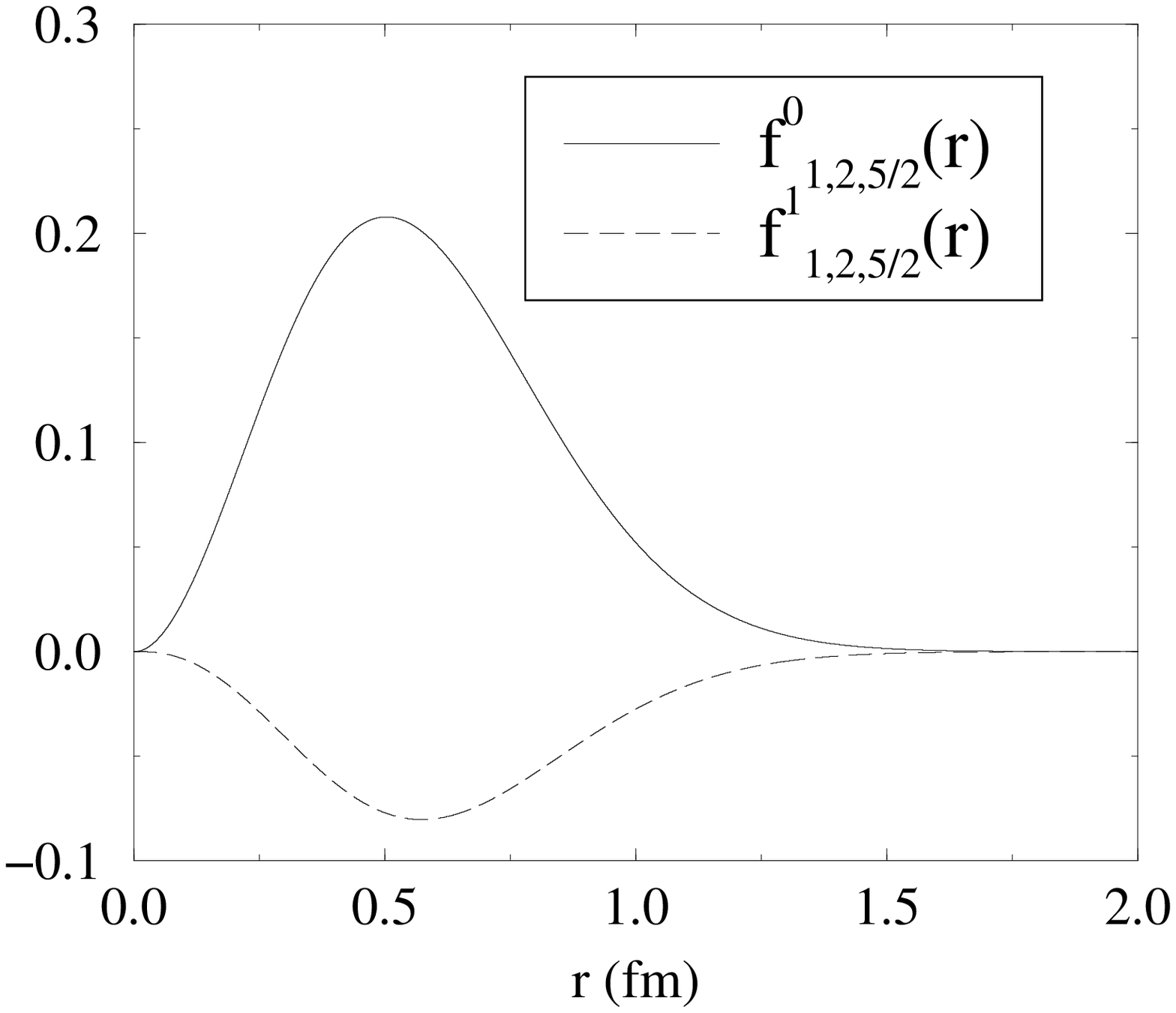} &
\epsfxsize=4.5cm
\epsfysize=3.5cm
\hskip -5mm \epsfbox{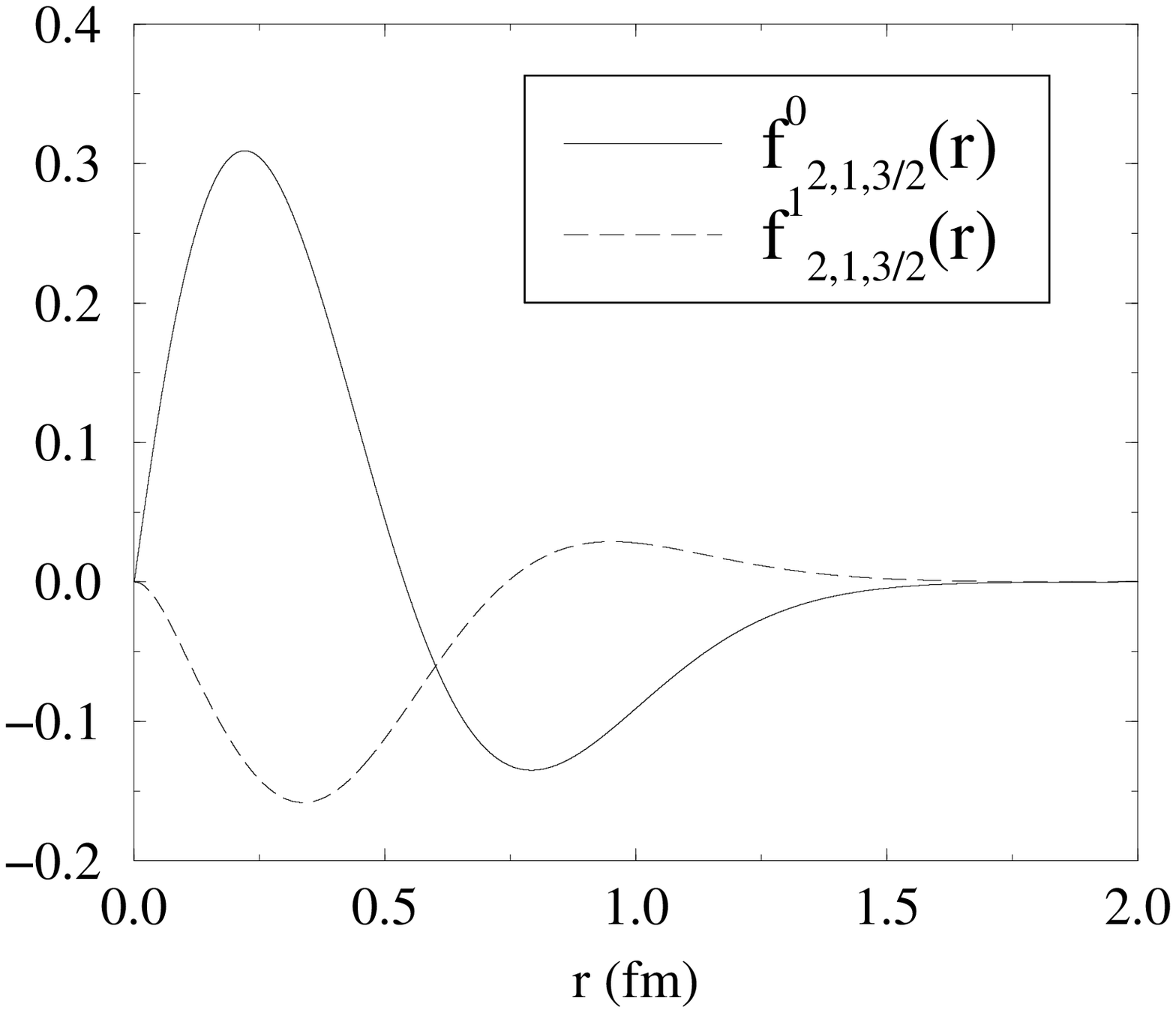} \\
\epsfxsize=4.5cm
\epsfysize=3.5cm
\hskip -5mm \epsfbox{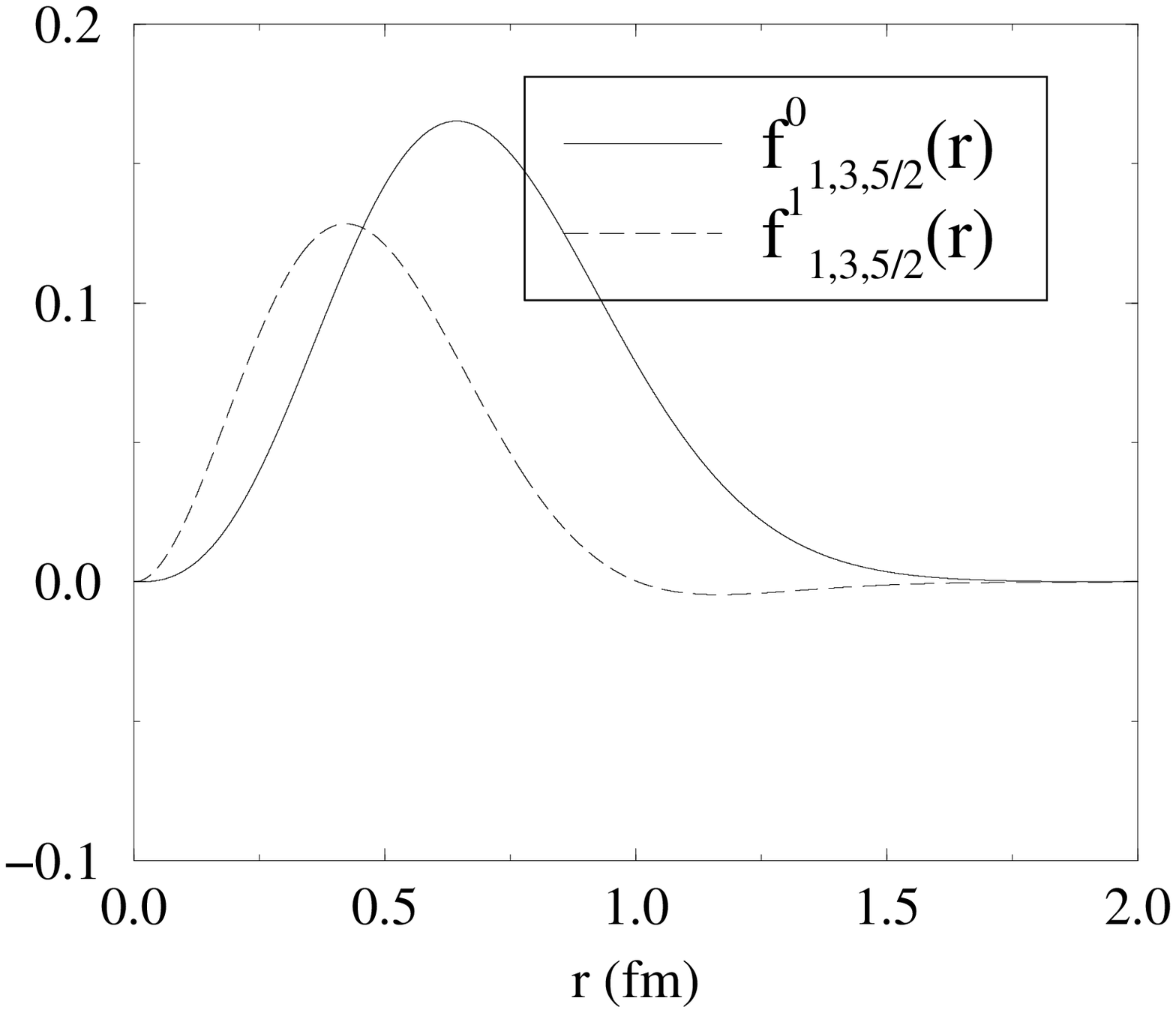} &
\epsfxsize=4.5cm
\epsfysize=3.5cm
\hskip -5mm \epsfbox{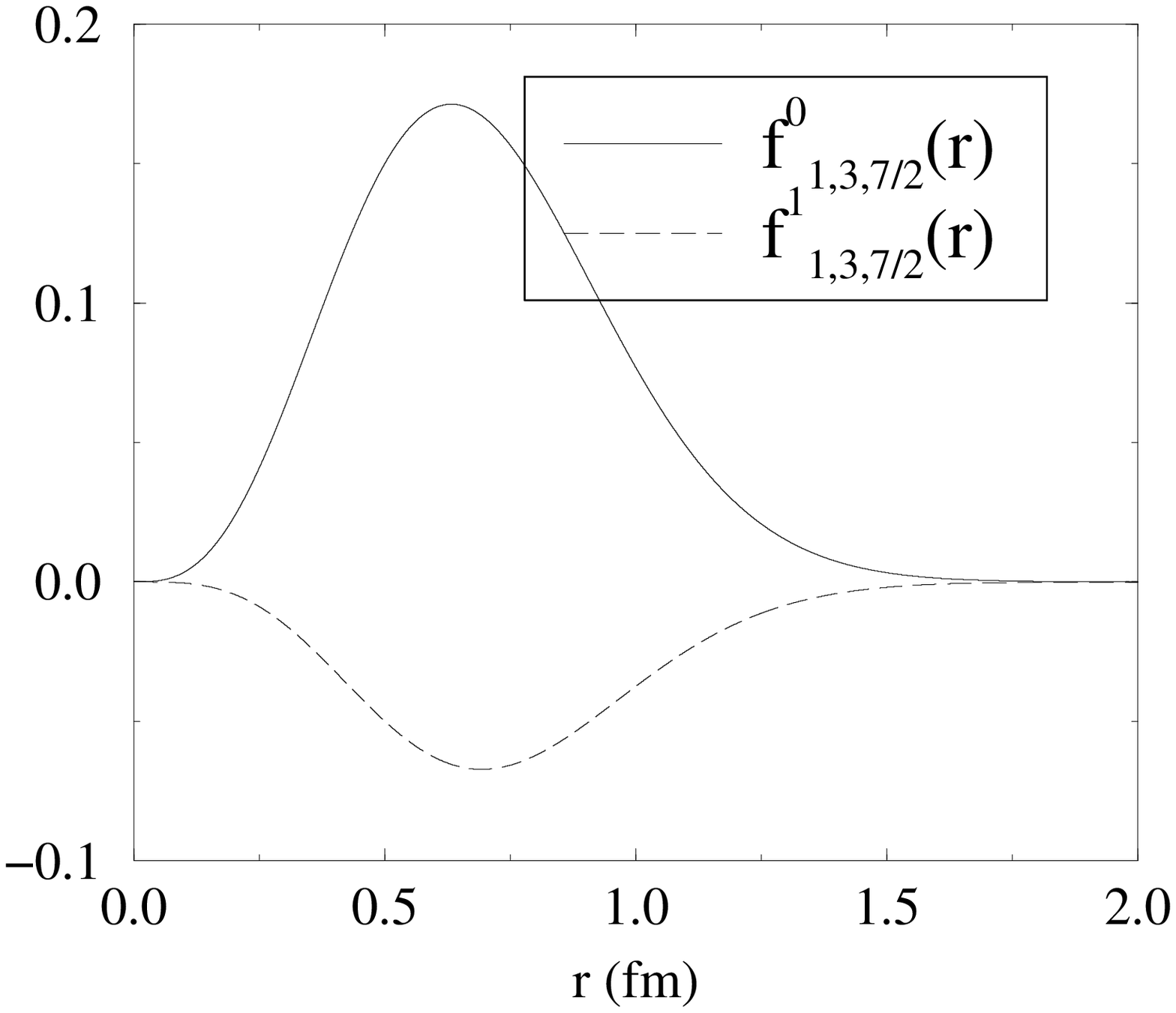} & \\
\end{tabular}
\end{center}
\caption{Radial wavefunctions for some excited states (for non-strange
mesons).
The continuum (dashed) line refers to the $f^0(r)$ ($f^1(r)$) function. 
These plots do not include the mixing contribution.\label{fig_wf}}
\end{figure}

\subsection{Determination of the parameters and predictions}

The nine parameters of our model, Eq.(\ref{params}), are determined
numerically
as follows: We define a function $\mathcal{F}$ of the input parameters that
finds eigenvalues and eigenfunctions of Eq.~(\ref{dirac}) using a fourth
order Runge-Kutta formula, corrects the energy levels by including the $1/m_h$
perturbative corrections (including mixing effects) and returns 
\begin{equation}
\chi ^2=\sum_{\text{observed states}} \left( \frac{%
E_{n,\ell ,j,J}^{\text{phys.}}-m_{n,\ell ,j,J}}
{\delta m_{n,\ell ,j,J}}\right) ^2
\end{equation}
where $E_{n,\ell ,j,J}^{\text{phys.}}$ are the computed energy levels 
and $m_{n,\ell ,j,J} \pm \delta m$ 
are the measured masses (with their experimental error) 
of the corresponding particles. 

We then minimize $\mathcal F $ in its nine dimensional domain. We repeat 
this procedure with different sets of starting parameters until we are 
confident that we have found the absolute minimum.
The experimental data used for the ``observed states'' in the fit are
reported in the third column of 
Tables~\ref{table_spectrum1}-\ref{table_spectrum4}.

Our best fit gives the following values for the parameters 
\begin{equation}
\begin{tabular}{|l|l|}
\hline
$\alpha _s$ & 0.339 \\ 
$\lambda $ & 2.823 GeV \\
$b$ & 0.257 GeV$^2$ \\ 
$m_u$ & 0.071 GeV \\
$m_s$ & 0.216 GeV \\
$m_c$ & 1.511 GeV \\ 
$M_c$ & 1.292 GeV \\ 
$m_b$ & 4.655 GeV \\ 
$M_b$ & 4.685 GeV \\ \hline
\end{tabular}
\label{results}
\end{equation}
The corresponding predicted spectrum is reported in the fifth column of 
Tables~\ref{table_spectrum1}-\ref{table_spectrum4}. 
The best fit parameters reported here differ slightly from those reported in 
Ref.~\cite{DiPierro} because we use here most recent value for
$m_{n,\ell,j,J}$.

We remark that $m_u$ and $m_s$ are mass parameters which differ 
from the constuent quark masses for an overall undetermined constant shift.

As a consistency check of our results we observe that 
the mass splitting $m_s-m_u\simeq 140$MeV comes out in agreement with naive
expectations based on Gell-Mann-Okubo type relations. This difference also 
agrees, within 1\%, with the corresponding splitting 
determined in ref.~\cite{roberts} and used in ref.~\cite{GR} as input 
for their calculations.

Remarkably $m_b\simeq M_b$ with much better agreement than expected.
Moreover $\lambda ^{-1}\simeq 0.06$fm is smaller than any other length scale 
involved in the problem, as was required.

Figure~\ref{fig_wf} shows some of the computed radial wavefunctions for 
non-strange mesons, $f_{n,\ell,j}^0(r)$ and $f_{n,\ell ,j}^1(r)$. These 
wavefunctions do not include $1/m_h$ corrections, therefore they are the same
for $D$ and $B$ mesons.  The corresponding wavefunctions for strange mesons
are very similar.

\begin{figure}
\begin{center}
\epsfxsize=8cm
\epsfysize=8cm
\epsfbox{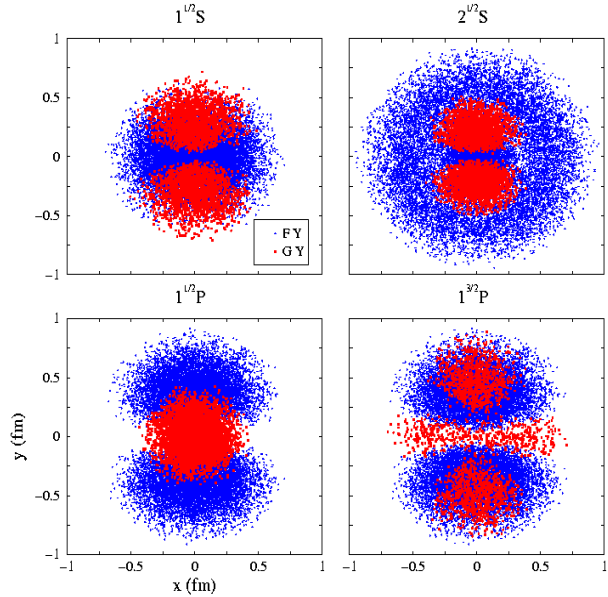}
\end{center}
\caption{Orbitals for some excited B mesons.\label{fig_density}}
\end{figure}

Figure~\ref{fig_density} shows density plots for each
couple of independent spin components of some of the computed light-quark
wavefunctions, $\left| f_{n,\ell ,j}^0Y_0^\ell \right| ^2$ in black (blue) 
and $\left| f_{n,\ell ,j}^1 Y_0^{2j-\ell }\right| ^2$ in gray (red). They
represent the
analogous, in the heavy meson systems, of the orbitals of the hydrogen atom.

\subsection{Comparison with experiment}

\begin{table*}
\begin{center}
\begin{tabular}{|clll|clll|} 
\hline
\multicolumn{4}{|c|}{charmed meson masses (MeV)} 
& \multicolumn{4}{c|}{bottom meson masses (MeV)} \\
\cline{1-4}\cline{5-8}
               & model          & exp &[Ref.]                   & 
               & model          & exp &[Ref.]                   \\
\hline
\hline
$D^*-D$        & 137~[a]        & 141,142 &\cite{PDG00}         & 
$B^*-B$        & 45~~[a]        & 46 &\cite{PDG00}              \\
\hline
$D_0^*-D$      & 512            &     &                           & 
$B_0^*-B$      & 427            &     &                           \\
$D_1^*-D$       & 622            & 596(53) &\cite{CLEO-1}        & 
$B_1^*-B$       & 463            & 391(16) &\cite{L3}[b]\!\!     \\
$D_1-D$        & 549~[a]        & 558(2) &\cite{PDG00}          & 
$B_1-B$        & 421            & 459(9) &\cite{OPAL-1}[b]\!\!     \\
               &                &        &                        & 
$B_1-B$        & 421            & 431(20) &\cite{CDF}[b]\!\!      \\
$D_2^*-D$      & 592~[a]        & 594(2) &\cite{PDG00}          & 
$B_2^*-B$      & 435            & 489(8) &\cite{L3}[b]\!\!     \\
               &                &        &                        & 
               &                & 460(13) &\cite{ALEPH}[b]\!\!     \\
               &                &         &                       & 
($B^{**}-B$)   &                & 418(9) &\cite{PDG00}[c]     \\ 
\hline
$D'-D$         & 721            &         &                       & 
$B'-B$         & 607            &          &                      \\
$D^{*\prime}-D$     & 824            & 772(6) &\cite{DELPHI}         & 
$B^{*\prime}-B$     & 641            &        &                        \\
               &                & 
\!\!\!\!\! not seen &\cite{CLEO-2,OPAL-2}\!\!  & 
               &                &         &                       \\
\hline
\hline
$D_s-D$        & 97~~[a]        & 99,104 &\cite{PDG00}          &
$B_s-B$        & 94~~[a]        & 90(2) &\cite{PDG00}           \\
$D_s^*-D_s$    & 148~[a]        & 144 &\cite{PDG00}             &
$B_s^*-B_s$    & 48~~[a]        & 46 &\cite{PDG00}              \\
\hline
$D_{s0}^*-D_s$ & 512            &     &                           &
$B_{s0}^*-B_s$ & 431            &     &                           \\
$D_{s1}^*-D_s$  & 640            &    &                            & 
$B_{s1}^*-B_s$  & 469            &    &                            \\
$D_{s1}-D_s$   & 570~[a]        & 566(1) &\cite{PDG00}          &
$B_{s1}-B_s$   & 432            &          &                      \\
$D_{s2}^*-D_s$ & 616~[a]        & 605(2) &\cite{PDG00}          &
$B_{s2}^*-B_s$ & 447            &           &                     \\
               &                &            &                    &
($B_s^{**}-B_s$)  &             & 484(15) &\cite{PDG00}[c]     \\
\hline
$D'_s-D_s$     & 735            &          &                      & 
$B'_s-B_s$     & 612            &          &                      \\
$D^{*\prime}_s-D_s$ & 841            &     &                           &
$B^{*\prime}_s-B_s$ & 646            &     &                           \\
\hline
\end{tabular}
\end{center}
\caption{The heavy-light spectrum compared to experiment. 
We report the difference between the excited 
state masses and the ground state ($D$ or $B$) in each case.  
\label{table_exp}}
[a]~Experimental input to model parameters fit. \newline
[b]~Theoretical estimates for some of the mass splittings have
been used as input. \newline
[c]~Experimental signal is a sum over resonances with $J=0,1,2$. 
\end{table*}

The comparison of our results to the present experimental information
on the excitation spectrum of the $(D, D_s, B, B_s)$ mesons is given
in Table~\ref{table_exp}. States which were used in the determining
our best fit parameters are so indicated.

Our model is in excellent agreement with the better established 
P waves in the  $D$ and $D_s$ systems. In particular the 
$D_1^*~(1^{\frac12}P_1)$ fits the
recent measurement of CLEO~\cite{CLEO-1}.
For the P-waves of the B meson systems the agreement with
preliminary measurements is somewhat less impressive.

However many of the existing experimental fits for 
individual masses of these states relied on patterns of masses
for the $j_l = 1/2$ states not found in our model.
For example, our relativistic quark model predict
$m_{1,1,\frac12,1} > m_{1,1,\frac32,1}$ for a 
relatively broad range of parameters consistent with 
light spectroscopy\footnote{This result is 
known in the literature as spin-orbit inversion. It was first predicted 
by Schnitzer~\cite{Schnitzer:1978kf} and later
by the models of Isgur~\cite{inversion1} and Ebert~\cite{inversion2}.}.
Also, we obtain a splitting for the $^{\frac12}P_J$ states 
more than twice as big as the splitting for the $^{\frac32}P_J$ states. 

Preliminary results from L3~\cite{L3} 
for the masses of P-wave $B$ meson excitations 
are:
\begin{eqnarray}
B_1^*: & m_{1,1,\frac12,1} = (5.670 \pm 0.010_{\text{stat}} 
\pm 0.013_{\text{syst}}) \text{GeV} & \text{(L3)} \\
B_2^*: & m_{1,1,\frac32,2} = (5.768 \pm 0.005_{\text{stat}} 
\pm 0.006_{\text{syst}}) \text{GeV} & \text{(L3)} \\
\end{eqnarray}
The L3 results were derived using the constraint 
that $m_{1,1,\frac12,1} - m_{1,1,\frac12,0} 
= m_{1,1,\frac32,2} - m_{1,1,\frac32,1} = 12$~MeV. 
This assumption is not realized in our model.  
Similar assumptions are needed for the extractions the masses of 
P-wave $B$ meson excitations from OPAL~\cite{OPAL-1} and
CDF~\cite{CDF}.
It would be interesting to reanalyze results using the pattern 
expected in this relativistic quark model.

Finally, the observation of the $D^{*\prime}$ by DELPHI~\cite{DELPHI} is
not consistent with searches by CLEO~\cite{CLEO-2} and OPAL~\cite{OPAL-2}.

\subsection{Regge trajectories}

\begin{figure}
\begin{center}
\epsfxsize=9cm
\epsfysize=9cm
\epsfbox{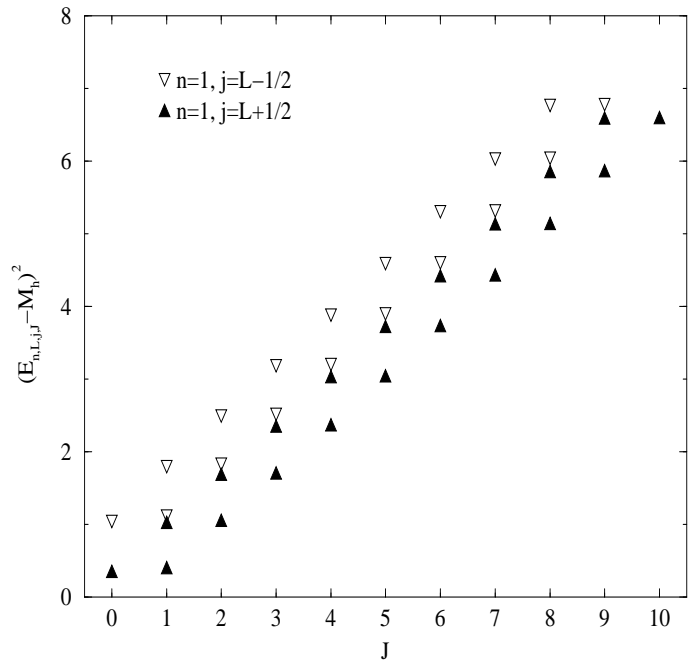}
\end{center}
\caption{Regge trajectories for some of our orbitally excited heavy mesons.
\label{fig_regge}}.
\end{figure} 

We find that $E_{n,\ell,j,J}-M_h$ all lie on 
Regge trajectories parameterized by $m J + q_{n,\ell,j}$ 
with $m\simeq 0.7$ (as shown in Figure~\ref{fig_regge}). 
This is a well understood phenomenon for light spectroscopy but, 
for light mesons, $m \simeq 2\pi b$ (where $b$ is the string tension). 
This is about a factor two bigger than we find. Our result can be explained in
the non relativistic limit or, alternatively, in
a simple and naive string picture: Ignoring the short distance behavior of 
the potential, we model the meson as a classical light quark attached through 
a string to the center of mass of the system. The energy of the system, $E$, 
(i.e. the energy of the string) is related to the 
classical angular momentum, $J$, by $E^2 = \pi b J$. 
The factor two in the light meson picture can be explained
with the fact that the latter rotate around a center of mass that is 
located at the 
middle of the string while for heavy-light mesons the center of 
mass coincides with one of the two ends of the string.  

This simple picture shows that the bulk of an heavy meson mass is 
dominated by the mass of the heavy quark plus the potential energy 
associated to the large distance interaction ($\propto b r$), 
and again gives support to our assumption that short distance 
behavior of the potential has a small contribution 
to the spectrum.

\section{Hadronic Transitions}

\subsection{Transition amplitudes}

We start by considering the most general hadronic transition of the form 
\begin{equation}
H^{\prime }\rightarrow H + x
\label{transition}
\end{equation}
where $H^{\prime }$ and $H$ are two heavy-light mesons containing the same
heavy quark, with wavefunctions $\Psi
_{n^{\prime },\ell ^{\prime },j^{\prime },J^{\prime },M^{\prime }}$
and $\Psi _{n,\ell ,j,J,M}$
respectively and $x$ can be any light meson with momentum $\mathbf{p}$. 
Although  we keep our formalism general,
in this paper we only compute numerically decays in which 
$x$ is a pseudoscalar 
meson belonging to the flavor octet ($\pi$, $K$, $\eta$).

In the context of the chiral quark model~\cite{GM84} this transition is
mediated by 
an effective interaction of the form
\begin{equation}
{\mathcal L}_{\text{int}} = \frac{g^8_A}{\sqrt{2} f_{x}} 
\bar q'_i X {\mathcal M}^{ij} q_j  + {\mathcal O}(\partial^2)
\label{lagrangian1}
\end{equation}
where $f_{x}$ can be identified with $f_\pi \simeq 130$MeV and
$g^8_A$ is an effective coupling. $X=\sla \partial \gamma^5$ is the spin
structure associated to the transition, $i,j$ are $SU(3)_{\text{flavor}}$
indices and
\begin{equation}
{\mathcal M} = \sqrt{2} \left(
\begin{tabular}{ccc}
$\frac1{\sqrt 2} \pi^0 + \frac1{\sqrt 6} \eta_8$ & $\pi^+$ & $K^+$ \\
$\pi^-$ & $-\frac1{\sqrt 2} \pi^0 + \frac1{\sqrt 6} \eta_8$ & $K^0$ \\
$K^-$ & $\bar K^0$ & $-\frac2{\sqrt 6} \eta_8$  
\end{tabular}
\right)
\label{calm}
\end{equation}
is the usual $SU(3)_{L+R}$ invariant representation of the 
pseudoscalar mesons. 

The transition mediated by this Lagrangian is associated to the following 
matrix element
\begin{equation}
I^{H'H{x}}(\mathbf{p})=\frac{i g^8_A\zeta}
{\sqrt{2} f_{x}} \int \overline{\Psi }%
_{n,\ell ,j,J,M}(z)Xe^{i\mathbf{p}\cdot \mathbf{z}}\Psi _{n^{\prime },\ell
^{\prime },j^{\prime },J^{\prime },M^{\prime }}(z)\text{d}^3z  \label{th2}
\end{equation}
where  $\zeta $ is a coefficient that characterize the flavor structure of the
decay. A list of all possible cases has been derived from Eq.~(\ref{calm}) and
is reported in Table~\ref{table3}. The physical $\eta$ and $\eta'$ are
of course mixtures of the ideal $\eta_8$ and $\eta_1$. In particular, 
$\eta = \eta_8 \cos(\theta_p) - \eta_1 \sin(\theta_p)$ where $\theta_p \approx
-10.1\deg $ with a large uncertainty. In this work we ignore this 
mixing and we assume $\theta_p=0$. The corrective multiplicative factor for a 
different choice for $\theta_p$ can be derived by the reader using Table~\ref{table3}.

\begin{table}
\begin{center}
\begin{tabular}{|l|c||l|c|} \hline
$H'\rightarrow H+x$ & $\zeta $ & 
$H'\rightarrow H+x$ &$\zeta $  \\ \hline
$B^0\rightarrow B^0+\pi ^0$ & $1$ & 
$B_s\rightarrow B_s+\eta_8 $ & $-2/\sqrt{3}$  \\
$B^{\pm }\rightarrow B^{\pm }+\pi ^0$ &  $1$ & 
$B_s\rightarrow B^0+K$ & $\sqrt{2}$   \\ 
$B^{\pm}\rightarrow B^0+\pi ^{\pm}$ & $\sqrt{2}$ &  
$B_s\rightarrow B^{\pm}+K^{\mp}$ & $\sqrt{2}$  \\
$B^0\rightarrow B^{\pm}+\pi ^{\mp}$ & $\sqrt{2}$  &
$B^0\rightarrow B_s+\bar K$ & $\sqrt{2}$     \\ 
$B^0\rightarrow B^0+\eta_8 $ &  $1/\sqrt{3}$  & 
$B^{\pm}\rightarrow B_s+K^{\pm}$ & $\sqrt{2}$  \\ 
$B^{\pm }\rightarrow B^{\pm }+\eta_8 $ & $1/\sqrt{3}$ & 
$B_q\rightarrow B_q+\eta_1 $ & $\sqrt{\frac23} + {\mathcal O}(\frac1{N_c}) $
\\ \hline
\end{tabular}
\end{center}
\caption{List of decay channels for $B$ (or $D$) mesons with the 
corresponding flavor factor $\zeta$.
\label{table3}}
\end{table}
The exponential in Eq.~(\ref{th2}) can be expanded in products 
of spherical harmonics and spherical Bessel functions, thus giving 
\begin{equation}
I^{H'H{x}}(\mathbf{p})=\sum_{\ell _x,m_x}Y_{m_x}^{\ell
_x*}(\widehat{p})C_{J,M;\ell _x,m_x}^{J^{\prime },M^{\prime }}\mathcal{A}%
_{\ell _x}^{H'Hx}(X,p) 
\label{hprimetoh}
\end{equation}
Eq.~(\ref{hprimetoh}) implicitly defines the 
transition amplitude,
$\mathcal{A}_{\ell _x}^{H^{\prime }\rightarrow Hx}(X,p)$,
for a $x$ in a given eigenstate $\ell_x$
of its angular momentum. By projecting the matrix $X$ on the 
basis presented in the Appendix~A, the transition amplitude can be 
rewritten as a linear combination of terms, each
factorized into a radial part and a spin dependent part 
\begin{equation}
\mathcal{A}_{\ell _x}^{H'Hx}(X,p)=\frac{ig^8_A\zeta}{\sqrt{2} f_x} 
\sum_{ab=\{0,1\}}\sum_{k} c_{\ell _x}^{ab,k}(X)\int_0^\infty f_{n^{\prime
},\ell ^{\prime },j^{\prime }}^{a}(r)j_k(rp)f_{n,\ell ,j}^{b}(r)r^2\text{d}r 
\label{amplitude}
\end{equation}
The coefficients $c_{\ell _x}^{ab,k}(X)$ depend on the quantum numbers
of the mother and the daughter heavy mesons. 
Their explicit expression is given in the Appendix~A.
The integrals are computed numerically. 

One can extend our analysis for octet pseudoscalar transitions
to the approximately flavor singlet $\eta'$.
In the large $N_c$ limit the $\eta_1$ 
combines with the octet to form a nonet. 
In that case the effective interaction in the Lagrangian takes the form
\begin{equation}
{\mathcal L}_{\text{int}} = \frac{g^1_A  \zeta}{\sqrt{2} f_{\eta_1}} 
\bar q'_i X \eta_1 q_i  + {\mathcal O}(\partial^2)
\label{nonetl}
\end{equation}
with $X=\sla p$ and $g^1_A \simeq g^8_A$.
This symmetry is badly broken in QCD. However it is reasonable to
assume that in these transitions that the form (Eq.~(\ref{nonetl})) 
still holds.
If one further assumes that the spatial wavefunctions of $\pi$ 
and $\eta '$ are the approximately equal, 
one obtains $f_{\eta '} \simeq f_\pi$. 
Hence the coefficient $\zeta$ can be set to 
$\sqrt{2/3}$ both for heavy strange and non strange decaying heavy mesons.

The situation for decays in which $x$ is a light
vector mesons ($\rho$, $\omega$, $K^\ast$) is different.
When compared to the pseudoscalar mesons, they have a different spin 
coupling to the quarks  ($X=\sla \epsilon$ where $\epsilon_\mu$ is the 
polarization vector of the meson), a different effective coupling 
($g_V \neq g^8_A$), and a different wavefunction ($f_\rho \neq f_\pi$).
With these replacements Eq.~(\ref{amplitude}) remains valid for decays with 
emission of light vector mesons.\footnote{It is possible to relate these 
the pseudoscalar and vector couplings and coefficients within the context 
of an approximate $SU(6)_W$ symmetry for the low lying states.} 
The detailed study of these vector
meson transitions is deferred to a future paper.

\subsection{Partial widths}

The partial width for the transition in Eq.~(\ref{transition}) 
(for a light meson $x$ emitted with total 
momentum $p$ and angular momentum $\ell_x$) is given by
\begin{equation}
\Gamma_{x}(H' \rightarrow H + x;~\ell_x) = \frac{p}{8 \pi^2} \frac{2 J +1}{2
J' +1} 
  \frac{m_H}{m_{H'}} \left| {\mathcal A}^{H'Hx}_{\ell_x}(X,p) \right|^2
\label{use_amplitude}
\end{equation}
where $m_{H'}$ and $m_H$ are the masses of the mother and 
daughter heavy mesons respectively. 

The total hadronic decay width (via a pseudoscalar meson transition) 
is defined simply as the sum of the partial widths:
\begin{equation}
\Gamma_{\mathcal M}^{H'} = \sum_{H}~ \sum_{x=\{\pi,\eta_8,K\}} 
~\sum_{\ell_x} ~\Gamma_{x}(H' \rightarrow H + x;~\ell_x) \;
\end{equation}
Transitions involving excited states very near 
their kinematic threshold for an allowed decay (e.g. where 
the light pseudoscalar momentum is less than 100~MeV) 
are extremely sensitive to our calculated mass values.
This is particularly true for the allowed transitions within the 
$1S$ multiplets.  In these cases, even the small mass differences between the
the charged and neutral states are important. Using the physical  
masses for the various $D$ mesons~\cite{PDG00}, the individual pion transitions
are shown in Table~\ref{tbl:decayS}.

\begin{table}
\begin{center}
\begin{tabular}{|l|crr|} \hline 
channel & $l_{\pi}$ & $p_{\pi}$ {\rm (MeV)} & $\Gamma_{\pi}/(g_A^8)^2 {\rm
(keV)}$ \\
\hline \hline
$D^{\star 0}(1^{\frac12}S_1) \rightarrow D^0(1^{\frac12}S_0) + \pi^0$ 
    & $1$ & $42.8 \pm .2$ & $62 \pm 1$ \phantom{aa} \\
$D^{\star +}(1^{\frac12}S_1) \rightarrow D^0(1^{\frac12}S_0) + \pi^{+}$ 
    & $1$ & $39.4 \pm .5$ & $97 \pm 3$ \phantom{aa}\\
$D^{\star +}(1^{\frac12}S_1) \rightarrow D^{+}(1^{\frac12}S_0) + \pi^0$ 
    & $1$ & $38.1 \pm .3$ & $44 \pm 1$ \phantom{aa}\\
\hline
\end{tabular}
\end{center}
\caption{List of partial decay rates for $1S$ $D$ mesons. 
The measured $D^*$ and $D$ masses~\cite{PDG00} are used in these results.
\label{tbl:decayS}}
\end{table}

After removing these phase space uncertainties,
reasonable variations in our model parameters gave variations
of about 10\% in the overall hadronic widths.
The listed branching ratios with emission of a $\pi$ are flavor blind and 
sum over the final state pion charge (i.e. they have 
been computed with $\zeta=\sqrt{3}$). 
Each exclusive decay can be deduced by correcting for this factor 
using Table~\ref{table3} to determine the relative strength of the
charged and neutral decays. In addition, a small phase space correction
should be included appropriate to the slight difference in the
masses of the various charge states.
These rates are shown in Table~\ref{tbl:DdecayP} for the
$D$ and $D_s$ mesons and in Table~\ref{tbl:BdecayP} for the
$B$ and $B_s$ mesons. 
For the $1^{\frac32}P_{(1,2)}$ $B_s$ states, our model predicts that 
they are below threshold for $K$ transitions to the 
the $1^{\frac12}S_{(0,1)}$ $B$ states. However, this is very sensitive to the
details of the model. So, for completeness, we note the partial rates divided
by 
the appropriate phase space factor at $p_K = 0$ in Table~\ref{tbl:BnodecayP}. 

\begin{table*}
\begin{center}
\begin{tabular}{|llc|crr|}
\hline
 $H' ({n'}^{j'}\ell_{J'})$ & $H (n^j\ell_J)$ & $x$ & $\ell_x$ & $p_x$ & 
$\Gamma_x/(g^8_A)^2$ \\ \hline
\hline
$D (1^{\frac12}P_0)$   & $D (1^{\frac12}S_0)$ & $\pi$ & 0 & 437 &
$189$  \\
\hline
$D (1^{\frac32}P_1)$   & $D (1^{\frac12}S_1)$ & $\pi$ & 0 & 355 & $(\ast)\,
1.7$  \\
                       & $D (1^{\frac12}S_1)$ & $\pi$ & 2 & 355 &
$14.5$  \\
\hline 
$D (1^{\frac32}P_2)$   & $D (1^{\frac12}S_0)$ & $\pi$ & 2 & 506 &
$24.6$  \\ 
                       & $D (1^{\frac12}S_1)$ & $\pi$ & 2 & 394 &
$13.7$  \\ 
\hline 
$D (1^{\frac12}P_1)$   & $D (1^{\frac12}S_1)$ & $\pi$ & 0 & 420 &
$181$  \\ 
\hline
\hline 
$D_s (1^{\frac12}P_0)$ & $D (1^{\frac12}S_0)$ & $K$      & 0 & 325 &
$236$  \\
\hline 
$D_s (1^{\frac32}P_1)$ & $D (1^{\frac12}S_1)$ & $K$      & 0 & 175 & $(\ast)\,
1.89$ \\
                       & $D (1^{\frac12}S_1)$ & $K$      & 2 & 175 &
$0.3$  \\
\hline 
$D_s (1^{\frac32}P_2)$ & $D (1^{\frac12}S_0)$ & $K$      & 2 & 442 &
$8.9$  \\
                       & $D (1^{\frac12}S_1)$ & $K$      & 2 & 264 &
$1.4$  \\
                       & $D_s (1^{\frac12}S_0)$ & $\eta$ & 2 & 248 &
$0.4$  \\
\hline 
$D_s (1^{\frac12}P_1)$ & $D (1^{\frac12}S_1)$ & $K$      & 0 & 302 &
$224$  \\
\hline 
\end{tabular}
\end{center}
\caption{The heavy-light 1P state hadronic transition rates for $D$ and $D_s$
mesons.
$H' \rightarrow H + x$. Decays denoted with an $(\ast)$ 
are allowed only because of the order $1/m_h$ mixing of states.
$p_x$ and $\Gamma_x/(g^8_A)^2$ are in MeV.
\label{tbl:DdecayP}}
\end{table*}

\begin{table*}
\begin{center}
\begin{tabular}{|llc|crr|}
\hline
 $H' ({n'}^{j'}\ell_{J'})$ & $H (n^j\ell_J)$ & $x$ & $\ell_x$ & $p_x$ & 
$\Gamma_x/(g^8_A)^2$ \\ \hline
\hline
$B (1^{\frac12}P_0)$   & $B (1^{\frac12}S_0)$ & $\pi$ & 0 & 388 &
$186$  \\
\hline
$B (1^{\frac32}P_1)$   & $B (1^{\frac12}S_1)$ & $\pi$ & 0 & 338 & $(\ast)\,
0.5$  \\
                       & $B (1^{\frac12}S_1)$ & $\pi$ & 2 & 338 &
$13.1$  \\
\hline
$B (1^{\frac32}P_2)$   & $B (1^{\frac12}S_0)$ & $\pi$ & 2 & 396 &
$10.6$  \\
                       & $B (1^{\frac12}S_1)$ & $\pi$ & 2 & 352 &
$9.5$  \\
\hline
$B (1^{\frac12}P_1)$   & $B (1^{\frac12}S_1)$ & $\pi$ & 0 & 381 &
$180$  \\
\hline
\hline
$B_s (1^{\frac12}P_0)$ & $B (1^{\frac12}S_0)$ & $K$   & 0 & 170 &
$159$  \\
\hline
$B_s (1^{\frac12}P_1)$ & $B (1^{\frac12}S_1)$ & $K$   & 0 & 153 &
$143$  \\
\hline
\end{tabular}
\end{center}
\caption{The 1P state hadronic transition rates for $B$ and $B_s$ systems.
$H' \rightarrow H + x$. Decays denoted with an $(\ast)$
are allowed only because of the order $1/m_h$ mixing of states.
Values for $p_x$ and $\Gamma_x/(g^8_A)^2$ are in MeV.
\label{tbl:BdecayP}}
\end{table*}

\begin{table}
\begin{center}
\begin{tabular}{|llc|cr|} \hline
 $H' ({n'}^{j'}\ell_{J'})$ & $H (n^j\ell_J)$ & $x$ & $\ell_x$ & 
 $\Gamma_x/(g_A^8)^2 \times (100/p_x)^{(2l_x+1)} $ \\ \hline
\hline
$B_s (1^{\frac32}P_1)$ & $B(1^{\frac12}S_1)$ & $K$ & $0$ 
                       & $(\ast)\, 4.92\times 10^{-1}$\phantom{hello} \\ 
\hline
$B_s (1^{\frac32}P_1)$ & $B(1^{\frac12}S_1)$ & $K$ & $2$ 
                       &          $2.38\times 10^{-2}$\phantom{hello} \\ 
\hline
$B_s (1^{\frac32}P_2)$ & $B(1^{\frac12}S_0)$ & $K$ & $2$ 
                       &          $9.48\times 10^{-3}$\phantom{hello} \\ 
\hline
$B_s (1^{\frac32}P_2)$ & $B(1^{\frac12}S_1)$ & $K$ & $2$ 
                       &          $1.43\times 10^{-2}$\phantom{hello} \\ 
\hline
\end{tabular}
\end{center}
\caption{Decay rates for $1^{\frac32}P_{(1,2)}$ $B_s$ mesons 
with phase space dependence divided out.
These states are very near the 
kinematical threshold in our model.  
Values for $p_x$ and $\Gamma_x/(g^8_A)^2$ are in MeV.
\label{tbl:BnodecayP}}
\end{table}

A list of the allowed transitions for other low-lying excited states 
is reported in Appendix~B.  

\subsection{Comparison to lattice results} 

\begin{figure} 
\begin{center}
\epsfxsize=7.5cm
\epsfysize=7.5cm
\epsfbox{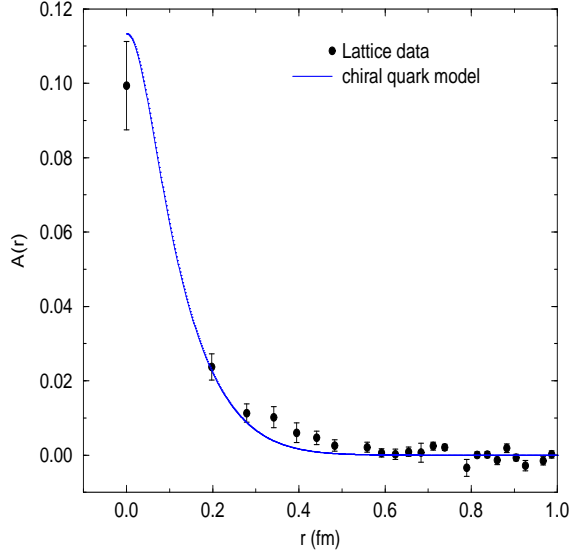}
\end{center}
\caption{Comparison between a prediction of our model and the lattice QCD
result~\cite{me1}.\label{lattice}}
\end{figure}

As one more consistency check of our model we compare our prediction for the
transition\footnote{%
Even if this transition is kinematically forbidden it is physically 
relevant to the $B \rightarrow \pi + \ell + \bar \nu$ exclusive decay 
under the  assumption of vector meson dominance.} 
$B\rightarrow B^\ast+\pi$ with model independent results coming from
lattice simulations. We define 

\begin{equation}
\mathcal{A}^{B B^\ast\pi }(r)=\frac 13\sum_{\mu =1,2,3}\int
\left\langle B ^\ast \right| A_\mu (\mathbf{r})\left| B \right\rangle \text{d}
\Omega _r  \label{abbpi1}
\end{equation}
with $A_\mu(\mathbf{r}) =\bar q(t,\mathbf{r}) \gamma_\mu 
\gamma^5 q(t,\mathbf{r})$ at $t=0$.
The same matrix element can be expressed in terms of our 
radial wavefunction and, in the limit $p_\pi \rightarrow 0$,
\begin{equation}
\mathcal{A}^{B B^\ast\pi }(r) = -g^8_A \Big[ (f^0_{1,0,\frac12}(r))^2 
-\frac13
(f^1_{1,0,\frac12}(r))^2 \Big] + {\mathcal O}(p_\pi) 
\label{abbpi2}
\end{equation}
$g^8_A$ is the effective coupling of the transition as defined 
in Eq.~(\ref{lagrangian1})\footnote{%
Note from Eq.~(\ref{abbpi2}) that in the non relativistic limit 
the coupling $g^8_A$ coincides with the coupling constant $g$ 
that appears in the Heavy Meson Chiral 
Lagrangian~\cite{casalbuoni97}. The lattice result for 
this quantity is $g=0.42 \pm 0.09$.}
In the chiral quark model $g^8_A$ is an effective parameter and it is 
has to be given as input. On the other side, in the context of lattice 
computations, the matrix element in Eq.~(\ref{abbpi1}) follows directly 
from first principles (the QCD Lagrangian) and can be computed 
explicitly. 
By fitting the lattice results of Ref.~\cite{me1} with our prediction 
for $\mathcal{A}^{B B^\ast\pi }(r)$ as function of $g^8_A$ we are able to 
determine\footnote{%
We set to zero terms of the order ${\mathcal O}(p_\pi)$ for consistency
with the lattice computation.}
\begin{equation}
g^8_A = Z_A g^{8\,\text{(lattice)}}_A = 0.53 \pm 0.11
\end{equation}
where $g^{8\,\text{(lattice)}}_A$ is the naive lattice result extracted 
from the fit and $Z_A=0.78$ is lattice matching factor discussed in 
Ref.~\cite{me1}. The error includes the statistical error due to 
the simulation and the fits ($\simeq 10\%$), 
and an estimate of the systematic error in the matching 
coefficient and in the the chiral extrapolation.
This is a preliminary result, as those of Ref.~\cite{me1} are, 
because of the small lattice size and the poor chiral extrapolation.
In any case our result is in agreement with lattice determination
of Ref.~\cite{dong}, $g_A^8 = 0.61\pm0.13$, and with experimental results 
from nucleon and hyperon $\beta$ decays, $g_A^8=0.58\pm0.02$,~\cite{close}.
A more precise lattice determination is possible and may be carried out in 
the near future.

Apart for the overall normalization given by $g^8_A$, the 
radial dependence of the function 
$\mathcal{A}_{B^{*}B\pi }(r)$
is predicted independently by our model and by the lattice 
computation. In Figure~\ref{lattice} 
we present a comparison between our 
analytical result, with adjusted normalization, and the 
lattice data. We believe this comparison provides 
a satisfactory consistency check of the two methods\footnote{
It also suggests a possible use of chiral quark model 
wavefunctions to isolate excited states contributions
in lattice correlation functions.}.

\subsection{Comparison to experiment} 

The total width of the $D^{*+}$ meson has recently been measured by 
the CLEO Collaboration~\cite{D*width}. They obtain $\Gamma = 
96\pm 4_{\rm stat}\pm 22_{\rm syst} {\rm (keV)} $. 
Combining this measurement
with the well-known branching ratios for the various pionic transitions, gives
a measurement of chiral coupling constant. We obtain 
\begin{equation}
g^8_A = 0.82 \pm 0.09 
\end{equation}
within our model.  

\begin{table*}
\begin{center}
\begin{tabular}{|lllll|} 
\hline
state &  width (exp) & [Ref.]  & $\Gamma/(g^8_A)^2$ (model) &  $g^8_A$ \\
\hline
\hline
$D^+(1^{\frac12}S_1)$ & $96\pm 4\pm 22$ {\rm (keV)} & \cite{D*width}  
     & $143$~{\rm (keV)}~[a] & $0.82 \pm 0.09$ \\
\hline
\hline
$D^{0}(1^{\frac32}P_1) $ & $18.9^{+4.6}_{-3.5}$ & & $16$ &
$1.09^{+0.12}_{-0.11}$ \\
$D^{+}(1^{\frac32}P_1) $ & $28\pm 8$ & & $16$ & $1.32^{+0.18}_{-0.27}$ \\
$D^{+}(1^{\frac12}P_1) $ & $290^{+101}_{-79}\pm 26\pm 36$ & \cite{CLEO-1} &
$181$ 
& $1.27 \pm 0.22$ \\
$D^{0}(1^{\frac32}P_2) $ & $23\pm 5$ & & $38$ & $0.77 \pm 0.08$ \\
$D^{+}(1^{\frac32}P_2) $ & $25\pm 8$ & & $38$ & $0.81^{+.012}_{-0.14}$ \\
\hline
\hline
$D_s(1^{\frac32}P_1) $ & $\le 2.3$  & & $2.0$  & $\le 1.07$ \\
$D_s(1^{\frac32}P_2) $ & $15\pm 5$  & & $10.9$ & $1.17^{+0.18}_{-0.11}$ \\
\hline
\hline
$B (1^{\frac12}P_1) $ & $73\pm 44$ & \cite{L3}[b]\!\!  & $180$ &
$0.64^{+0.17}_{-0.21}$ \\
$B (1^{\frac32}P_1) $ & ${18^{+15}_{-13}}~^{+29}_{-23}$ &\cite{OPAL-1}[b]\!\! &
$14.0$ & $1.13^{+0.77}_{-1.13}$ \\
$B (1^{\frac32}P_2) $ & $41\pm 43$ & \cite{L3}[b]\!\! & $20.0$ &
$1.43^{+0.61}_{-1.43}$ \\
\hline
\end{tabular}
\end{center}
\caption{The $1S$ and $1P$ heavy-light hadronic transition rates 
compared to experiment.  All widths are in MeV unless otherwise indicated.
Experimental values from PDG~\cite{PDG00} unless otherwise indicated.  
\label{tbl:width_exp}}
[a]~Theoretical value corrected for phase space observed mass(see
Table~\ref{tbl:decayS})
and the $1.6\%$ branching ratio to $D^+ + \gamma$~\cite{PDG00}. \newline
[b]~Experimental results depend strongly on model dependent assumptions.
\newline
\end{table*}

Within the chiral quark model,  
the coupling determined from any transition should agree. (i.e the
coupling is independent of particular initial or final heavy-light states.)
Table~\ref{tbl:width_exp} lists the various determinations of $g^8_A$
from existing data.  Within the existing large uncertainties the various 
determinations are consistent.

\section{Conclusions}

We have computed the spectrum and hadronic decay width of the excited 
$D$, $D_s$, $B$ and $B_s$ mesons using a relativistic quark model 
for the masses and wavefunction of the heavy-light mesons.
This work is based on that of refs.~\cite{isgur,roberts,GR} but 
departs from these previous works because we choose a simpler form 
for the potential and determined its parameters exclusively from fitting
the experimental heavy-light spectrum.
Moreover we computed all corrections within the model to order $1/m_h$, 
including mixing between nearby states with the same $J^P$.  

Our spectrum results agree very well with the existing data in the
$D$ and $D_s$ systems.  The agreement in the $B$ and $B_s$ systems
is also fairly good but the experimental situation for the P states
is not yet completely resolved. 

For example, our model predicts a 
spin-orbit inversion  for the excited P-wave
states in these systems. This agrees with the recent 
CLEO~\cite{CLEO-1} results for the $D$ mesons; but is in disagreement 
with preliminary L3 results~\cite{L3} for the $B$ mesons.  
However,
in our model we find that the splitting between the $j_l=1/2$ P-wave
states ($J=0,1$) more than twice as large as the splitting of 
the $j_l=3/2$ P-wave states ($J=1,2$). which is inconsistent with
the assumption of equal splitting used by $L3$~\cite{L3} in their analysis.

The information on the spectrum and the wavefunction was used to
compute a complete list of allowed hadronic decays for these  
excited mesons into lower energy states with emission of a light 
pseudoscalar meson ($\eta$, $\pi$ or $K$) and their relative branching ratios.

We also compared our model prediction for  
$\langle B^\ast | A_\mu^\ast(r) | B \rangle$ with lattice results. 
We find good agreement between the shapes of the decay amplitudes.
We used this comparison to extract a preliminary lattice 
determination of $g_A^8$, the effective coupling of the chiral quark model. 
We find the value $g_A^8 = 0.53 \pm 0.11$.
Using this value for $g_A^8$ to translate the total hadronic decay widths 
in physical units, we would conclude that the widths of the $D$ meson $P$
waves are 
consistently below experimental findings.  Of course, 
the lattice results have not yet been extrapolated to the continuum.

Comparison of our results with a recent experiment measurement of
the $D^{*+}$ width~\cite{D*width} yields 
$g^8_A = 0.82 \pm 0.09$. 
Using this value, we find much better overall agreement 
with experimental findings.

Finally, there is also a whole set of hadronic decays with 
emission of light vector mesons 
which we did not include in our study. 
For the higher excited states
these transitions give an important contribution to the total 
physical widths. We intend to study these transitions in a future
paper. 

\section*{Acknowledgment}

We wish to acknowledge the authors of Ref.~\cite{me1} from which we extracted
the lattice data of Fig.~\ref{lattice}. One of us, D.P., thanks 
G.~Chiodini for stimulating discussions on experimental issues.

This work was performed at Fermilab, a U.S. Department of Energy Lab (operated
by the University Research Association, Inc.) under contract
DE-AC02-76CHO3000.

\newpage

\section*{Appendix A}

In order to be able to factorize the expression for the transition 
amplitude, Eq.~(\ref{hprimetoh}) 
into radial and angular parts, it is convenient to adopt the 
following basis for the $\Gamma$ matrices
\begin{eqnarray}
&\Gamma^{00}_\mu = \left(
\begin{tabular}{cc}
$\sigma_\mu$ & 0 \\ 0 & 0
\end{tabular}
\right),\qquad
&\Gamma^{01}_\mu = \left(
\begin{tabular}{cc}
0 & $\sigma_\mu$ \\ 0 & 0
\end{tabular}
\right), \\
&\Gamma^{10}_\mu = \left(
\begin{tabular}{cc}
0 & 0 \\
$\sigma_\mu$ & 0 
\end{tabular}
\right),\qquad
&\Gamma^{11}_\mu = \left(
\begin{tabular}{cc}
0 & 0 \\
0 & $\sigma_\mu$ 
\end{tabular}
\right)
\end{eqnarray}
where $\sigma_0$ is the $2\times2$ identity and $\sigma_i$ are the usual Pauli
matrices.

In this basis we obtain simple expressions for the $c^{ab,k}_{\ell_x}(X)$ 
coefficients\footnote{%
In this Appendix we follow the notation of Elbaz~\cite{elbaz} where
\begin{equation}
[a_1 ... a_n] = \sqrt{(2a_1+1)...(2a_n+1)}
\end{equation}}
\begin{eqnarray}
c_{\ell _x}^{00,k}(X) &=&  c_0 {\text tr}(\gamma^0 X\Gamma^{00}_\mu) \langle
j', m', \ell' | \sigma^\mu Y_{k,m_k} | j,m,\ell \rangle  \\ 
c_{\ell _x}^{01,k}(X) &=&  c_0 {\text tr}(\gamma^0 X\Gamma^{01}_\mu) \langle
j', m', \ell' | \sigma^\mu Y_{k,m_k} | j,m,2j-\ell \rangle \\ 
c_{\ell _x}^{10,k}(X) &=&  c_0 {\text tr}(\gamma^0 X\Gamma^{10}_\mu) \langle
j', m', 2j'-\ell' | \sigma^\mu Y_{k,m_k} | j,m,\ell \rangle  \\
c_{\ell _x}^{11,k}(X) &=&  c_0 {\text tr}(\gamma^0 X\Gamma^{11}_\mu) \langle
j', m', 2j'-\ell' | \sigma^\mu Y_{k,m_k} | j,m,2j-\ell \rangle  
\end{eqnarray}
where
\begin{equation}
c_0 = 4\pi(-i)^{\ell_x}
\frac{\sum_s C^{J',M'}_{j',M'-s;\frac12,s} C^{J,M}_{j,M-s;\frac12,s}}
{C^{J',M'}_{\ell_x,m_x;j,M}} \frac12
\end{equation}

For the particular case $X=\sla p \gamma^5$ the angular dependence from the 
$\mathbf p$ vector disappears and we obtain the following explicit 
expression
\begin{eqnarray}
c_{\ell _x}^{00,k}(\sla p \gamma^5) &=& +c_2 |p| \langle j', \ell' ||
T^{(k)}_{\ell_x} || j,\ell \rangle  \\
c_{\ell _x}^{01,k}(\sla p \gamma^5) &=& +c_1 p_0 \langle j', \ell' || Y_k || j,
2j-\ell \rangle  \\
c_{\ell _x}^{10,k}(\sla p \gamma^5) &=& -c_1 p_0 \langle j', 2j'-\ell' || Y_k
|| j,\ell \rangle  \\
c_{\ell _x}^{11,k}(\sla p \gamma^5) &=& -c_2 |p| \langle j',  2j'-\ell' ||
T^{(k)}_{\ell_x} || j,2j-\ell \rangle  
\end{eqnarray}
with
\begin{eqnarray}
c_1&=&4\pi i(-i)^{\ell_x}
(-)^{j^{\prime }-\frac12+J'+\ell'-\ell+\ell_x}[J]\left\{ 
\begin{array}{lll}
\ell _x & j & j^{\prime } \\ 
\frac{1}{2} & J^{\prime } & J
\end{array}
\right\} \delta_{k,\ell_x} \\
c_2&=&4\pi (-i)^{k}(-)^{j^{\prime }-\frac12 +J'-k} [J k] \left\{ 
\begin{array}{lll}
\ell _x & j & j^{\prime } \\
\frac{1}{2} & J^{\prime } & J
\end{array}
\right\} 
\left( 
\begin{array}{lll}
1 & k & \ell _{\pi } \\ 
0 & 0 & 0
\end{array}
\right)
\end{eqnarray}
Explicit expressions for the Wigner-Ekkart reduced tensors are 
\begin{equation}
\left\langle \ell^{\prime }\hspace{0.1cm}j^{\prime }\right\| Y_{k}\left\|
\ell%
\hspace{0.1cm}j\right\rangle =\frac{[j\hspace{0.1cm}j^{\prime }\hspace{0.1cm}%
\ell\hspace{0.1cm}\ell^{\prime }\hspace{0.1cm}k]}{\sqrt{4\pi
}}(-)^{j+\ell^{\prime
}-\ell+1/2}\left( 
\begin{array}{lll}
\ell^{\prime } & k & \ell \\ 
0 & 0 & 0
\end{array}
\right) \left\{ 
\begin{array}{lll}
k & \ell^{\prime } & \ell \\ 
\frac{1}{2} & j & j^{\prime }
\end{array}
\right\}   \label{reducedY}
\end{equation}
and
\begin{equation}
\left\langle \ell^{\prime }\hspace{0.1cm}j^{\prime }\right\|
T_{\ell_x}^{(k)}\left\|
\ell\hspace{0.1cm}j\right\rangle =\sqrt{\frac{3}{2\pi }}[j\hspace{0.1cm}
j^{\prime }\hspace{0.1cm}\ell\hspace{0.1cm}\ell^{\prime
}\hspace{0.1cm}k\hspace{
0.1cm}\ell_x](-)^{\ell^{\prime }}\left( 
\begin{array}{lll}
\ell^{\prime } & k & \ell \\ 
0 & 0 & 0
\end{array}
\right) \left\{ 
\begin{array}{lll}
\frac{1}{2} & \ell^{\prime } & j^{\prime } \\ 
\frac{1}{2} & \ell & j \\ 
1 & k & \ell_x
\end{array}
\right\} 
\end{equation}
Our expression for the spin structure of the decay amplitudes, reported in 
this Appendix, disagree with Ref.~\cite{GR} in the overall
phase factor. This factor does not affect their results but is relevant
in case of mixing.

\section*{Appendix B}

In this appendix we list the hadronic transitions 
$H' \rightarrow H + x$ for the low-lying excited states 
not discussed in Section 3.2.
The following Tables list the transitions in 
decending order according to:
\begin{itemize}
\item The flavor of the decaying heavy meson $H'$; 
\item Its mass; 
\item The decay channel.
\end{itemize}

Transitions with a branching ratio less than 
1\% are not reported.  Some decays are marked with an asterisk. 
These are decays that apparently do
not conserve the heavy quark spin ($|j'-j| \le \ell_x \le |j'+j|$) but,
we remind the reader, our initial and final states are physical states
and therefore such decays become allowed because of mixing effects.


\twocolumn
{\footnotesize
\vskip 5mm 
\begin{tabular}{l}

\begin{tabular}{lr}
$H'=D (2^{\frac12}S_0)$ & 
$m = 2.589$ GeV 
\end{tabular} \\ 

\begin{tabular}{|ccrrr|} \hline
$H (n^j\ell_J)$ & $x$ & $\ell_x$ & $p_x$ & $\Gamma_x/(g_A^8)^2$\\ \hline
\smash{$D (1^{\frac12}S_1)$} & \smash{$\pi$} & 1 & 504 &  $14.5$ \hskip 5mm \\
\smash{$D (1^{\frac12}P_0)$} & \smash{$\pi$} & 0 & 154 &  $7.0$ \hskip 5mm \\ 
\hline \end{tabular}
\end{tabular}

\vskip 5mm 
\begin{tabular}{l}

\begin{tabular}{lr}
$H'=D (2^{\frac12}S_1)$ & 
$m = 2.692$ GeV 
\end{tabular} \\ 

\begin{tabular}{|ccrrr|} \hline
$H (n^j\ell_J)$ & $x$ & $\ell_x$ & $p_x$ & $\Gamma_x/(g_A^8)^2$\\ \hline
\smash{$D (1^{\frac12}S_0)$} & \smash{$\eta$} & 1 & 518 &  $1.4$ \hskip 5mm \\
\smash{$D (1^{\frac12}S_0)$} & \smash{$\pi$} & 1 & 688 &  $39.9$ \hskip 5mm \\
\smash{$D (1^{\frac12}S_1)$} & \smash{$\pi$} & 1 & 587 &  $30.4$ \hskip 5mm \\
\smash{$D (1^{\frac32}P_1)$} & \smash{$\pi$} & 2 & 225 &  $1.9$ \hskip 5mm \\ 
\smash{$D (1^{\frac12}P_1)$} & \smash{$\pi$} & 0 & 141 &  $6.2$ \hskip 5mm \\ 
\smash{$D_s (1^{\frac12}S_0)$} & \smash{$K$} & 1 & 460 &  $5.6$ \hskip 5mm \\ 
\hline \end{tabular}
\end{tabular}

\vskip 5mm 
\begin{tabular}{l}

\begin{tabular}{lr}
$H'=D (1^{\frac52}D_2)$ & 
$m = 2.775$ GeV 
\end{tabular} \\ 

\begin{tabular}{|ccrrr|} \hline
$H (n^j\ell_J)$ & $x$ & $\ell_x$ & $p_x$ & $\Gamma_x/(g_A^8)^2$\\ \hline
\smash{$D (1^{\frac12}S_1)$} & \smash{$\pi$} & 3 & 652 &  $20.1$ \hskip 5mm \\
\smash{$D (1^{\frac12}P_0)$} & \smash{$\pi$} & 2 & 347 &  $7.3$ \hskip 5mm \\ 
\smash{$D (1^{\frac32}P_1)$} & \smash{$\pi$} & 2 & 308 &  $4.3$ \hskip 5mm \\ 
\smash{$D (1^{\frac32}P_2)$} & \smash{$\pi$} & 2 & 266 &  $1.4$ \hskip 5mm \\ 
\smash{$D (1^{\frac12}P_1)$} & \smash{$\pi$} & 2 & 236 &  $1.3$ \hskip 5mm \\ 
\smash{$D_s (1^{\frac12}S_1)$} & \smash{$K$} & 3 & 387 &  $0.5$ \hskip 5mm \\ 
\hline \end{tabular}
\end{tabular}

\vskip 5mm 
\begin{tabular}{l}

\begin{tabular}{lr}
$H'=D (1^{\frac32}D_1)$ & 
$m = 2.795$ GeV 
\end{tabular} \\ 

\begin{tabular}{|ccrrr|} \hline
$H (n^j\ell_J)$ & $x$ & $\ell_x$ & $p_x$ & $\Gamma_x/(g_A^8)^2$\\ \hline
\smash{$D (1^{\frac12}S_0)$} & \smash{$\eta$} & 1 & 620 &  $4.0$ \hskip 5mm \\
\smash{$D (1^{\frac12}S_0)$} & \smash{$\pi$} & 1 & 764 &  $18.6$ \hskip 5mm \\
\smash{$D (1^{\frac12}S_1)$} & \smash{$\pi$} & 1 & 668 &  $6.8$ \hskip 5mm \\ 
\smash{$D (1^{\frac32}P_1)$} & \smash{$\pi$} & 0 & 328 &  $87.2$ \hskip 5mm \\
\smash{$D (1^{\frac32}P_1)$} & \smash{$\pi$} & 2 & 328 &  $2.4$ \hskip 5mm \\ 
\smash{$D (1^{\frac32}P_2)$} & \smash{$\pi$} & 2 & 286 &  $1.4$ \hskip 5mm \\ 
\smash{$D_s (1^{\frac12}S_0)$} & \smash{$K$} & 1 & 566 &  $15.0$ \hskip 5mm \\
\smash{$D_s (1^{\frac12}S_1)$} & \smash{$K$} & 1 & 412 &  $3.3$ \hskip 5mm \\ 
\hline \end{tabular}
\end{tabular}

\vskip 5mm 
\begin{tabular}{l}

\begin{tabular}{lr}
$H'=D (1^{\frac52}D_3)$ & 
$m = 2.799$ GeV 
\end{tabular} \\ 

\begin{tabular}{|ccrrr|} \hline
$H (n^j\ell_J)$ & $x$ & $\ell_x$ & $p_x$ & $\Gamma_x/(g_A^8)^2$\\ \hline
\smash{$D (1^{\frac12}S_0)$} & \smash{$\eta$} & 3 & 624 &  $0.7$ \hskip 5mm \\
\smash{$D (1^{\frac12}S_0)$} & \smash{$\pi$} & 3 & 767 &  $18.0$ \hskip 5mm \\
\smash{$D (1^{\frac12}S_1)$} & \smash{$\pi$} & 3 & 671 &  $13.2$ \hskip 5mm \\
\smash{$D (1^{\frac32}P_1)$} & \smash{$\pi$} & 2 & 331 &  $2.5$ \hskip 5mm \\ 
\smash{$D (1^{\frac32}P_2)$} & \smash{$\pi$} & 2 & 290 &  $5.2$ \hskip 5mm \\ 
\smash{$D (1^{\frac12}P_1)$} & \smash{$\pi$} & 2 & 261 &  $3.5$ \hskip 5mm \\ 
\smash{$D_s (1^{\frac12}S_0)$} & \smash{$K$} & 3 & 570 &  $2.1$ \hskip 5mm \\ 
\hline \end{tabular}
\end{tabular}

\vskip 5mm 
\begin{tabular}{l}

\begin{tabular}{lr}
$H'=D (1^{\frac32}D_2)$ & 
$m = 2.833$ GeV 
\end{tabular} \\ 

\begin{tabular}{|ccrrr|} \hline
$H (n^j\ell_J)$ & $x$ & $\ell_x$ & $p_x$ & $\Gamma_x/(g_A^8)^2$\\ \hline
\smash{$D (1^{\frac12}S_1)$} & \smash{$\eta$} & 1 & 528 &  $4.4$ \hskip 5mm \\
\smash{$D (1^{\frac12}S_1)$} & \smash{$\pi$} & 1 & 697 &  $22.9$ \hskip 5mm \\
\smash{$D (1^{\frac12}P_0)$} & \smash{$\pi$} & 2 & 400 &  $1.5$ \hskip 5mm \\ 
\smash{$D (1^{\frac32}P_1)$} & \smash{$\pi$} & 2 & 363 &  $3.6$ \hskip 5mm \\ 
\smash{$D (1^{\frac32}P_2)$} & \smash{$\pi$} & 0 & 323 &  $87.0$ \hskip 5mm \\
\smash{$D (1^{\frac32}P_2)$} & \smash{$\pi$} & 2 & 323 &  $3.2$ \hskip 5mm \\ 
\smash{$D_s (1^{\frac12}S_1)$} & \smash{$K$} & 1 & 456 &  $13.0$ \hskip 5mm \\
\hline \end{tabular}
\end{tabular}

\vskip 5mm 
\begin{tabular}{l}

\begin{tabular}{lr}
$H'=D (2^{\frac12}P_0)$ & 
$m = 2.949$ GeV 
\end{tabular} \\ 

\begin{tabular}{|ccrrr|} \hline
$H (n^j\ell_J)$ & $x$ & $\ell_x$ & $p_x$ & $\Gamma_x/(g_A^8)^2$\\ \hline
\smash{$D (1^{\frac12}S_0)$} & \smash{$\eta$} & 0 & 756 &  $11.7$ \hskip 5mm
\\ 
\smash{$D (1^{\frac12}S_0)$} & \smash{$\pi$} & 0 & 875 &  $88.0$ \hskip 5mm \\
\smash{$D (1^{\frac32}P_1)$} & \smash{$\pi$} & 1 & 467 &  $15.2$ \hskip 5mm \\
\smash{$D (1^{\frac12}P_1)$} & \smash{$\pi$} & 1 & 403 &  $60.6$ \hskip 5mm \\
\smash{$D (2^{\frac12}S_0)$} & \smash{$\pi$} & 0 & 311 &  $51.9$ \hskip 5mm \\
\smash{$D_s (1^{\frac12}S_0)$} & \smash{$K$} & 0 & 706 &  $66.6$ \hskip 5mm \\
\hline \end{tabular}
\end{tabular}

\vskip 5mm 
\begin{tabular}{l}

\begin{tabular}{lr}
$H'=D (2^{\frac32}P_1)$ & 
$m = 2.995$ GeV 
\end{tabular} \\ 

\begin{tabular}{|ccrrr|} \hline
$H (n^j\ell_J)$ & $x$ & $\ell_x$ & $p_x$ & $\Gamma_x/(g_A^8)^2$\\ \hline
\smash{$D (1^{\frac12}S_1)$} & \smash{$\eta$} & 2 & 685 &  $2.7$ \hskip 5mm \\
\smash{$D (1^{\frac12}S_1)$} & \smash{$\pi$} & 2 & 818 &  $62.5$ \hskip 5mm \\
\smash{$D (1^{\frac12}P_0)$} & \smash{$\pi$} & 1 & 540 &  $1.5$ \hskip 5mm \\ 
\smash{$D (1^{\frac32}P_1)$} & \smash{$\pi$} & 1 & 506 &  $5.9$ \hskip 5mm \\ 
\smash{$D (1^{\frac32}P_2)$} & \smash{$\pi$} & 3 & 470 &  $4.0$ \hskip 5mm \\ 
\smash{$D (1^{\frac12}P_1)$} & \smash{$\pi$} & 1 & 444 &  $3.9$ \hskip 5mm \\ 
\smash{$D (2^{\frac12}S_1)$} & \smash{$\pi$} & 2 & 255 &  $5.1$ \hskip 5mm \\ 
\smash{$D (1^{\frac32}D_1)$} & \smash{$\pi$} & 0 & 138 &  $3.8$ \hskip 5mm \\ 
\smash{$D_s (1^{\frac12}S_1)$} & \smash{$K$} & 2 & 620 &  $9.9$ \hskip 5mm \\ 
\hline \end{tabular}
\end{tabular}

\vskip 5mm 
\begin{tabular}{l}

\begin{tabular}{lr}
$H'=D (2^{\frac32}P_2)$ & 
$m = 3.035$ GeV 
\end{tabular} \\ 

\begin{tabular}{|ccrrr|} \hline
$H (n^j\ell_J)$ & $x$ & $\ell_x$ & $p_x$ & $\Gamma_x/(g_A^8)^2$\\ \hline
\smash{$D (1^{\frac12}S_0)$} & \smash{$\eta$} & 2 & 828 &  $3.3$ \hskip 5mm \\
\smash{$D (1^{\frac12}S_0)$} & \smash{$\pi$} & 2 & 936 &  $53.4$ \hskip 5mm \\
\smash{$D (1^{\frac12}S_1)$} & \smash{$\eta$} & 2 & 721 &  $2.3$ \hskip 5mm \\
\smash{$D (1^{\frac12}S_1)$} & \smash{$\pi$} & 2 & 848 &  $47.1$ \hskip 5mm \\
\smash{$D (1^{\frac32}P_1)$} & \smash{$\pi$} & 3 & 541 &  $4.3$ \hskip 5mm \\ 
\smash{$D (1^{\frac32}P_2)$} & \smash{$\pi$} & 1 & 505 &  $4.2$ \hskip 5mm \\ 
\smash{$D (1^{\frac32}P_2)$} & \smash{$\pi$} & 3 & 505 &  $2.2$ \hskip 5mm \\ 
\smash{$D (1^{\frac12}P_1)$} & \smash{$\pi$} & 1 & 480 &  $6.1$ \hskip 5mm \\ 
\smash{$D (2^{\frac12}S_0)$} & \smash{$\pi$} & 2 & 393 &  $10.8$ \hskip 5mm \\
\smash{$D (2^{\frac12}S_1)$} & \smash{$\pi$} & 2 & 296 &  $5.8$ \hskip 5mm \\ 
\smash{$D (1^{\frac32}D_2)$} & \smash{$\pi$} & 0 & 142 &  $3.9$ \hskip 5mm \\ 
\smash{$D_s (1^{\frac12}S_0)$} & \smash{$K$} & 2 & 779 &  $15.3$ \hskip 5mm \\
\smash{$D_s (1^{\frac12}S_1)$} & \smash{$K$} & 2 & 658 &  $8.7$ \hskip 5mm \\ 
\hline \end{tabular}
\end{tabular}

\vskip 5mm 
\begin{tabular}{l}

\begin{tabular}{lr}
$H'=D (2^{\frac12}P_1)$ & 
$m = 3.045$ GeV 
\end{tabular} \\ 

\begin{tabular}{|ccrrr|} \hline
$H (n^j\ell_J)$ & $x$ & $\ell_x$ & $p_x$ & $\Gamma_x/(g_A^8)^2$\\ \hline
\smash{$D (1^{\frac12}S_1)$} & \smash{$\eta$} & 0 & 730 &  $10.9$ \hskip 5mm
\\ 
\smash{$D (1^{\frac12}S_1)$} & \smash{$\pi$} & 0 & 855 &  $85.4$ \hskip 5mm \\
\smash{$D (1^{\frac12}P_0)$} & \smash{$\pi$} & 1 & 582 &  $61.0$ \hskip 5mm \\
\smash{$D (1^{\frac32}P_1)$} & \smash{$\pi$} & 1 & 549 &  $6.6$ \hskip 5mm \\ 
\smash{$D (1^{\frac32}P_2)$} & \smash{$\pi$} & 1 & 513 &  $19.9$ \hskip 5mm \\
\smash{$D (1^{\frac12}P_1)$} & \smash{$\pi$} & 1 & 488 &  $61.5$ \hskip 5mm \\
\smash{$D (2^{\frac12}S_1)$} & \smash{$\pi$} & 0 & 306 &  $50.6$ \hskip 5mm \\
\smash{$D_s (1^{\frac12}S_1)$} & \smash{$K$} & 0 & 667 &  $59.3$ \hskip 5mm \\
\hline \end{tabular}
\end{tabular}

\vskip 5mm 
\begin{tabular}{l}

\begin{tabular}{lr}
$H'=D (1^{\frac72}F_3)$ & 
$m = 3.074$ GeV 
\end{tabular} \\ 

\begin{tabular}{|ccrrr|} \hline
$H (n^j\ell_J)$ & $x$ & $\ell_x$ & $p_x$ & $\Gamma_x/(g_A^8)^2$\\ \hline
\smash{$D (1^{\frac12}S_1)$} & \smash{$\eta$} & 4 & 754 &  $0.5$ \hskip 5mm \\
\smash{$D (1^{\frac12}S_1)$} & \smash{$\pi$} & 4 & 875 &  $13.1$ \hskip 5mm \\
\smash{$D (1^{\frac12}P_0)$} & \smash{$\pi$} & 3 & 606 &  $11.4$ \hskip 5mm \\
\smash{$D (1^{\frac32}P_1)$} & \smash{$\pi$} & 3 & 573 &  $7.3$ \hskip 5mm \\ 
\smash{$D (1^{\frac32}P_2)$} & \smash{$\pi$} & 3 & 538 &  $3.8$ \hskip 5mm \\ 
\smash{$D (1^{\frac12}P_1)$} & \smash{$\pi$} & 3 & 513 &  $5.0$ \hskip 5mm \\ 
\smash{$D (1^{\frac52}D_2)$} & \smash{$\pi$} & 2 & 251 &  $2.9$ \hskip 5mm \\ 
\smash{$D (1^{\frac32}D_1)$} & \smash{$\pi$} & 2 & 230 &  $3.3$ \hskip 5mm \\ 
\smash{$D_s (1^{\frac12}S_1)$} & \smash{$K$} & 4 & 693 &  $1.5$ \hskip 5mm \\ 
\hline \end{tabular}
\end{tabular}

\vskip 5mm 
\begin{tabular}{l}

\begin{tabular}{lr}
$H'=D (1^{\frac72}F_4)$ & 
$m = 3.091$ GeV 
\end{tabular} \\ 

\begin{tabular}{|ccrrr|} \hline
$H (n^j\ell_J)$ & $x$ & $\ell_x$ & $p_x$ & $\Gamma_x/(g_A^8)^2$\\ \hline
\smash{$D (1^{\frac12}S_0)$} & \smash{$\eta$} & 4 & 872 &  $0.6$ \hskip 5mm \\
\smash{$D (1^{\frac12}S_0)$} & \smash{$\pi$} & 4 & 974 &  $9.8$ \hskip 5mm \\ 
\smash{$D (1^{\frac12}S_1)$} & \smash{$\pi$} & 4 & 887 &  $7.8$ \hskip 5mm \\ 
\smash{$D (1^{\frac32}P_1)$} & \smash{$\pi$} & 3 & 587 &  $4.4$ \hskip 5mm \\ 
\smash{$D (1^{\frac32}P_2)$} & \smash{$\pi$} & 3 & 552 &  $9.9$ \hskip 5mm \\ 
\smash{$D (1^{\frac12}P_1)$} & \smash{$\pi$} & 3 & 527 &  $9.6$ \hskip 5mm \\ 
\smash{$D (1^{\frac52}D_2)$} & \smash{$\pi$} & 2 & 268 &  $0.6$ \hskip 5mm \\ 
\smash{$D (1^{\frac52}D_3)$} & \smash{$\pi$} & 2 & 244 &  $2.9$ \hskip 5mm \\ 
\smash{$D (1^{\frac32}D_2)$} & \smash{$\pi$} & 2 & 208 &  $2.4$ \hskip 5mm \\ 
\smash{$D_s (1^{\frac12}S_0)$} & \smash{$K$} & 4 & 824 &  $2.0$ \hskip 5mm \\ 
\smash{$D_s (1^{\frac12}S_1)$} & \smash{$K$} & 4 & 707 &  $0.9$ \hskip 5mm \\ 
\hline \end{tabular}
\end{tabular}

\vskip 5mm 
\begin{tabular}{l}

\begin{tabular}{lr}
$H'=D (1^{\frac52}F_2)$ & 
$m = 3.101$ GeV 
\end{tabular} \\ 

\begin{tabular}{|ccrrr|} \hline
$H (n^j\ell_J)$ & $x$ & $\ell_x$ & $p_x$ & $\Gamma_x/(g_A^8)^2$\\ \hline
\smash{$D (1^{\frac12}S_0)$} & \smash{$\pi$} & 2 & 981 &  $3.7$ \hskip 5mm \\ 
\smash{$D (1^{\frac12}S_1)$} & \smash{$\pi$} & 2 & 895 &  $1.8$ \hskip 5mm \\ 
\smash{$D (1^{\frac32}P_1)$} & \smash{$\eta$} & 1 & 362 &  $3.7$ \hskip 5mm \\
\smash{$D (1^{\frac32}P_1)$} & \smash{$\pi$} & 1 & 595 &  $22.2$ \hskip 5mm \\
\smash{$D (1^{\frac32}P_1)$} & \smash{$\pi$} & 3 & 595 &  $1.4$ \hskip 5mm \\ 
\smash{$D (1^{\frac32}P_2)$} & \smash{$\pi$} & 1 & 560 &  $2.2$ \hskip 5mm \\ 
\smash{$D (1^{\frac32}P_2)$} & \smash{$\pi$} & 3 & 560 &  $2.1$ \hskip 5mm \\ 
\smash{$D (1^{\frac12}P_1)$} & \smash{$\pi$} & 3 & 536 &  $2.6$ \hskip 5mm \\ 
\smash{$D (1^{\frac52}D_2)$} & \smash{$\pi$} & 0 & 279 &  $56.8$ \hskip 5mm \\
\smash{$D (1^{\frac52}D_2)$} & \smash{$\pi$} & 2 & 279 &  $3.4$ \hskip 5mm \\ 
\smash{$D_s (1^{\frac12}S_0)$} & \smash{$K$} & 2 & 832 &  $3.1$ \hskip 5mm \\ 
\smash{$D_s (1^{\frac12}S_1)$} & \smash{$K$} & 2 & 716 &  $1.2$ \hskip 5mm \\ 
\smash{$D_s (1^{\frac32}P_1)$} & \smash{$K$} & 1 & 250 &  $6.0$ \hskip 5mm \\ 
\hline \end{tabular}
\end{tabular}

\vskip 5mm 
\begin{tabular}{l}

\begin{tabular}{lr}
$H'=D (1^{\frac52}F_3)$ & 
$m = 3.123$ GeV 
\end{tabular} \\ 

\begin{tabular}{|ccrrr|} \hline
$H (n^j\ell_J)$ & $x$ & $\ell_x$ & $p_x$ & $\Gamma_x/(g_A^8)^2$\\ \hline
\smash{$D (1^{\frac12}S_1)$} & \smash{$\pi$} & 2 & 910 &  $4.9$ \hskip 5mm \\ 
\smash{$D (1^{\frac12}P_0)$} & \smash{$\pi$} & 3 & 645 &  $4.3$ \hskip 5mm \\ 
\smash{$D (1^{\frac32}P_1)$} & \smash{$\pi$} & 3 & 613 &  $3.2$ \hskip 5mm \\ 
\smash{$D (1^{\frac32}P_2)$} & \smash{$\eta$} & 1 & 332 &  $3.4$ \hskip 5mm \\
\smash{$D (1^{\frac32}P_2)$} & \smash{$\pi$} & 1 & 579 &  $23.7$ \hskip 5mm \\
\smash{$D (1^{\frac32}P_2)$} & \smash{$\pi$} & 3 & 579 &  $2.6$ \hskip 5mm \\ 
\smash{$D (1^{\frac12}P_1)$} & \smash{$\pi$} & 3 & 554 &  $1.2$ \hskip 5mm \\ 
\smash{$D (1^{\frac52}D_2)$} & \smash{$\pi$} & 2 & 300 &  $1.2$ \hskip 5mm \\ 
\smash{$D (1^{\frac52}D_3)$} & \smash{$\pi$} & 0 & 277 &  $56.7$ \hskip 5mm \\
\smash{$D (1^{\frac52}D_3)$} & \smash{$\pi$} & 2 & 277 &  $3.6$ \hskip 5mm \\ 
\smash{$D_s (1^{\frac12}S_1)$} & \smash{$K$} & 2 & 735 &  $3.3$ \hskip 5mm \\ 
\smash{$D_s (1^{\frac32}P_2)$} & \smash{$K$} & 1 & 203 &  $3.7$ \hskip 5mm \\ 
\hline \end{tabular}
\end{tabular}

\vskip 5mm 
\begin{tabular}{l}

\begin{tabular}{lr}
$H'=D (3^{\frac12}S_0)$ & 
$m = 3.141$ GeV 
\end{tabular} \\ 

\begin{tabular}{|ccrrr|} \hline
$H (n^j\ell_J)$ & $x$ & $\ell_x$ & $p_x$ & $\Gamma_x/(g_A^8)^2$\\ \hline
\smash{$D (1^{\frac12}S_1)$} & \smash{$\eta$} & 1 & 811 &  $1.7$ \hskip 5mm \\
\smash{$D (1^{\frac12}S_1)$} & \smash{$\pi$} & 1 & 923 &  $44.5$ \hskip 5mm \\
\smash{$D (1^{\frac12}P_0)$} & \smash{$\pi$} & 0 & 660 &  $2.9$ \hskip 5mm \\ 
\smash{$D (1^{\frac32}P_2)$} & \smash{$\pi$} & 2 & 594 &  $15.1$ \hskip 5mm \\
\smash{$D (2^{\frac12}S_1)$} & \smash{$\pi$} & 1 & 396 &  $16.3$ \hskip 5mm \\
\smash{$D (2^{\frac12}P_0)$} & \smash{$\pi$} & 0 & 128 &  $6.5$ \hskip 5mm \\ 
\smash{$D_s (1^{\frac12}S_1)$} & \smash{$K$} & 1 & 751 &  $8.7$ \hskip 5mm \\ 
\smash{$D_s (1^{\frac12}P_0)$} & \smash{$K$} & 0 & 384 &  $1.4$ \hskip 5mm \\ 
\hline \end{tabular}
\end{tabular}

\vskip 5mm 
\begin{tabular}{l}

\begin{tabular}{lr}
$H'=D (3^{\frac12}S_1)$ & 
$m = 3.226$ GeV 
\end{tabular} \\ 

\begin{tabular}{|ccrrr|} \hline
$H (n^j\ell_J)$ & $x$ & $\ell_x$ & $p_x$ & $\Gamma_x/(g_A^8)^2$\\ \hline
\smash{$D (1^{\frac12}S_0)$} & \smash{$\eta$} & 1 & 976 &  $3.2$ \hskip 5mm \\
\smash{$D (1^{\frac12}S_0)$} & \smash{$\pi$} & 1 & 1066 &  $50.2$ \hskip 5mm
\\ 
\smash{$D (1^{\frac12}S_1)$} & \smash{$\pi$} & 1 & 984 &  $53.5$ \hskip 5mm \\
\smash{$D (1^{\frac32}P_1)$} & \smash{$\pi$} & 2 & 697 &  $23.1$ \hskip 5mm \\
\smash{$D (1^{\frac32}P_2)$} & \smash{$\pi$} & 2 & 663 &  $17.0$ \hskip 5mm \\
\smash{$D (2^{\frac12}S_0)$} & \smash{$\pi$} & 1 & 560 &  $41.3$ \hskip 5mm \\
\smash{$D (2^{\frac12}S_1)$} & \smash{$\pi$} & 1 & 473 &  $33.5$ \hskip 5mm \\
\smash{$D (2^{\frac12}P_1)$} & \smash{$\pi$} & 0 & 113 &  $5.6$ \hskip 5mm \\ 
\smash{$D_s (1^{\frac12}S_0)$} & \smash{$K$} & 1 & 930 &  $17.6$ \hskip 5mm \\
\smash{$D_s (1^{\frac12}S_1)$} & \smash{$K$} & 1 & 822 &  $13.4$ \hskip 5mm \\
\hline \end{tabular}
\end{tabular}

\vskip 5mm 
\begin{tabular}{l}

\begin{tabular}{lr}
$H'=D_s (2^{\frac12}S_0)$ & 
$m = 2.700$ GeV 
\end{tabular} \\ 

\begin{tabular}{|ccrrr|} \hline
$H (n^j\ell_J)$ & $x$ & $\ell_x$ & $p_x$ & $\Gamma_x/(g_A^8)^2$\\ \hline
\smash{$D (1^{\frac12}S_1)$} & \smash{$K$} & 1 & 424 &  $3.12$ \hskip 5mm \\ 
\smash{$D_s (1^{\frac12}S_1)$} & \smash{$\eta$} & 1 & 187 &  $0.04$ \hskip 5mm
\\ 
\hline \end{tabular}
\end{tabular}

\vskip 5mm 
\begin{tabular}{l}

\begin{tabular}{lr}
$H'=D_s (2^{\frac12}S_1)$ & 
$m = 2.806$ GeV 
\end{tabular} \\ 

\begin{tabular}{|ccrrr|} \hline
$H (n^j\ell_J)$ & $x$ & $\ell_x$ & $p_x$ & $\Gamma_x/(g_A^8)^2$\\ \hline
\smash{$D (1^{\frac12}S_0)$} & \smash{$K$} & 1 & 661 &  $21.1$ \hskip 5mm \\ 
\smash{$D (1^{\frac12}S_1)$} & \smash{$K$} & 1 & 539 &  $12.2$ \hskip 5mm \\ 
\smash{$D_s (1^{\frac12}S_0)$} & \smash{$\eta$} & 1 & 540 &  $6.2$ \hskip 5mm
\\ 
\smash{$D_s (1^{\frac12}S_1)$} & \smash{$\eta$} & 1 & 371 &  $1.5$ \hskip 5mm
\\ 
\hline \end{tabular}
\end{tabular}

\vskip 5mm 
\begin{tabular}{l}

\begin{tabular}{lr}
$H'=D_s (1^{\frac52}D_2)$ & 
$m = 2.900$ GeV 
\end{tabular} \\ 

\begin{tabular}{|ccrrr|} \hline
$H (n^j\ell_J)$ & $x$ & $\ell_x$ & $p_x$ & $\Gamma_x/(g_A^8)^2$\\ \hline
\smash{$D (1^{\frac12}S_1)$} & \smash{$K$} & 3 & 629 &  $10.6$ \hskip 5mm \\ 
\smash{$D_s (1^{\frac12}S_1)$} & \smash{$\eta$} & 3 & 486 &  $1.4$ \hskip 5mm
\\ 
\hline \end{tabular}
\end{tabular}

\vskip 5mm 
\begin{tabular}{l}

\begin{tabular}{lr}
$H'=D_s (1^{\frac32}D_1)$ & 
$m = 2.913$ GeV 
\end{tabular} \\ 

\begin{tabular}{|ccrrr|} \hline
$H (n^j\ell_J)$ & $x$ & $\ell_x$ & $p_x$ & $\Gamma_x/(g_A^8)^2$\\ \hline
\smash{$D (1^{\frac12}S_0)$} & \smash{$K$} & 1 & 752 &  $26.1$ \hskip 5mm \\ 
\smash{$D (1^{\frac12}S_1)$} & \smash{$K$} & 1 & 641 &  $10.7$ \hskip 5mm \\ 
\smash{$D (1^{\frac32}P_1)$} & \smash{$K$} & 0 & 47 &  $39.7$ \hskip 5mm \\ 
\smash{$D_s (1^{\frac12}S_0)$} & \smash{$\eta$} & 1 & 644 &  $15.2$ \hskip 5mm
\\ 
\smash{$D_s (1^{\frac12}S_1)$} & \smash{$\eta$} & 1 & 501 &  $4.5$ \hskip 5mm
\\ 
\hline \end{tabular}
\end{tabular}

\vskip 5mm 
\begin{tabular}{l}

\begin{tabular}{lr}
$H'=D_s (1^{\frac52}D_3)$ & 
$m = 2.925$ GeV 
\end{tabular} \\ 

\begin{tabular}{|ccrrr|} \hline
$H (n^j\ell_J)$ & $x$ & $\ell_x$ & $p_x$ & $\Gamma_x/(g_A^8)^2$\\ \hline
\smash{$D (1^{\frac12}S_0)$} & \smash{$K$} & 3 & 762 &  $11.4$ \hskip 5mm \\ 
\smash{$D (1^{\frac12}S_1)$} & \smash{$K$} & 3 & 652 &  $7.3$ \hskip 5mm \\ 
\smash{$D_s (1^{\frac12}S_0)$} & \smash{$\eta$} & 3 & 656 &  $3.1$ \hskip 5mm
\\ 
\smash{$D_s (1^{\frac12}S_1)$} & \smash{$\eta$} & 3 & 514 &  $1.1$ \hskip 5mm
\\ 
\hline \end{tabular}
\end{tabular}

\vskip 5mm 
\begin{tabular}{l}

\begin{tabular}{lr}
$H'=D_s (1^{\frac32}D_2)$ & 
$m = 2.953$ GeV 
\end{tabular} \\ 

\begin{tabular}{|ccrrr|} \hline
$H (n^j\ell_J)$ & $x$ & $\ell_x$ & $p_x$ & $\Gamma_x/(g_A^8)^2$\\ \hline
\smash{$D (1^{\frac12}S_1)$} & \smash{$K$} & 1 & 677 &  $35.0$ \hskip 5mm \\ 
\smash{$D_s (1^{\frac12}S_1)$} & \smash{$\eta$} & 1 & 543 &  $16.4$ \hskip 5mm
\\ 
\hline \end{tabular}
\end{tabular}

\vskip 5mm 
\begin{tabular}{l}

\begin{tabular}{lr}
$H'=D_s (2^{\frac12}P_0)$ & 
$m = 3.067$ GeV 
\end{tabular} \\ 

\begin{tabular}{|ccrrr|} \hline
$H (n^j\ell_J)$ & $x$ & $\ell_x$ & $p_x$ & $\Gamma_x/(g_A^8)^2$\\ \hline
\smash{$D (1^{\frac12}S_0)$} & \smash{$K$} & 0 & 875 &  $74.1$ \hskip 5mm \\ 
\smash{$D (1^{\frac32}P_1)$} & \smash{$K$} & 1 & 377 &  $32.3$ \hskip 5mm \\ 
\smash{$D (1^{\frac12}P_1)$} & \smash{$K$} & 1 & 270 &  $17.8$ \hskip 5mm \\ 
\smash{$D_s (1^{\frac12}S_0)$} & \smash{$\eta$} & 0 & 780 &  $49.1$ \hskip 5mm
\\ 
\hline \end{tabular}
\end{tabular}

\vskip 5mm 
\begin{tabular}{l}

\begin{tabular}{lr}
$H'=D_s (2^{\frac32}P_1)$ & 
$m = 3.114$ GeV 
\end{tabular} \\ 

\begin{tabular}{|ccrrr|} \hline
$H (n^j\ell_J)$ & $x$ & $\ell_x$ & $p_x$ & $\Gamma_x/(g_A^8)^2$\\ \hline
\smash{$D (1^{\frac12}S_1)$} & \smash{$K$} & 0 & 813 & $(\ast)\, $ $0.7$
\hskip 5mm \\ 
\smash{$D (1^{\frac12}S_1)$} & \smash{$K$} & 2 & 813 &  $36.6$ \hskip 5mm \\ 
\smash{$D (1^{\frac12}P_0)$} & \smash{$K$} & 1 & 482 &  $5.5$ \hskip 5mm \\ 
\smash{$D (1^{\frac32}P_1)$} & \smash{$K$} & 1 & 436 &  $1.6$ \hskip 5mm \\ 
\smash{$D (1^{\frac32}P_2)$} & \smash{$K$} & 3 & 383 &  $0.9$ \hskip 5mm \\ 
\smash{$D (1^{\frac12}P_1)$} & \smash{$K$} & 1 & 343 &  $4.5$ \hskip 5mm \\ 
\smash{$D_s (1^{\frac12}S_1)$} & \smash{$\eta$} & 2 & 700 &  $11.1$ \hskip 5mm
\\ 
\smash{$D_s (1^{\frac12}P_0)$} & \smash{$\eta$} & 1 & 275 &  $1.6$ \hskip 5mm
\\ 
\hline \end{tabular}
\end{tabular}

\vskip 5mm 
\begin{tabular}{l}

\begin{tabular}{lr}
$H'=D_s (2^{\frac32}P_2)$ & 
$m = 3.157$ GeV 
\end{tabular} \\ 

\begin{tabular}{|ccrrr|} \hline
$H (n^j\ell_J)$ & $x$ & $\ell_x$ & $p_x$ & $\Gamma_x/(g_A^8)^2$\\ \hline
\smash{$D (1^{\frac12}S_0)$} & \smash{$K$} & 2 & 943 &  $34.9$ \hskip 5mm \\ 
\smash{$D (1^{\frac12}S_1)$} & \smash{$K$} & 2 & 848 &  $28.9$ \hskip 5mm \\ 
\smash{$D (1^{\frac32}P_1)$} & \smash{$K$} & 3 & 485 &  $1.8$ \hskip 5mm \\ 
\smash{$D (1^{\frac12}P_1)$} & \smash{$K$} & 1 & 400 &  $11.0$ \hskip 5mm \\ 
\smash{$D (2^{\frac12}S_0)$} & \smash{$K$} & 2 & 255 &  $1.3$ \hskip 5mm \\ 
\smash{$D_s (1^{\frac12}S_0)$} & \smash{$\eta$} & 2 & 854 &  $14.8$ \hskip 5mm
\\ 
\smash{$D_s (1^{\frac12}S_1)$} & \smash{$\eta$} & 2 & 738 &  $9.5$ \hskip 5mm
\\ 
\hline \end{tabular}
\end{tabular}

\vskip 5mm 
\begin{tabular}{l}

\begin{tabular}{lr}
$H'=D_s (2^{\frac12}P_1)$ & 
$m = 3.165$ GeV 
\end{tabular} \\ 

\begin{tabular}{|ccrrr|} \hline
$H (n^j\ell_J)$ & $x$ & $\ell_x$ & $p_x$ & $\Gamma_x/(g_A^8)^2$\\ \hline
\smash{$D (1^{\frac12}S_1)$} & \smash{$K$} & 0 & 854 &  $72.1$ \hskip 5mm \\ 
\smash{$D (1^{\frac12}P_0)$} & \smash{$K$} & 1 & 537 &  $42.6$ \hskip 5mm \\ 
\smash{$D (1^{\frac32}P_1)$} & \smash{$K$} & 1 & 494 &  $12.6$ \hskip 5mm \\ 
\smash{$D (1^{\frac32}P_2)$} & \smash{$K$} & 1 & 446 &  $41.3$ \hskip 5mm \\ 
\smash{$D (1^{\frac12}P_1)$} & \smash{$K$} & 1 & 410 &  $28.7$ \hskip 5mm \\ 
\smash{$D_s (1^{\frac12}S_1)$} & \smash{$\eta$} & 0 & 745 &  $45.1$ \hskip 5mm
\\ 
\smash{$D_s (1^{\frac12}P_0)$} & \smash{$\eta$} & 1 & 357 &  $8.1$ \hskip 5mm
\\ 
\smash{$D_s (1^{\frac32}P_1)$} & \smash{$\eta$} & 1 & 280 &  $2.7$ \hskip 5mm
\\ 
\smash{$D_s (1^{\frac32}P_2)$} & \smash{$\eta$} & 1 & 184 &  $3.4$ \hskip 5mm
\\ 
\hline \end{tabular}
\end{tabular}

\vskip 5mm 
\begin{tabular}{l}

\begin{tabular}{lr}
$H'=D_s (1^{\frac72}F_3)$ & 
$m = 3.203$ GeV 
\end{tabular} \\ 

\begin{tabular}{|ccrrr|} \hline
$H (n^j\ell_J)$ & $x$ & $\ell_x$ & $p_x$ & $\Gamma_x/(g_A^8)^2$\\ \hline
\smash{$D (1^{\frac12}S_1)$} & \smash{$K$} & 4 & 884 &  $9.2$ \hskip 5mm \\ 
\smash{$D (1^{\frac12}P_0)$} & \smash{$K$} & 3 & 575 &  $5.5$ \hskip 5mm \\ 
\smash{$D (1^{\frac32}P_1)$} & \smash{$K$} & 3 & 534 &  $3.2$ \hskip 5mm \\ 
\smash{$D (1^{\frac32}P_2)$} & \smash{$K$} & 3 & 489 &  $1.5$ \hskip 5mm \\ 
\smash{$D (1^{\frac12}P_1)$} & \smash{$K$} & 3 & 456 &  $1.6$ \hskip 5mm \\ 
\smash{$D_s (1^{\frac12}S_1)$} & \smash{$\eta$} & 4 & 778 &  $2.2$ \hskip 5mm
\\ 
\smash{$D_s (1^{\frac12}P_0)$} & \smash{$\eta$} & 3 & 409 &  $0.5$ \hskip 5mm
\\ 
\hline \end{tabular}
\end{tabular}

\vskip 5mm 
\begin{tabular}{l}

\begin{tabular}{lr}
$H'=D_s (1^{\frac72}F_4)$ & 
$m = 3.220$ GeV 
\end{tabular} \\ 

\begin{tabular}{|ccrrr|} \hline
$H (n^j\ell_J)$ & $x$ & $\ell_x$ & $p_x$ & $\Gamma_x/(g_A^8)^2$\\ \hline
\smash{$D (1^{\frac12}S_0)$} & \smash{$K$} & 4 & 990 &  $7.2$ \hskip 5mm \\ 
\smash{$D (1^{\frac12}S_1)$} & \smash{$K$} & 4 & 897 &  $5.6$ \hskip 5mm \\ 
\smash{$D (1^{\frac32}P_1)$} & \smash{$K$} & 3 & 552 &  $2.0$ \hskip 5mm \\ 
\smash{$D (1^{\frac32}P_2)$} & \smash{$K$} & 3 & 508 &  $4.0$ \hskip 5mm \\ 
\smash{$D (1^{\frac12}P_1)$} & \smash{$K$} & 3 & 475 &  $3.3$ \hskip 5mm \\ 
\smash{$D_s (1^{\frac12}S_0)$} & \smash{$\eta$} & 4 & 904 &  $2.4$ \hskip 5mm
\\ 
\smash{$D_s (1^{\frac12}S_1)$} & \smash{$\eta$} & 4 & 793 &  $1.4$ \hskip 5mm
\\ 
\hline \end{tabular}
\end{tabular}

\vskip 5mm 
\begin{tabular}{l}

\begin{tabular}{lr}
$H'=D_s (1^{\frac52}F_2)$ & 
$m = 3.224$ GeV 
\end{tabular} \\ 

\begin{tabular}{|ccrrr|} \hline
$H (n^j\ell_J)$ & $x$ & $\ell_x$ & $p_x$ & $\Gamma_x/(g_A^8)^2$\\ \hline
\smash{$D (1^{\frac12}S_0)$} & \smash{$K$} & 2 & 993 &  $5.4$ \hskip 5mm \\ 
\smash{$D (1^{\frac12}S_1)$} & \smash{$K$} & 2 & 900 &  $3.1$ \hskip 5mm \\ 
\smash{$D (1^{\frac32}P_1)$} & \smash{$K$} & 1 & 556 &  $39.1$ \hskip 5mm \\ 
\smash{$D (1^{\frac32}P_1)$} & \smash{$K$} & 3 & 556 &  $0.9$ \hskip 5mm \\ 
\smash{$D (1^{\frac32}P_2)$} & \smash{$K$} & 1 & 512 &  $4.0$ \hskip 5mm \\ 
\smash{$D (1^{\frac32}P_2)$} & \smash{$K$} & 3 & 512 &  $1.1$ \hskip 5mm \\ 
\smash{$D (1^{\frac12}P_1)$} & \smash{$K$} & 3 & 480 &  $0.9$ \hskip 5mm \\ 
\smash{$D_s (1^{\frac12}S_0)$} & \smash{$\eta$} & 2 & 907 &  $3.1$ \hskip 5mm
\\ 
\smash{$D_s (1^{\frac12}S_1)$} & \smash{$\eta$} & 2 & 796 &  $1.5$ \hskip 5mm
\\ 
\smash{$D_s (1^{\frac32}P_1)$} & \smash{$\eta$} & 1 & 372 &  $14.0$ \hskip 5mm
\\ 
\smash{$D_s (1^{\frac32}P_2)$} & \smash{$\eta$} & 1 & 302 &  $0.9$ \hskip 5mm
\\ 
\hline \end{tabular}
\end{tabular}

\vskip 5mm 
\begin{tabular}{l}

\begin{tabular}{lr}
$H'=D_s (1^{\frac52}F_3)$ & 
$m = 3.247$ GeV 
\end{tabular} \\ 

\begin{tabular}{|ccrrr|} \hline
$H (n^j\ell_J)$ & $x$ & $\ell_x$ & $p_x$ & $\Gamma_x/(g_A^8)^2$\\ \hline
\smash{$D (1^{\frac12}S_1)$} & \smash{$K$} & 2 & 918 &  $8.2$ \hskip 5mm \\ 
\smash{$D (1^{\frac12}P_0)$} & \smash{$K$} & 3 & 619 &  $2.2$ \hskip 5mm \\ 
\smash{$D (1^{\frac32}P_1)$} & \smash{$K$} & 3 & 580 &  $2.0$ \hskip 5mm \\ 
\smash{$D (1^{\frac32}P_2)$} & \smash{$K$} & 1 & 536 &  $42.2$ \hskip 5mm \\ 
\smash{$D (1^{\frac32}P_2)$} & \smash{$K$} & 3 & 536 &  $1.5$ \hskip 5mm \\ 
\smash{$D_s (1^{\frac12}S_1)$} & \smash{$\eta$} & 2 & 815 &  $3.9$ \hskip 5mm
\\ 
\smash{$D_s (1^{\frac32}P_2)$} & \smash{$\eta$} & 1 & 339 &  $12.6$ \hskip 5mm
\\ 
\hline \end{tabular}
\end{tabular}

\vskip 5mm 
\begin{tabular}{l}

\begin{tabular}{lr}
$H'=D_s (3^{\frac12}S_0)$ & 
$m = 3.259$ GeV 
\end{tabular} \\ 

\begin{tabular}{|ccrrr|} \hline
$H (n^j\ell_J)$ & $x$ & $\ell_x$ & $p_x$ & $\Gamma_x/(g_A^8)^2$\\ \hline
\smash{$D (1^{\frac12}S_1)$} & \smash{$K$} & 1 & 927 &  $19.0$ \hskip 5mm \\ 
\smash{$D (1^{\frac12}P_0)$} & \smash{$K$} & 0 & 630 &  $2.1$ \hskip 5mm \\ 
\smash{$D (1^{\frac32}P_2)$} & \smash{$K$} & 2 & 549 &  $5.1$ \hskip 5mm \\ 
\smash{$D (2^{\frac12}S_1)$} & \smash{$K$} & 1 & 254 &  $1.0$ \hskip 5mm \\ 
\smash{$D_s (1^{\frac12}S_1)$} & \smash{$\eta$} & 1 & 825 &  $6.8$ \hskip 5mm
\\ 
\smash{$D_s (1^{\frac12}P_0)$} & \smash{$\eta$} & 0 & 478 &  $0.9$ \hskip 5mm
\\ 
\hline \end{tabular}
\end{tabular}

\vskip 5mm 
\begin{tabular}{l}

\begin{tabular}{lr}
$H'=D_s (3^{\frac12}S_1)$ & 
$m = 3.345$ GeV 
\end{tabular} \\ 

\begin{tabular}{|ccrrr|} \hline
$H (n^j\ell_J)$ & $x$ & $\ell_x$ & $p_x$ & $\Gamma_x/(g_A^8)^2$\\ \hline
\smash{$D (1^{\frac12}S_0)$} & \smash{$K$} & 1 & 1080 &  $28.0$ \hskip 5mm \\ 
\smash{$D (1^{\frac12}S_1)$} & \smash{$K$} & 1 & 993 &  $26.6$ \hskip 5mm \\ 
\smash{$D (1^{\frac32}P_1)$} & \smash{$K$} & 2 & 675 &  $12.2$ \hskip 5mm \\ 
\smash{$D (1^{\frac32}P_2)$} & \smash{$K$} & 2 & 635 &  $8.1$ \hskip 5mm \\ 
\smash{$D (1^{\frac12}P_1)$} & \smash{$K$} & 0 & 607 &  $1.8$ \hskip 5mm \\ 
\smash{$D (2^{\frac12}S_0)$} & \smash{$K$} & 1 & 507 &  $18.8$ \hskip 5mm \\ 
\smash{$D (2^{\frac12}S_1)$} & \smash{$K$} & 1 & 385 &  $8.4$ \hskip 5mm \\ 
\smash{$D_s (1^{\frac12}S_0)$} & \smash{$\eta$} & 1 & 1001 &  $13.6$ \hskip
5mm \\ 
\smash{$D_s (1^{\frac12}S_1)$} & \smash{$\eta$} & 1 & 896 &  $10.6$ \hskip 5mm
\\ 
\smash{$D_s (1^{\frac32}P_1)$} & \smash{$\eta$} & 2 & 523 &  $1.7$ \hskip 5mm
\\ 
\smash{$D_s (2^{\frac12}S_0)$} & \smash{$\eta$} & 1 & 309 &  $1.7$ \hskip 5mm
\\ 
\hline \end{tabular}
\end{tabular}

\vskip 5mm 
\begin{tabular}{l}

\begin{tabular}{lr}
$H'=B (2^{\frac12}S_0)$ & 
$m = 5.886$ GeV 
\end{tabular} \\ 

\begin{tabular}{|ccrrr|} \hline
$H (n^j\ell_J)$ & $x$ & $\ell_x$ & $p_x$ & $\Gamma_x/(g_A^8)^2$\\ \hline
\smash{$B (1^{\frac12}S_1)$} & \smash{$\pi$} & 1 & 519 &  $21.9$ \hskip 5mm \\
\smash{$B (1^{\frac12}P_0)$} & \smash{$\pi$} & 0 & 114 &  $4.9$ \hskip 5mm \\ 
\hline \end{tabular}
\end{tabular}

\vskip 5mm 
\begin{tabular}{l}

\begin{tabular}{lr}
$H'=B (2^{\frac12}S_1)$ & 
$m = 5.920$ GeV 
\end{tabular} \\ 

\begin{tabular}{|ccrrr|} \hline
$H (n^j\ell_J)$ & $x$ & $\ell_x$ & $p_x$ & $\Gamma_x/(g_A^8)^2$\\ \hline
\smash{$B (1^{\frac12}S_0)$} & \smash{$\pi$} & 1 & 591 &  $19.2$ \hskip 5mm \\
\smash{$B (1^{\frac12}S_1)$} & \smash{$\pi$} & 1 & 550 &  $22.9$ \hskip 5mm \\
\smash{$B (1^{\frac32}P_1)$} & \smash{$\pi$} & 2 & 167 &  $0.5$ \hskip 5mm \\ 
\smash{$B (1^{\frac12}P_1)$} & \smash{$\pi$} & 0 & 109 &  $4.6$ \hskip 5mm \\ 
\hline \end{tabular}
\end{tabular}

\vskip 5mm 
\begin{tabular}{l}

\begin{tabular}{lr}
$H'=B (1^{\frac52}D_2)$ & 
$m = 5.985$ GeV 
\end{tabular} \\ 

\begin{tabular}{|ccrrr|} \hline
$H (n^j\ell_J)$ & $x$ & $\ell_x$ & $p_x$ & $\Gamma_x/(g_A^8)^2$\\ \hline
\smash{$B (1^{\frac12}S_1)$} & \smash{$\pi$} & 3 & 611 &  $17.4$ \hskip 5mm \\
\smash{$B (1^{\frac32}P_1)$} & \smash{$\pi$} & 2 & 244 &  $1.8$ \hskip 5mm \\ 
\smash{$B (1^{\frac12}P_0)$} & \smash{$\pi$} & 2 & 237 &  $1.6$ \hskip 5mm \\ 
\smash{$B (1^{\frac32}P_2)$} & \smash{$\pi$} & 2 & 228 &  $0.7$ \hskip 5mm \\ 
\smash{$B (1^{\frac12}P_1)$} & \smash{$\pi$} & 2 & 195 &  $0.5$ \hskip 5mm \\ 
\hline \end{tabular}
\end{tabular}

\vskip 5mm 
\begin{tabular}{l}

\begin{tabular}{lr}
$H'=B (1^{\frac52}D_3)$ & 
$m = 5.993$ GeV 
\end{tabular} \\ 

\begin{tabular}{|ccrrr|} \hline
$H (n^j\ell_J)$ & $x$ & $\ell_x$ & $p_x$ & $\Gamma_x/(g_A^8)^2$\\ \hline
\smash{$B (1^{\frac12}S_0)$} & \smash{$\pi$} & 3 & 658 &  $11.1$ \hskip 5mm \\
\smash{$B (1^{\frac12}S_1)$} & \smash{$\pi$} & 3 & 618 &  $10.6$ \hskip 5mm \\
\smash{$B (1^{\frac32}P_1)$} & \smash{$\pi$} & 2 & 252 &  $0.6$ \hskip 5mm \\ 
\smash{$B (1^{\frac32}P_2)$} & \smash{$\pi$} & 2 & 237 &  $2.2$ \hskip 5mm \\ 
\smash{$B (1^{\frac12}P_1)$} & \smash{$\pi$} & 2 & 204 &  $1.3$ \hskip 5mm \\ 
\hline \end{tabular}
\end{tabular}

\vskip 5mm 
\begin{tabular}{l}

\begin{tabular}{lr}
$H'=B (1^{\frac32}D_1)$ & 
$m = 6.025$ GeV 
\end{tabular} \\ 

\begin{tabular}{|ccrrr|} \hline
$H (n^j\ell_J)$ & $x$ & $\ell_x$ & $p_x$ & $\Gamma_x/(g_A^8)^2$\\ \hline
\smash{$B (1^{\frac12}S_0)$} & \smash{$\eta$} & 1 & 475 &  $2.8$ \hskip 5mm \\
\smash{$B (1^{\frac12}S_0)$} & \smash{$\pi$} & 1 & 687 &  $18.2$ \hskip 5mm \\
\smash{$B (1^{\frac12}S_1)$} & \smash{$\pi$} & 1 & 647 &  $7.5$ \hskip 5mm \\ 
\smash{$B (1^{\frac32}P_1)$} & \smash{$\pi$} & 0 & 286 &  $81.4$ \hskip 5mm \\
\smash{$B (1^{\frac32}P_1)$} & \smash{$\pi$} & 2 & 286 &  $1.4$ \hskip 5mm \\ 
\smash{$B_s (1^{\frac12}S_0)$} & \smash{$K$} & 1 & 403 &  $7.2$ \hskip 5mm \\ 
\smash{$B_s (1^{\frac12}S_1)$} & \smash{$K$} & 1 & 330 &  $2.0$ \hskip 5mm \\ 
\hline \end{tabular}
\end{tabular}

\vskip 5mm 
\begin{tabular}{l}

\begin{tabular}{lr}
$H'=B (1^{\frac32}D_2)$ & 
$m = 6.037$ GeV 
\end{tabular} \\ 

\begin{tabular}{|ccrrr|} \hline
$H (n^j\ell_J)$ & $x$ & $\ell_x$ & $p_x$ & $\Gamma_x/(g_A^8)^2$\\ \hline
\smash{$B (1^{\frac12}S_1)$} & \smash{$\eta$} & 1 & 431 &  $3.3$ \hskip 5mm \\
\smash{$B (1^{\frac12}S_1)$} & \smash{$\pi$} & 1 & 658 &  $23.9$ \hskip 5mm \\
\smash{$B (1^{\frac32}P_1)$} & \smash{$\pi$} & 2 & 299 &  $1.3$ \hskip 5mm \\ 
\smash{$B (1^{\frac32}P_2)$} & \smash{$\pi$} & 0 & 284 &  $81.1$ \hskip 5mm \\
\smash{$B (1^{\frac32}P_2)$} & \smash{$\pi$} & 2 & 284 &  $2.0$ \hskip 5mm \\ 
\smash{$B_s (1^{\frac12}S_1)$} & \smash{$K$} & 1 & 350 &  $7.1$ \hskip 5mm \\ 
\hline \end{tabular}
\end{tabular}

\vskip 5mm 
\begin{tabular}{l}

\begin{tabular}{lr}
$H'=B (2^{\frac12}P_0)$ & 
$m = 6.163$ GeV 
\end{tabular} \\ 

\begin{tabular}{|ccrrr|} \hline
$H (n^j\ell_J)$ & $x$ & $\ell_x$ & $p_x$ & $\Gamma_x/(g_A^8)^2$\\ \hline
\smash{$B (1^{\frac12}S_0)$} & \smash{$\eta$} & 0 & 643 &  $9.9$ \hskip 5mm \\
\smash{$B (1^{\frac12}S_0)$} & \smash{$\pi$} & 0 & 810 &  $97.4$ \hskip 5mm \\
\smash{$B (1^{\frac32}P_1)$} & \smash{$\pi$} & 1 & 425 &  $21.1$ \hskip 5mm \\
\smash{$B (1^{\frac12}P_1)$} & \smash{$\pi$} & 1 & 383 &  $52.9$ \hskip 5mm \\
\smash{$B (2^{\frac12}S_0)$} & \smash{$\pi$} & 0 & 233 &  $39.6$ \hskip 5mm \\
\smash{$B_s (1^{\frac12}S_0)$} & \smash{$K$} & 0 & 576 &  $51.0$ \hskip 5mm \\
\hline \end{tabular}
\end{tabular}

\vskip 5mm 
\begin{tabular}{l}

\begin{tabular}{lr}
$H'=B (2^{\frac32}P_1)$ & 
$m = 6.175$ GeV 
\end{tabular} \\ 

\begin{tabular}{|ccrrr|} \hline
$H (n^j\ell_J)$ & $x$ & $\ell_x$ & $p_x$ & $\Gamma_x/(g_A^8)^2$\\ \hline
\smash{$B (1^{\frac12}S_1)$} & \smash{$\eta$} & 2 & 606 &  $1.5$ \hskip 5mm \\
\smash{$B (1^{\frac12}S_1)$} & \smash{$\pi$} & 2 & 782 &  $59.6$ \hskip 5mm \\
\smash{$B (1^{\frac32}P_1)$} & \smash{$\pi$} & 1 & 437 &  $1.6$ \hskip 5mm \\ 
\smash{$B (1^{\frac12}P_0)$} & \smash{$\pi$} & 1 & 431 &  $2.4$ \hskip 5mm \\ 
\smash{$B (1^{\frac32}P_2)$} & \smash{$\pi$} & 3 & 423 &  $2.6$ \hskip 5mm \\ 
\smash{$B (1^{\frac12}P_1)$} & \smash{$\pi$} & 1 & 395 &  $3.4$ \hskip 5mm \\ 
\smash{$B (2^{\frac12}S_1)$} & \smash{$\pi$} & 2 & 209 &  $2.3$ \hskip 5mm \\ 
\smash{$B (1^{\frac32}D_1)$} & \smash{$\pi$} & 0 & 55 &  $1.4$ \hskip 5mm \\ 
\smash{$B_s (1^{\frac12}S_1)$} & \smash{$K$} & 2 & 534 &  $4.6$ \hskip 5mm \\ 
\hline \end{tabular}
\end{tabular}

\vskip 5mm 
\begin{tabular}{l}

\begin{tabular}{lr}
$H'=B (2^{\frac32}P_2)$ & 
$m = 6.188$ GeV 
\end{tabular} \\ 

\begin{tabular}{|ccrrr|} \hline
$H (n^j\ell_J)$ & $x$ & $\ell_x$ & $p_x$ & $\Gamma_x/(g_A^8)^2$\\ \hline
\smash{$B (1^{\frac12}S_0)$} & \smash{$\eta$} & 2 & 671 &  $1.2$ \hskip 5mm \\
\smash{$B (1^{\frac12}S_0)$} & \smash{$\pi$} & 2 & 832 &  $36.0$ \hskip 5mm \\
\smash{$B (1^{\frac12}S_1)$} & \smash{$\eta$} & 2 & 621 &  $1.1$ \hskip 5mm \\
\smash{$B (1^{\frac12}S_1)$} & \smash{$\pi$} & 2 & 793 &  $39.7$ \hskip 5mm \\
\smash{$B (1^{\frac32}P_1)$} & \smash{$\pi$} & 3 & 449 &  $2.2$ \hskip 5mm \\ 
\smash{$B (1^{\frac32}P_2)$} & \smash{$\pi$} & 3 & 436 &  $1.2$ \hskip 5mm \\ 
\smash{$B (1^{\frac12}P_1)$} & \smash{$\pi$} & 1 & 408 &  $5.7$ \hskip 5mm \\ 
\smash{$B (2^{\frac12}S_0)$} & \smash{$\pi$} & 2 & 261 &  $2.4$ \hskip 5mm \\ 
\smash{$B (2^{\frac12}S_1)$} & \smash{$\pi$} & 2 & 224 &  $1.9$ \hskip 5mm \\ 
\smash{$B (1^{\frac32}D_2)$} & \smash{$\pi$} & 0 & 57 &  $1.5$ \hskip 5mm \\ 
\smash{$B_s (1^{\frac12}S_0)$} & \smash{$K$} & 2 & 605 &  $4.2$ \hskip 5mm \\ 
\smash{$B_s (1^{\frac12}S_1)$} & \smash{$K$} & 2 & 550 &  $3.4$ \hskip 5mm \\ 
\hline \end{tabular}
\end{tabular}

\vskip 5mm 
\begin{tabular}{l}

\begin{tabular}{lr}
$H'=B (2^{\frac12}P_1)$ & 
$m = 6.194$ GeV 
\end{tabular} \\ 

\begin{tabular}{|ccrrr|} \hline
$H (n^j\ell_J)$ & $x$ & $\ell_x$ & $p_x$ & $\Gamma_x/(g_A^8)^2$\\ \hline
\smash{$B (1^{\frac12}S_1)$} & \smash{$\eta$} & 0 & 629 &  $9.2$ \hskip 5mm \\
\smash{$B (1^{\frac12}S_1)$} & \smash{$\pi$} & 0 & 799 &  $93.2$ \hskip 5mm \\
\smash{$B (1^{\frac32}P_1)$} & \smash{$\pi$} & 1 & 455 &  $6.1$ \hskip 5mm \\ 
\smash{$B (1^{\frac12}P_0)$} & \smash{$\pi$} & 1 & 449 &  $30.6$ \hskip 5mm \\
\smash{$B (1^{\frac32}P_2)$} & \smash{$\pi$} & 1 & 442 &  $22.6$ \hskip 5mm \\
\smash{$B (1^{\frac12}P_1)$} & \smash{$\pi$} & 1 & 414 &  $39.5$ \hskip 5mm \\
\smash{$B (2^{\frac12}S_1)$} & \smash{$\pi$} & 0 & 231 &  $38.7$ \hskip 5mm \\
\smash{$B_s (1^{\frac12}S_1)$} & \smash{$K$} & 0 & 557 &  $46.3$ \hskip 5mm \\
\hline \end{tabular}
\end{tabular}

\vskip 5mm 
\begin{tabular}{l}

\begin{tabular}{lr}
$H'=B (1^{\frac72}F_3)$ & 
$m = 6.220$ GeV 
\end{tabular} \\ 

\begin{tabular}{|ccrrr|} \hline
$H (n^j\ell_J)$ & $x$ & $\ell_x$ & $p_x$ & $\Gamma_x/(g_A^8)^2$\\ \hline
\smash{$B (1^{\frac12}S_1)$} & \smash{$\eta$} & 4 & 659 &  $0.3$ \hskip 5mm \\
\smash{$B (1^{\frac12}S_1)$} & \smash{$\pi$} & 4 & 822 &  $11.7$ \hskip 5mm \\
\smash{$B (1^{\frac32}P_1)$} & \smash{$\pi$} & 3 & 481 &  $3.5$ \hskip 5mm \\ 
\smash{$B (1^{\frac12}P_0)$} & \smash{$\pi$} & 3 & 475 &  $3.8$ \hskip 5mm \\ 
\smash{$B (1^{\frac32}P_2)$} & \smash{$\pi$} & 3 & 468 &  $1.9$ \hskip 5mm \\ 
\smash{$B (1^{\frac12}P_1)$} & \smash{$\pi$} & 3 & 440 &  $2.1$ \hskip 5mm \\ 
\smash{$B (1^{\frac52}D_2)$} & \smash{$\pi$} & 2 & 186 &  $0.8$ \hskip 5mm \\ 
\smash{$B (1^{\frac32}D_1)$} & \smash{$\pi$} & 2 & 135 &  $0.3$ \hskip 5mm \\ 
\smash{$B_s (1^{\frac12}S_1)$} & \smash{$K$} & 4 & 588 &  $0.5$ \hskip 5mm \\ 
\hline \end{tabular}
\end{tabular}

\vskip 5mm 
\begin{tabular}{l}

\begin{tabular}{lr}
$H'=B (1^{\frac72}F_4)$ & 
$m = 6.226$ GeV 
\end{tabular} \\ 

\begin{tabular}{|ccrrr|} \hline
$H (n^j\ell_J)$ & $x$ & $\ell_x$ & $p_x$ & $\Gamma_x/(g_A^8)^2$\\ \hline
\smash{$B (1^{\frac12}S_0)$} & \smash{$\pi$} & 4 & 865 &  $7.0$ \hskip 5mm \\ 
\smash{$B (1^{\frac12}S_1)$} & \smash{$\pi$} & 4 & 827 &  $6.7$ \hskip 5mm \\ 
\smash{$B (1^{\frac32}P_1)$} & \smash{$\pi$} & 3 & 486 &  $1.5$ \hskip 5mm \\ 
\smash{$B (1^{\frac32}P_2)$} & \smash{$\pi$} & 3 & 473 &  $4.7$ \hskip 5mm \\ 
\smash{$B (1^{\frac12}P_1)$} & \smash{$\pi$} & 3 & 445 &  $4.4$ \hskip 5mm \\ 
\smash{$B (1^{\frac52}D_3)$} & \smash{$\pi$} & 2 & 183 &  $0.8$ \hskip 5mm \\ 
\smash{$B_s (1^{\frac12}S_0)$} & \smash{$K$} & 4 & 647 &  $0.5$ \hskip 5mm \\ 
\smash{$B_s (1^{\frac12}S_1)$} & \smash{$K$} & 4 & 594 &  $0.3$ \hskip 5mm \\ 
\hline \end{tabular}
\end{tabular}

\vskip 5mm 
\begin{tabular}{l}

\begin{tabular}{lr}
$H'=B (1^{\frac52}F_2)$ & 
$m = 6.264$ GeV 
\end{tabular} \\ 

\begin{tabular}{|ccrrr|} \hline
$H (n^j\ell_J)$ & $x$ & $\ell_x$ & $p_x$ & $\Gamma_x/(g_A^8)^2$\\ \hline
\smash{$B (1^{\frac12}S_0)$} & \smash{$\pi$} & 2 & 898 &  $3.7$ \hskip 5mm \\ 
\smash{$B (1^{\frac12}S_1)$} & \smash{$\pi$} & 2 & 860 &  $2.0$ \hskip 5mm \\ 
\smash{$B (1^{\frac32}P_1)$} & \smash{$\pi$} & 1 & 522 &  $18.9$ \hskip 5mm \\
\smash{$B (1^{\frac32}P_2)$} & \smash{$\pi$} & 1 & 509 &  $2.0$ \hskip 5mm \\ 
\smash{$B (1^{\frac32}P_2)$} & \smash{$\pi$} & 3 & 509 &  $1.3$ \hskip 5mm \\ 
\smash{$B (1^{\frac12}P_1)$} & \smash{$\pi$} & 3 & 482 &  $1.3$ \hskip 5mm \\ 
\smash{$B (1^{\frac52}D_2)$} & \smash{$\pi$} & 0 & 236 &  $51.2$ \hskip 5mm \\
\smash{$B (1^{\frac52}D_2)$} & \smash{$\pi$} & 2 & 236 &  $1.6$ \hskip 5mm \\ 
\smash{$B_s (1^{\frac12}S_0)$} & \smash{$K$} & 2 & 688 &  $1.9$ \hskip 5mm \\ 
\smash{$B_s (1^{\frac12}S_1)$} & \smash{$K$} & 2 & 637 &  $0.9$ \hskip 5mm \\ 
\hline \end{tabular}
\end{tabular}

\vskip 5mm 
\begin{tabular}{l}

\begin{tabular}{lr}
$H'=B (1^{\frac52}F_3)$ & 
$m = 6.271$ GeV 
\end{tabular} \\ 

\begin{tabular}{|ccrrr|} \hline
$H (n^j\ell_J)$ & $x$ & $\ell_x$ & $p_x$ & $\Gamma_x/(g_A^8)^2$\\ \hline
\smash{$B (1^{\frac12}S_1)$} & \smash{$\eta$} & 2 & 714 &  $1.0$ \hskip 5mm \\
\smash{$B (1^{\frac12}S_1)$} & \smash{$\pi$} & 2 & 866 &  $5.2$ \hskip 5mm \\ 
\smash{$B (1^{\frac32}P_1)$} & \smash{$\pi$} & 3 & 529 &  $1.2$ \hskip 5mm \\ 
\smash{$B (1^{\frac12}P_0)$} & \smash{$\pi$} & 3 & 523 &  $1.0$ \hskip 5mm \\ 
\smash{$B (1^{\frac32}P_2)$} & \smash{$\pi$} & 1 & 516 &  $20.4$ \hskip 5mm \\
\smash{$B (1^{\frac32}P_2)$} & \smash{$\pi$} & 3 & 516 &  $1.5$ \hskip 5mm \\ 
\smash{$B (1^{\frac52}D_3)$} & \smash{$\pi$} & 0 & 235 &  $51.0$ \hskip 5mm \\
\smash{$B (1^{\frac52}D_3)$} & \smash{$\pi$} & 2 & 235 &  $1.7$ \hskip 5mm \\ 
\smash{$B_s (1^{\frac12}S_1)$} & \smash{$K$} & 2 & 644 &  $2.3$ \hskip 5mm \\ 
\hline \end{tabular}
\end{tabular}

\vskip 5mm 
\begin{tabular}{l}

\begin{tabular}{lr}
$H'=B (3^{\frac12}S_0)$ & 
$m = 6.320$ GeV 
\end{tabular} \\ 

\begin{tabular}{|ccrrr|} \hline
$H (n^j\ell_J)$ & $x$ & $\ell_x$ & $p_x$ & $\Gamma_x/(g_A^8)^2$\\ \hline
\smash{$B (1^{\frac12}S_1)$} & \smash{$\eta$} & 1 & 766 &  $1.1$ \hskip 5mm \\
\smash{$B (1^{\frac12}S_1)$} & \smash{$\pi$} & 1 & 908 &  $49.5$ \hskip 5mm \\
\smash{$B (1^{\frac12}P_0)$} & \smash{$\pi$} & 0 & 569 &  $1.8$ \hskip 5mm \\ 
\smash{$B (1^{\frac32}P_2)$} & \smash{$\pi$} & 2 & 562 &  $11.0$ \hskip 5mm \\
\smash{$B (2^{\frac12}S_1)$} & \smash{$\pi$} & 1 & 363 &  $9.5$ \hskip 5mm \\ 
\smash{$B (2^{\frac12}P_0)$} & \smash{$\pi$} & 0 & 72 &  $3.5$ \hskip 5mm \\ 
\smash{$B_s (1^{\frac12}S_1)$} & \smash{$K$} & 1 & 697 &  $5.1$ \hskip 5mm \\ 
\hline \end{tabular}
\end{tabular}

\vskip 5mm 
\begin{tabular}{l}

\begin{tabular}{lr}
$H'=B (3^{\frac12}S_1)$ & 
$m = 6.347$ GeV 
\end{tabular} \\ 

\begin{tabular}{|ccrrr|} \hline
$H (n^j\ell_J)$ & $x$ & $\ell_x$ & $p_x$ & $\Gamma_x/(g_A^8)^2$\\ \hline
\smash{$B (1^{\frac12}S_0)$} & \smash{$\pi$} & 1 & 970 &  $31.3$ \hskip 5mm \\
\smash{$B (1^{\frac12}S_1)$} & \smash{$\pi$} & 1 & 932 &  $43.0$ \hskip 5mm \\
\smash{$B (1^{\frac32}P_1)$} & \smash{$\pi$} & 2 & 600 &  $9.4$ \hskip 5mm \\ 
\smash{$B (1^{\frac32}P_2)$} & \smash{$\pi$} & 2 & 587 &  $7.9$ \hskip 5mm \\ 
\smash{$B (1^{\frac12}P_1)$} & \smash{$\pi$} & 0 & 560 &  $1.7$ \hskip 5mm \\ 
\smash{$B (2^{\frac12}S_0)$} & \smash{$\pi$} & 1 & 423 &  $9.3$ \hskip 5mm \\ 
\smash{$B (2^{\frac12}S_1)$} & \smash{$\pi$} & 1 & 390 &  $10.8$ \hskip 5mm \\
\smash{$B (2^{\frac12}P_1)$} & \smash{$\pi$} & 0 & 64 &  $3.0$ \hskip 5mm \\ 
\smash{$B_s (1^{\frac12}S_0)$} & \smash{$K$} & 1 & 775 &  $5.0$ \hskip 5mm \\ 
\smash{$B_s (1^{\frac12}S_1)$} & \smash{$K$} & 1 & 726 &  $5.2$ \hskip 5mm \\ 
\hline \end{tabular}
\end{tabular}

\vskip 5mm 
\begin{tabular}{l}

\begin{tabular}{lr}
$H'=B_s (2^{\frac12}S_0)$ & 
$m = 5.985$ GeV 
\end{tabular} \\ 

\begin{tabular}{|ccrrr|} \hline
$H (n^j\ell_J)$ & $x$ & $\ell_x$ & $p_x$ & $\Gamma_x/(g_A^8)^2$\\ \hline
\smash{$B (1^{\frac12}S_1)$} & \smash{$K$} & 1 & 416 &  $3.15$ \hskip 5mm \\ 
\hline \end{tabular}
\end{tabular}

\vskip 5mm 
\begin{tabular}{l}

\begin{tabular}{lr}
$H'=B_s (2^{\frac12}S_1)$ & 
$m = 6.019$ GeV 
\end{tabular} \\ 

\begin{tabular}{|ccrrr|} \hline
$H (n^j\ell_J)$ & $x$ & $\ell_x$ & $p_x$ & $\Gamma_x/(g_A^8)^2$\\ \hline
\smash{$B (1^{\frac12}S_0)$} & \smash{$K$} & 1 & 517 &  $5.6$ \hskip 5mm \\ 
\smash{$B (1^{\frac12}S_1)$} & \smash{$K$} & 1 & 462 &  $4.9$ \hskip 5mm \\ 
\smash{$B_s (1^{\frac12}S_0)$} & \smash{$\eta$} & 1 & 325 &  $0.4$ \hskip 5mm
\\ 
\hline \end{tabular}
\end{tabular}

\vskip 5mm 
\begin{tabular}{l}

\begin{tabular}{lr}
$H'=B_s (1^{\frac52}D_2)$ & 
$m = 6.095$ GeV 
\end{tabular} \\ 

\begin{tabular}{|ccrrr|} \hline
$H (n^j\ell_J)$ & $x$ & $\ell_x$ & $p_x$ & $\Gamma_x/(g_A^8)^2$\\ \hline
\smash{$B (1^{\frac12}S_1)$} & \smash{$K$} & 3 & 555 &  $6.62$ \hskip 5mm \\ 
\smash{$B_s (1^{\frac12}S_1)$} & \smash{$\eta$} & 3 & 370 &  $0.30$ \hskip 5mm
\\ 
\hline \end{tabular}
\end{tabular}

\vskip 5mm 
\begin{tabular}{l}

\begin{tabular}{lr}
$H'=B_s (1^{\frac52}D_3)$ & 
$m = 6.103$ GeV 
\end{tabular} \\ 

\begin{tabular}{|ccrrr|} \hline
$H (n^j\ell_J)$ & $x$ & $\ell_x$ & $p_x$ & $\Gamma_x/(g_A^8)^2$\\ \hline
\smash{$B (1^{\frac12}S_0)$} & \smash{$K$} & 3 & 614 &  $5.03$ \hskip 5mm \\ 
\smash{$B (1^{\frac12}S_1)$} & \smash{$K$} & 3 & 564 &  $4.17$ \hskip 5mm \\ 
\smash{$B_s (1^{\frac12}S_0)$} & \smash{$\eta$} & 3 & 453 &  $0.46$ \hskip 5mm
\\ 
\smash{$B_s (1^{\frac12}S_1)$} & \smash{$\eta$} & 3 & 383 &  $0.21$ \hskip 5mm
\\ 
\hline \end{tabular}
\end{tabular}

\vskip 5mm 
\begin{tabular}{l}

\begin{tabular}{lr}
$H'=B_s (1^{\frac32}D_1)$ & 
$m = 6.127$ GeV 
\end{tabular} \\ 

\begin{tabular}{|ccrrr|} \hline
$H (n^j\ell_J)$ & $x$ & $\ell_x$ & $p_x$ & $\Gamma_x/(g_A^8)^2$\\ \hline
\smash{$B (1^{\frac12}S_0)$} & \smash{$K$} & 1 & 641 &  $26.9$ \hskip 5mm \\ 
\smash{$B (1^{\frac12}S_1)$} & \smash{$K$} & 1 & 592 &  $11.5$ \hskip 5mm \\ 
\smash{$B_s (1^{\frac12}S_0)$} & \smash{$\eta$} & 1 & 486 &  $10.1$ \hskip 5mm
\\ 
\smash{$B_s (1^{\frac12}S_1)$} & \smash{$\eta$} & 1 & 420 &  $3.4$ \hskip 5mm
\\ 
\hline \end{tabular}
\end{tabular}

\vskip 5mm 
\begin{tabular}{l}

\begin{tabular}{lr}
$H'=B_s (1^{\frac32}D_2)$ & 
$m = 6.140$ GeV 
\end{tabular} \\ 

\begin{tabular}{|ccrrr|} \hline
$H (n^j\ell_J)$ & $x$ & $\ell_x$ & $p_x$ & $\Gamma_x/(g_A^8)^2$\\ \hline
\smash{$B (1^{\frac12}S_1)$} & \smash{$K$} & 1 & 607 &  $36.2$ \hskip 5mm \\ 
\smash{$B_s (1^{\frac12}S_1)$} & \smash{$\eta$} & 1 & 438 &  $11.5$ \hskip 5mm
\\ 
\hline \end{tabular}
\end{tabular}

\vskip 5mm 
\begin{tabular}{l}

\begin{tabular}{lr}
$H'=B_s (2^{\frac12}P_0)$ & 
$m = 6.264$ GeV 
\end{tabular} \\ 

\begin{tabular}{|ccrrr|} \hline
$H (n^j\ell_J)$ & $x$ & $\ell_x$ & $p_x$ & $\Gamma_x/(g_A^8)^2$\\ \hline
\smash{$B (1^{\frac12}S_0)$} & \smash{$K$} & 0 & 785 &  $77.7$ \hskip 5mm \\ 
\smash{$B (1^{\frac32}P_1)$} & \smash{$K$} & 1 & 262 &  $20.6$ \hskip 5mm \\ 
\smash{$B (1^{\frac12}P_1)$} & \smash{$K$} & 1 & 164 &  $3.8$ \hskip 5mm \\ 
\smash{$B_s (1^{\frac12}S_0)$} & \smash{$\eta$} & 0 & 652 &  $40.5$ \hskip 5mm
\\ 
\hline \end{tabular}
\end{tabular}

\vskip 5mm 
\begin{tabular}{l}

\begin{tabular}{lr}
$H'=B_s (2^{\frac32}P_1)$ & 
$m = 6.278$ GeV 
\end{tabular} \\ 

\begin{tabular}{|ccrrr|} \hline
$H (n^j\ell_J)$ & $x$ & $\ell_x$ & $p_x$ & $\Gamma_x/(g_A^8)^2$\\ \hline
\smash{$B (1^{\frac12}S_1)$} & \smash{$K$} & 0 & 755 & $(\ast)\, $ $0.7$
\hskip 5mm \\ 
\smash{$B (1^{\frac12}S_1)$} & \smash{$K$} & 2 & 755 &  $28.7$ \hskip 5mm \\ 
\smash{$B (1^{\frac12}P_0)$} & \smash{$K$} & 1 & 278 &  $3.3$ \hskip 5mm \\ 
\smash{$B (1^{\frac12}P_1)$} & \smash{$K$} & 1 & 201 &  $1.4$ \hskip 5mm \\ 
\smash{$B_s (1^{\frac12}S_1)$} & \smash{$\eta$} & 2 & 614 &  $5.8$ \hskip 5mm
\\ 
\hline \end{tabular}
\end{tabular}

\vskip 5mm 
\begin{tabular}{l}

\begin{tabular}{lr}
$H'=B_s (2^{\frac32}P_2)$ & 
$m = 6.292$ GeV 
\end{tabular} \\ 

\begin{tabular}{|ccrrr|} \hline
$H (n^j\ell_J)$ & $x$ & $\ell_x$ & $p_x$ & $\Gamma_x/(g_A^8)^2$\\ \hline
\smash{$B (1^{\frac12}S_0)$} & \smash{$K$} & 2 & 813 &  $19.2$ \hskip 5mm \\ 
\smash{$B (1^{\frac12}S_1)$} & \smash{$K$} & 2 & 769 &  $19.8$ \hskip 5mm \\ 
\smash{$B (1^{\frac12}P_1)$} & \smash{$K$} & 1 & 232 &  $4.2$ \hskip 5mm \\ 
\smash{$B_s (1^{\frac12}S_0)$} & \smash{$\eta$} & 2 & 683 &  $4.8$ \hskip 5mm
\\ 
\smash{$B_s (1^{\frac12}S_1)$} & \smash{$\eta$} & 2 & 630 &  $4.2$ \hskip 5mm
\\ 
\hline \end{tabular}
\end{tabular}

\vskip 5mm 
\begin{tabular}{l}

\begin{tabular}{lr}
$H'=B_s (2^{\frac12}P_1)$ & 
$m = 6.296$ GeV 
\end{tabular} \\ 

\begin{tabular}{|ccrrr|} \hline
$H (n^j\ell_J)$ & $x$ & $\ell_x$ & $p_x$ & $\Gamma_x/(g_A^8)^2$\\ \hline
\smash{$B (1^{\frac12}S_1)$} & \smash{$K$} & 0 & 773 &  $73.8$ \hskip 5mm \\ 
\smash{$B (1^{\frac32}P_1)$} & \smash{$K$} & 1 & 319 &  $7.0$ \hskip 5mm \\ 
\smash{$B (1^{\frac12}P_0)$} & \smash{$K$} & 1 & 309 &  $9.3$ \hskip 5mm \\ 
\smash{$B (1^{\frac32}P_2)$} & \smash{$K$} & 1 & 296 &  $24.5$ \hskip 5mm \\ 
\smash{$B (1^{\frac12}P_1)$} & \smash{$K$} & 1 & 241 &  $6.0$ \hskip 5mm \\ 
\smash{$B_s (1^{\frac12}S_1)$} & \smash{$\eta$} & 0 & 634 &  $37.1$ \hskip 5mm
\\ 
\hline \end{tabular}
\end{tabular}

\vskip 5mm 
\begin{tabular}{l}

\begin{tabular}{lr}
$H'=B_s (1^{\frac72}F_3)$ & 
$m = 6.332$ GeV 
\end{tabular} \\ 

\begin{tabular}{|ccrrr|} \hline
$H (n^j\ell_J)$ & $x$ & $\ell_x$ & $p_x$ & $\Gamma_x/(g_A^8)^2$\\ \hline
\smash{$B (1^{\frac12}S_1)$} & \smash{$K$} & 4 & 808 &  $7.07$ \hskip 5mm \\ 
\smash{$B (1^{\frac32}P_1)$} & \smash{$K$} & 3 & 375 &  $0.51$ \hskip 5mm \\ 
\smash{$B (1^{\frac12}P_0)$} & \smash{$K$} & 3 & 366 &  $0.50$ \hskip 5mm \\ 
\smash{$B (1^{\frac32}P_2)$} & \smash{$K$} & 3 & 354 &  $0.22$ \hskip 5mm \\ 
\smash{$B (1^{\frac12}P_1)$} & \smash{$K$} & 3 & 307 &  $0.14$ \hskip 5mm \\ 
\smash{$B_s (1^{\frac12}S_1)$} & \smash{$\eta$} & 4 & 674 &  $1.04$ \hskip 5mm
\\ 
\hline \end{tabular}
\end{tabular}

\vskip 5mm 
\begin{tabular}{l}

\begin{tabular}{lr}
$H'=B_s (1^{\frac72}F_4)$ & 
$m = 6.337$ GeV 
\end{tabular} \\ 

\begin{tabular}{|ccrrr|} \hline
$H (n^j\ell_J)$ & $x$ & $\ell_x$ & $p_x$ & $\Gamma_x/(g_A^8)^2$\\ \hline
\smash{$B (1^{\frac12}S_0)$} & \smash{$K$} & 4 & 857 &  $4.5$ \hskip 5mm \\ 
\smash{$B (1^{\frac12}S_1)$} & \smash{$K$} & 4 & 814 &  $4.1$ \hskip 5mm \\ 
\smash{$B (1^{\frac32}P_1)$} & \smash{$K$} & 3 & 383 &  $0.2$ \hskip 5mm \\ 
\smash{$B (1^{\frac32}P_2)$} & \smash{$K$} & 3 & 363 &  $0.6$ \hskip 5mm \\ 
\smash{$B (1^{\frac12}P_1)$} & \smash{$K$} & 3 & 317 &  $0.3$ \hskip 5mm \\ 
\smash{$B_s (1^{\frac12}S_0)$} & \smash{$\eta$} & 4 & 732 &  $0.8$ \hskip 5mm
\\ 
\smash{$B_s (1^{\frac12}S_1)$} & \smash{$\eta$} & 4 & 681 &  $0.6$ \hskip 5mm
\\ 
\hline \end{tabular}
\end{tabular}

\vskip 5mm 
\begin{tabular}{l}

\begin{tabular}{lr}
$H'=B_s (1^{\frac52}F_2)$ & 
$m = 6.369$ GeV 
\end{tabular} \\ 

\begin{tabular}{|ccrrr|} \hline
$H (n^j\ell_J)$ & $x$ & $\ell_x$ & $p_x$ & $\Gamma_x/(g_A^8)^2$\\ \hline
\smash{$B (1^{\frac12}S_0)$} & \smash{$K$} & 2 & 888 &  $6.0$ \hskip 5mm \\ 
\smash{$B (1^{\frac12}S_1)$} & \smash{$K$} & 2 & 845 &  $3.5$ \hskip 5mm \\ 
\smash{$B (1^{\frac32}P_1)$} & \smash{$K$} & 1 & 428 &  $29.6$ \hskip 5mm \\ 
\smash{$B (1^{\frac32}P_2)$} & \smash{$K$} & 1 & 409 &  $3.0$ \hskip 5mm \\ 
\smash{$B_s (1^{\frac12}S_0)$} & \smash{$\eta$} & 2 & 766 &  $2.4$ \hskip 5mm
\\ 
\smash{$B_s (1^{\frac12}S_1)$} & \smash{$\eta$} & 2 & 715 &  $1.3$ \hskip 5mm
\\ 
\smash{$B_s (1^{\frac32}P_1)$} & \smash{$\eta$} & 1 & 129 &  $0.9$ \hskip 5mm
\\ 
\hline \end{tabular}
\end{tabular}

\vskip 5mm 
\begin{tabular}{l}

\begin{tabular}{lr}
$H'=B_s (1^{\frac52}F_3)$ & 
$m = 6.376$ GeV 
\end{tabular} \\ 

\begin{tabular}{|ccrrr|} \hline
$H (n^j\ell_J)$ & $x$ & $\ell_x$ & $p_x$ & $\Gamma_x/(g_A^8)^2$\\ \hline
\smash{$B (1^{\frac12}S_1)$} & \smash{$K$} & 2 & 853 &  $9.0$ \hskip 5mm \\ 
\smash{$B (1^{\frac32}P_2)$} & \smash{$K$} & 1 & 420 &  $31.6$ \hskip 5mm \\ 
\smash{$B_s (1^{\frac12}S_1)$} & \smash{$\eta$} & 2 & 723 &  $3.3$ \hskip 5mm
\\ 
\hline \end{tabular}
\end{tabular}

\vskip 5mm 
\begin{tabular}{l}

\begin{tabular}{lr}
$H'=B_s (3^{\frac12}S_0)$ & 
$m = 6.421$ GeV 
\end{tabular} \\ 

\begin{tabular}{|ccrrr|} \hline
$H (n^j\ell_J)$ & $x$ & $\ell_x$ & $p_x$ & $\Gamma_x/(g_A^8)^2$\\ \hline
\smash{$B (1^{\frac12}S_1)$} & \smash{$K$} & 1 & 896 &  $16.5$ \hskip 5mm \\ 
\smash{$B (1^{\frac12}P_0)$} & \smash{$K$} & 0 & 489 &  $0.3$ \hskip 5mm \\ 
\smash{$B (1^{\frac32}P_2)$} & \smash{$K$} & 2 & 479 &  $1.7$ \hskip 5mm \\ 
\smash{$B_s (1^{\frac12}S_1)$} & \smash{$\eta$} & 1 & 770 &  $3.9$ \hskip 5mm
\\ 
\hline \end{tabular}
\end{tabular}

\vskip 5mm 
\begin{tabular}{l}

\begin{tabular}{lr}
$H'=B_s (3^{\frac12}S_1)$ & 
$m = 6.449$ GeV 
\end{tabular} \\ 

\begin{tabular}{|ccrrr|} \hline
$H (n^j\ell_J)$ & $x$ & $\ell_x$ & $p_x$ & $\Gamma_x/(g_A^8)^2$\\ \hline
\smash{$B (1^{\frac12}S_0)$} & \smash{$K$} & 1 & 963 &  $13.2$ \hskip 5mm \\ 
\smash{$B (1^{\frac12}S_1)$} & \smash{$K$} & 1 & 922 &  $16.0$ \hskip 5mm \\ 
\smash{$B (1^{\frac32}P_1)$} & \smash{$K$} & 2 & 531 &  $2.2$ \hskip 5mm \\ 
\smash{$B (1^{\frac32}P_2)$} & \smash{$K$} & 2 & 514 &  $1.7$ \hskip 5mm \\ 
\smash{$B_s (1^{\frac12}S_0)$} & \smash{$\eta$} & 1 & 848 &  $3.9$ \hskip 5mm
\\ 
\smash{$B_s (1^{\frac12}S_1)$} & \smash{$\eta$} & 1 & 800 &  $4.1$ \hskip 5mm
\\ 
\hline \end{tabular}
\end{tabular}

} 
\onecolumn


\end{document}